\documentclass[aps,prd,reprint,groupedaddress]{revtex4-1}

\usepackage{graphicx}	
\usepackage[usenames]{color}

\newcommand{\msun}{M_\odot}

\begin{document}

\title{Impact of binary interactions on the diffuse supernova neutrino background} 

\author{Shunsaku Horiuchi}
\email[]{horiuchi@vt.edu}
\affiliation{Center for Neutrino Physics, Department of Physics, Virginia Tech, Blacksburg, Virginia 24061, USA}
\author{Tomoya Kinugawa}
\affiliation{Institute for Cosmic Ray Research, The University of Tokyo, Kashiwa, Chiba, Japan}
\author{Tomoya Takiwaki}
\affiliation{National Astronomical Observatory of Japan, Mitaka, Tokyo 181-8588, Japan}
\author{Koh Takahashi}
\affiliation{Max Planck Institute for Gravitational Physics, D-14476 Potsdam, Germany}
\author{Kei Kotake}
\affiliation{Department of Applied Physics \& Research Institute of Stellar Explosive Phenomena, Fukuoka University, Fukuoka 814-0180, Japan}

\date{\today}

\begin{abstract}
Binary interactions, especially mass transfer and mergers, can strongly influence the evolution of massive stars and change their final properties and the occurrence of supernovae. Here, we investigate how binary interactions affect predictions of the diffuse flux of neutrinos. By performing stellar population syntheses including prescriptions for binary interactions, we show that the resulting detection rates of the diffuse supernova neutrino background is enhanced by 15\%--20\% compared to estimates without binary considerations. A source of significant uncertainty arises due to the presently sparse knowledge of the evolution of rapidly rotating carbon-oxygen cores, especially those created as a result of mergers near the white dwarf to core collapse boundary. The enhancement effect may be as small as a few percent if the effects of rotation in postmerger systems are neglected, or as large as 75\% if trends are extrapolated. Our estimates serve to highlight that binary effects can be important.
\end{abstract}


\maketitle

\section{Introduction}

A star with a sufficiently massive core ends its evolution with a core collapse, where its core, unable to sustain itself against gravity, collapses on dynamical time scales to a compact object. During this violent collapse a shock can be formed, which can lead to a core-collapse supernova explosion (CCSN) if the shock reaches the photosphere of the star (see, e.g., reviews \cite{Langanke:2002ab,Mezzacappa:2005ju,Kotake:2005zn,Woosley:2006ie,Burrows:2012ew,Janka:2012wk,Foglizzo:2015dma,Janka:2017vlw,Burrows:2020qrp}). The collapsed core emits a copious number of neutrinos of tens of MeV in energy, which carry away the majority of the gravitational binding energy liberated in the collapse. These neutrinos are therefore critical for understanding the energetics and evolution of the core from collapse to potential explosion, as well as provide an unique opportunity to study the system otherwise hidden from view by the stellar envelope. Neutrino detectors around the globe are poised to detect tens of thousands of neutrinos from the next core collapse occurring in the Milky Way galaxy, offering many opportunities for studying the properties of massive stars, core collapse, as well as neutrinos (see, e.g., reviews  \cite{Scholberg:2012id,Mirizzi:2015eza,Horiuchi:2017sku}). However, the occurrence rate of core collapse in the Milky Way is not very high, being a few per century \cite{Diehl:2006cf,Rozwadowska:2021lll}. 

A complementary strategy is to study extragalactic core collapses. Future Mton-class neutrino detectors will be sensitive to neutrinos from core collapses occurring in nearby galaxies \cite{Ando:2005ka,Kistler:2008us,Horiuchi:2013bc,Nakamura:2016kkl}. Another strategy is to search for the diffuse flux of neutrinos caused by past core collapses, called the diffuse supernova neutrino background (DSNB, see reviews \cite{Beacom:2010kk,Lunardini:2010ab}). The predicted DSNB is isotropic, constant in time, and its flux is roughly within a factor of two of the latest upper limits placed by Super-Kamiokande (Super-K) \cite{Bays:2011si}. Importantly, Super-K has recently completed a major upgrade---enhancing its detector volume with gadolinium \cite{Watanabe:2008ru} as proposed in Ref.\ \cite{Beacom:2003nk}---which allows a vastly improved differentiation between the DSNB and background signals. With this upgrade, the search for the DSNB with Super-K has transitioned from a background-limited one \cite{Bays:2011si} to a signal-limited one. 

Predictions of the DSNB have improved over the years \cite{Krauss:1983zn,Dar:1984aj,Totani:1995rg,Totani:1995dw,Malaney:1996ar,Hartmann:1997qe,Kaplinghat:1999xi,Ando:2002zj,Fukugita:2002qw,Strigari:2003ig,Iocco:2004wd,Strigari:2005hu,Lunardini:2005jf,Daigne:2005xi,Yuksel:2005ae,Horiuchi:2008jz,Lunardini:2009ya,Lien:2010yb,Keehn:2010pn,Vissani:2011kx,Lunardini:2012ne,Nakazato:2013maa,Mathews:2014qba,Yuksel:2012zy,Nakazato:2015rya,Hidaka:2016zei,Priya:2017bmm,Horiuchi:2017qja,Moller:2018kpn,Riya:2020wpw,Kresse:2020nto}. However, the impacts of stellar binaries have mostly been neglected. Recent observations of stars \cite{sana12,zapa20} show that the majority of massive stars have stellar companions and experience binary interactions. This includes massive stars which are the progenitors for CCSNe, and hence the majority of core collapses should be affected by binary interactions. Binary interaction, especially mass transfer, can significantly change the masses of core-collapse progenitors and strongly influence the type of CCSN \cite{Pod1992,Woosley:2002zz}. Furthermore, if stars in a binary merge together, a rapidly rotating star remains. This effect opens a novel channel for stars that are initially too low mass to undergo core collapse to become a viable core-collapse progenitor. Both this and mass transfer would affect the landscape of CCSNe and impact the DSNB. 

Recently, Ref.~\cite{Kresse:2020nto} made a first assessment of the impacts of binary interactions on the DSNB. The authors simulated the core collapse of progenitors stripped of its envelope due to mass loss \cite{Woosley2019} as a proxy for close-binary interactions. Due to mass stripping, these stars have less massive cores than their isolated counterparts, leading to lower neutrino emissions and larger ZAMS mass threshold for core collapse. Both of these result in lower DSNB predictions. However, the fraction of such hydrogen-stripped stars was determined by a fraction parameter, which was varied between 33\% to 100\% of the core-collapse progenitor population. Also, the impacts of binary mergers and other types of binary interactions were not included. 

In this paper, we quantify the impacts of binary interactions on the DSNB flux using binary population synthesis calculations which model the effects of both mass transfer and mergers. Based on these synthetic populations, we compute the DSNB flux and discuss consequences for current and future neutrino detectors. The paper is organized as follows. In Sec.~II, we summarize the important binary interactions and our binary population synthesis simulations. In Sec.~III, we make predictions for the DSNB and discuss detection rates at neutrino detectors. We finish with discussion and conclusions in Sec.~IV. Throughout, we adopt a $\Lambda$ cold dark matter cosmology with $\Omega_m = 0.3$, $\Omega_\Lambda = 0.7$, and $H_0 = 70$ km s$^{-1}$ Mpc$^{-1}$.

\section{Binary Treatment}\label{sec:binary}

Stars in binaries experience various binary interactions which can change the mass of the star during its evolution (e.g., Ref.\ \cite{2006epbm.book.....E}). In particular, stable mass transfer and the common envelope (CE) phase are important to understand the changing mass of CCSN progenitors (e.g., Refs.\ \cite{Pod1992,Pod2010}). In this section, we focus on binary effects which affect the evolution of CCSN progenitors.

\subsection{Binary effects / treatment}\label{sec:binary:effects}

If one of stars in a binary system fulfills its Roche lobe, some of the stellar material is transferred to the companion star. This process is called the Roche lobe overflow (RLOF). The radius of the Roche lobe around the donor star is approximately expressed as \citep{Eggleton_1983},
\begin{equation}
R_{\rm L,1}\simeq a\frac{0.49q_1^{2/3}}{0.6q_1^{2/3}+\ln(1+q_1^{1/3})},
\end{equation}
where $a$ and $q_1=M_1/M_2$ are the binary separation and the mass ratio, where $M_1$ is the donor and $M_2$ is the companion masses, respectively. The behavior of the mass transfer is determined by the response of the Roche lobe radius and the stellar radius of the donor star when it loses its material \citep{Paczynski_1976}. Assuming no mass is lost from the binary system during the mass transfer, i.e., the conservative mass transfer, the response of the Roche lobe radius is characterized by \citep{Tout_1997},
\begin{equation}
\zeta_{\rm L}\equiv \frac{d\ln R_{\rm L,1}}{d\ln M_1}\simeq 2.13 q_1-1.67
\label{eq:zeta_L}
\end{equation}
For $\zeta_{\rm L}<\zeta_\ast (\equiv d\ln R_{1,ad}/d\ln M_1)$, where $R_{1,ad}$ is the adiabatic radius of donor star, the radius of the donor star shrinks and becomes smaller than the Roche lobe radius after the mass of the donor star is transferred. In this case, stable mass transfer occurs.
On the other hand, for $\zeta_{\rm L}>\zeta_\ast$, the mass transfer would be unstable and the two stars would enter the CE phase. The value of $\zeta_\ast$ depends on the stellar envelope of the donor star. When the donor star is in the red giant phase with a convective envelope, $\zeta_\ast$ is given by,
\begin{equation}
    \zeta_\ast=-1+\frac{2}{3}\frac{M_1}{M_{\rm env,1}},
\end{equation}
where $M_{\rm env,1}$ is the envelope mass of the donor giant. 
When the donor star is in the main sequence (MS), giant phase with a radiative envelope \citep{Hjellming_1989}, naked-He MS, or naked-He giant star \citep{Ivanova_2002,Belczynski_2008}, we use $\zeta_\ast$ = 2.59, 6.85, 1.95 and 5.79, respectively. 

In the case of stable mass transfer ($\zeta_\ast>\zeta_{\rm L}$), we use the transfer rate which is expressed as \cite{Hurley_2002},
\begin{equation}
    \dot{M_1}=F(M_1)\left[\ln\left(\frac{R_1}{R_{\rm L,1}}\right)\right]^3
    ~M_{\odot}~\rm yr^{-1},
\end{equation}
where $R_1$ is the radius of the donor star, and
\begin{equation}
    F(M_1)=3\times10^{-6}\left\{{\rm min}\left[\left(\frac{M_1}{1M_{\odot}}\right),5.0\right]\right\}^2.
\end{equation}
Under our assumed conservative mass transfer, the accretion rate onto the companion star is the same as the mass loss rate of the donor star. This would not be valid if the accretion rate exceeds the Eddington limit of the companion star, but in reality this condition is not reached in any of our binary systems which are not compact objects.

In the case of the CE phase ($\zeta_\ast<\zeta_{\rm L}$), we use the $\alpha\lambda$ formalism \citep{Webbink1984}, 
\begin{equation}
\alpha\left(\frac{GM_{\rm{c,1}}M_2}{2a_{\rm{f}}}-\frac{GM_1M_2}{2a_{\rm{i}}}\right)=\frac{GM_{\rm{1}}M_{\rm{env,1}}}{\lambda R_1},
\label{eq:ce1}
\end{equation} 
for a binary of a mass-losing giant star and a companion star, where $a_{\rm i}$, and $M_{\rm c,1}$ are the initial binary separation just before the CE phase and the core mass of the mass-losing giant, respectively. In order to calculate the separation just after the CE phase, $a_{\rm f}$, we use Eq.~(\ref{eq:ce1}). Here, $\alpha$ is the efficiency parameter parametrizing how much of the orbital energy is transferred to the unbound envelope of the giant. The parameter $\lambda$ is for the envelope's binding energy of the giant. If the companion star is also a giant \cite{Nelemans2001}, Eq.~(\ref{eq:ce1}) is replaced by, 
\begin{eqnarray}
\alpha\left(\frac{GM_{\rm{c,1}}M_{c,2}}{2a_{\rm{f}}}-\frac{GM_1M_2}{2a_{\rm{i}}}\right)=&\frac{GM_{\rm{1}}M_{\rm{env,1}}}{\lambda R_1}\nonumber\\
&+\frac{GM_{\rm{2}}M_{\rm{env,2}}}{\lambda R_2},
\label{eq:ce2}
\end{eqnarray}
where $R_2$, $M_{\rm c,2}$, and $M_{\rm env,2}=M_2-M_{\rm c,2}$ are the radius, the core mass, and the envelope mass of the companion giant star, respectively. The CE parameters $\alpha$ and $\lambda$ are not understood very well \citep{Ivanova_2013}. We adopt the CE parameter values adopted in previous binary population studies, $\alpha\lambda=1$ and $0.1$ (e.g., Refs.\  \cite{Belczynski2007,Kinugawa2014}). 
Physically, the product $\alpha\lambda$ is related to how hard it is to unbind the envelope due to the CE. If $\alpha\lambda$ is small, a large orbital energy is need to unbind the envelope. 
Thus, it is easier to shrink the binary orbit and the binary more easily merges during the CE phase.

When the CE phase is over, we calculate the separation just after the CE phase $a_{\rm f}$, and check whether the binary has merged within the CE phase or not. If $a_{\rm f}$ is smaller than the sum of the remnant stellar radii, we consider the binary has merged. Additionally, when the post-MS star does not reach the Hayashi track or ignite helium burning, such a star, so-called a Hertzsprung gap star, may not have a clear core-envelope structure. In this case, we assume the binary merges \cite{Taam2000,Ivanova2004,Belczynski_2008}. If a binary merges before CCSNe, we treat the merged product as a rapidly rotating star.

\begin{figure}[t]
\includegraphics[width=1.1\hsize]{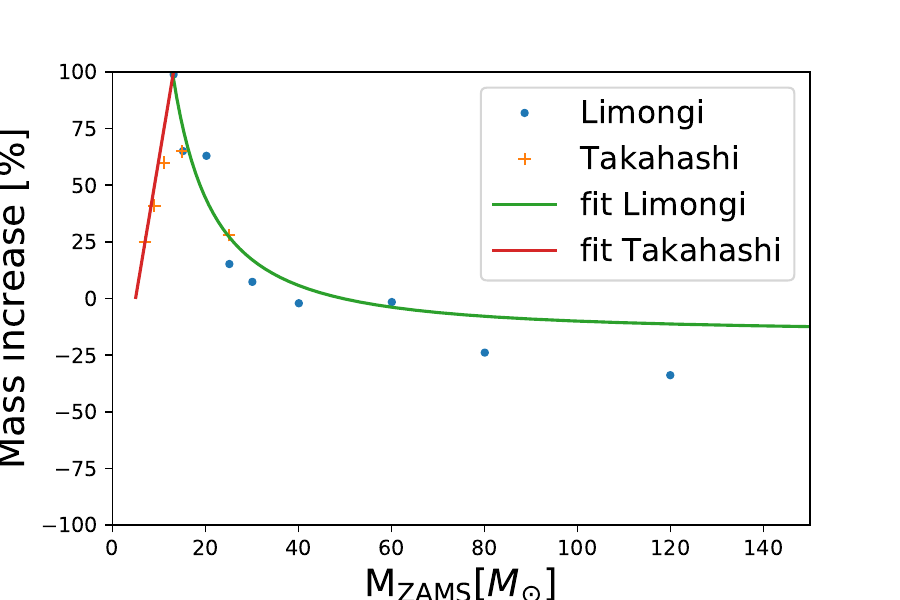}
\caption{The core mass increase due to rotational effects. The horizontal axis is the ZAMS mass. The vertical axis is the percentage of increase. Blue points are from the Limongi models \cite{Limongi2017}, while the orange points are for our fiducial model \cite{Takahashi2014}. The green line is the fitting to the Limongi models, while  the red line is the fitting to our fiducial models. We adopt the combination of the two fits for our fiducial calculation. 
}
\label{fig:coremass}
\end{figure}

Fast rotations can affect stellar evolution in various ways. Namely, the centrifugal force changes the pressure balance, the wind mass-loss rate can be enhanced, and the rotation induced instability can enhance the material mixing inside the star. Theoretical estimates of the integrated effect for solar-metallicity single stars are shown in Fig.~\ref{fig:coremass}, where the percentage increase of the carbon-oxygen (CO) core mass with respect to the nonrotating case is shown. For massive stars with $M_{\rm ZAMS}$ (the total mass at zero-age main sequence) $> 13 M_\odot$, we consider the results of fast rotators with an initial rotation velocity of $v_{\rm rot, ini}=300$ km s$^{-1}$ studied in Limongi et al.~\cite{Limongi2017}. For $M_{\rm ZAMS}<13 M_\odot$, models with $v_{\rm rot, ini}/v_{\rm Kep}$ = 0.6, where $v_{\rm Kep} = \sqrt{GM/R}$ is the Kepler rotation velocity, is newly calculated using the code described in Takahashi et al.~\cite{Takahashi2014}. The Takahashi models have initial rotation velocities of $v_{\rm rot, ini} \sim 500$ km s$^{-1}$. The chief effect of stellar rotation in stars with ZAMS masses of $\lesssim 20 M_\odot$ is the enhancement in material mixing. This leads to an increase of the core mass, but the efficiency is smaller for less massive models since the thermal (Kelvin-Helmholtz) timescale is relatively longer compared with the stellar lifetime. On the other hand, the enhancement of wind mass-loss due to the $\Omega$--$\Gamma$ effect becomes more significant for more massive stars with $\gtrsim 20 M_\odot$, which reduces the total mass, counterbalancing the core-mass increase due to rotation-induced mixing.

We assume that stellar mergers result in the formation of fast rotators; otherwise the stars are nonrotating. The evolution of the fast rotating merger remnant is approximately treated by enhancing the CO core mass with respect to the nonrotating counterpart having the same total mass.
From the results shown in Fig.~\ref{fig:coremass}, we derive simple fitting formulas as a function of the ZAMS mass,
\begin{equation}
    f_L=53.4M_{\rm ZAMS}^{-3/2}+0.847
\end{equation}
from Limongi's models and 
\begin{equation}
    f_T=0.123M_{\rm ZAMS}+0.392
\end{equation} 
from Takahashi's models, which are also shown by the green and red solid lines, respectively. Considering the large uncertainty in the theory of rotating stellar evolution, we consider following three schemes: 
\begin{itemize}
    \item \textit{Fiducial}: we apply $f_T$ for merger remnants with $M_{\rm ZAMS}<13M_{\odot}$ and $f_L$ for $M_{\rm ZAMS}>13M_{\odot}$. 
    \item \textit{Extrapolated}: we only apply $f_L$ for all merger remnants, i.e., for $M_{\rm ZAMS}<13M_{\odot}$ we perform an extrapolation of $f_L$.
    \item \textit{No rotation}: we do not increase the CO core mass of a merger remnant, hence, we ignore the effect of rotation.
\end{itemize}
Note that in all cases, if the CO core mass estimated by the above schemes exceed the total mass, we limit the CO core mass to the total stellar mass. 

\subsection{Population synthesis}

In order to calculate the binary effect for core-collapse progenitors, we use the binary population synthesis method developed by Ref.\ \cite{Hurley_2002}. Here, we give only a brief outline, and refer the reader to Refs.\ \cite{Hurley_2002,Kinugawa2014} for details. 

\begin{figure}[t]
\includegraphics[width=125mm]{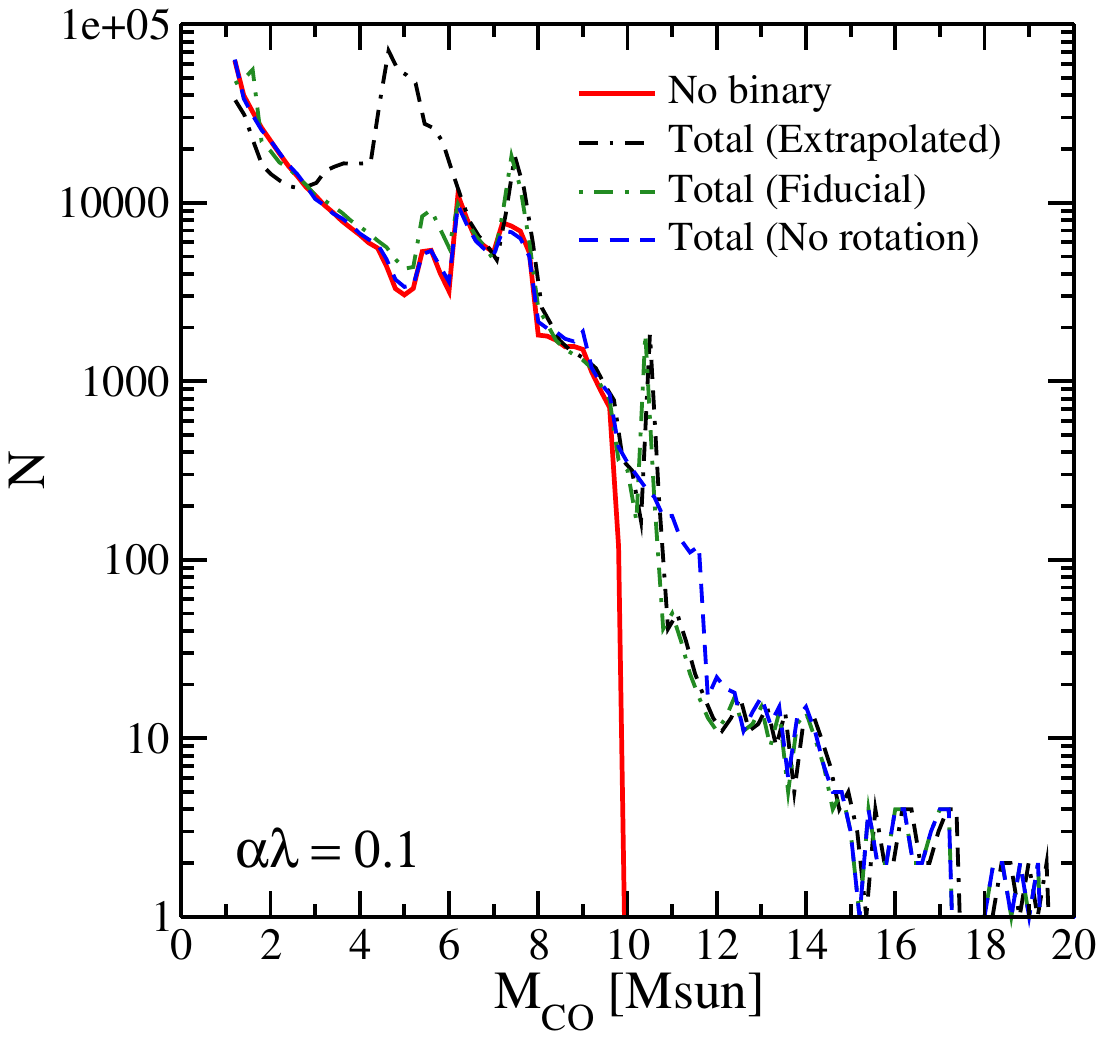}
\caption{Distributions of CO mass for our binary population synthesis compared to the case with no binary interactions (red solid). Three types of binary treatment with different postmerger evolution scenarios are shown: extending the study of Limongi et al (black dot-dashed-dashed) \cite{Limongi2017}, our fiducial model (green dot-dashed) \cite{Takahashi2014}, and a low version neglecting postmerger rotation (dashed blue). CE parameter is $\alpha\lambda=0.1$.}
\label{fig:COdistribution}
\end{figure}

First, we set initial binary parameters such as the primary mass $M_1$, mass ratio $M_2/M_1$, separation $a$, and eccentricity $e$, using the Monte Carlo method based on the initial distribution functions which are obtained by observations. We use a Salpeter IMF from $3\msun$ to $140\msun$ for the primary IMF \cite{Salpeter1955}, the flat distribution from $0.1\msun/M_1$ to 1 for the mass ratio \cite{Kobulnicky2007}, the logflat distribution from $a_{\rm min}$ to $10^6R_{\odot}$ for the separation  \cite{Abt_1983}, and the thermal distribution ($\propto e$) from 0 to 1 for the eccentricity \cite{Heggie_1975}, where $a_{\rm min}$ is the minimum separation where the binary cannot occur a mass transfer. Note that while the primary stars follow the Salpeter IMF by construction, after combining with the secondary stars the effective IMF is flatter. The power-law index of the combined stellar population is actually approximately ${-2.2}$.

Next, we calculate each stellar evolution and judge whether stars experience binary interactions such as tidal friction, Roche lobe overflow, and the CE phase, and change the parameters $M_1, M_2, a$, and $e$ in each time step accordingly \cite{Hurley2000,Hurley_2002,Kinugawa2014}. We calculate $10^6$ binaries for each run. We consider two CE parameters cases $\alpha\lambda=0.1$ and $\alpha\lambda=1$, and three schemes for rotating mergers stars (fiducial, extrapolated, and no rotation), for a total of 6 synthetic stellar populations. In addition, we perform a comparison set with binary separation large so binary effects do not operate. 

\begin{figure}[t]
\includegraphics[width=125mm]{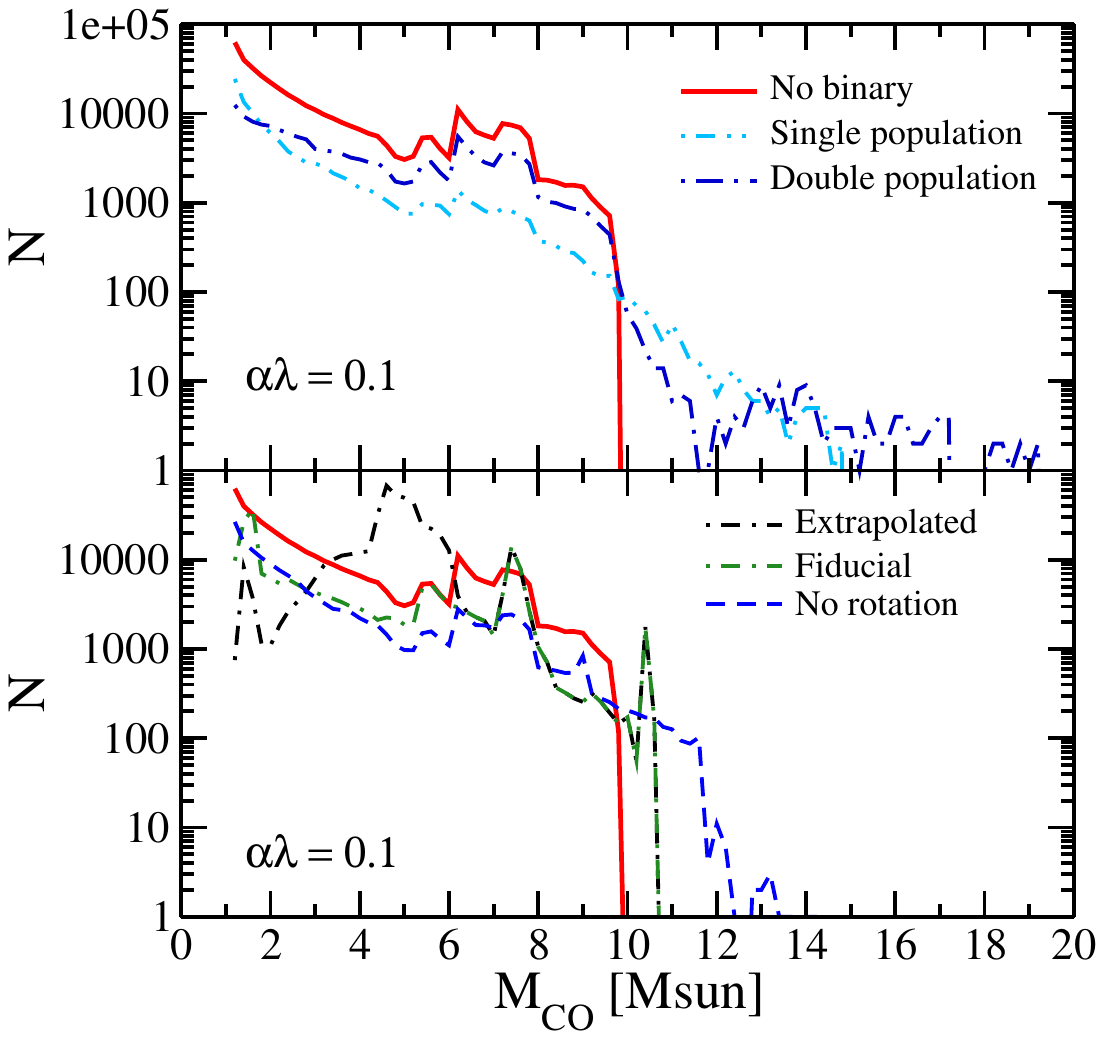}
\caption{Same as Fig.~\ref{fig:COdistribution}, but showing the breakdown into different binary evolution channels. In the top panel we show nonmerge populations: single (light-blue dot-dot-dashed) and double  systems (blue dot-dashed). In the bottom panel we show the merger channel, separately for the extending the study of Limongi et al (black dot-dashed-dashed), our fiducial model (green dot-dashed), and a low version neglecting postmerger rotation (dashed blue). Total numbers in each channel are summarized in Table \ref{tab:counts}.}
\label{fig:COdistribution2}
\end{figure}

\subsection{Effect on CCSN progenitors}\label{sec:binary:progenitors}

\begin{table*}[htb]
\vspace{1em}
\begin{tabular}{|l|l|l|l|l|l|}
\hline
                        & \multicolumn{2}{c|}{Merger}    & \multicolumn{2}{l|}{nonmerger}   &  Ratio wrt   \\ 
                        &   & (rotation) & Double & Single     &   no binary, $f_b$    \\ 
\hline \hline
No binary evolution                  
    & 0   & 0    & 122,600   & 171,002  &  1    \\ 
\hline
Binary $\alpha\lambda=0.1$  Extrapolated 
    & 155,235   & 315,722   & 75,723    & 109,276    &  1.76      \\ 
Binary $\alpha\lambda=0.1$  Fiducial  
    & 155,235   &  50,102  & 75,723   & 109,276     & 1.24      \\ 
Binary $\alpha\lambda=0.1$  No rotation 
    & 155,235   & 0        & 75,723   & 109,276     & 1.00      \\ 
\hline
Binary $\alpha\lambda=1$  Extrapolated  
    & 140,467   & 196,983   & 83,070    & 131,679   & 1.53 \\
Binary $\alpha\lambda=1$  Fiducial    
    & 140,467   & 39,869   & 83,070    & 131,679   & 1.24 \\
Binary $\alpha\lambda=1$  No rotation  
    & 140,467   & 0         & 83,070    & 131,679   & 1.05 \\
\hline 
\end{tabular}
\caption{Counts of core-collapse progenitor systems predicted by binary population syntheses, separated into the merger and nonmerger populations. For the merger population, we separately show numbers due to rapid rotation (column 3). For nonmerger population, we separately shown numbers of double and single systems. Note that these are system numbers, i.e., the number of progenitors in double is twice the quoted numbers. The final column shows the ratio of the total number of core-collapse progenitors with respect to the no binary evolution.}
\label{tab:counts}
\end{table*}

Figure \ref{fig:COdistribution} shows the binary effect for the CO core mass distributions compared with the no binary case. We see that the maximum CO core mass is significantly increased by the binary effect. This feature is seen in all rotation schemes of the merger remnants. This is because the highest mass increase is caused by mass accretion due to the Roche lobe overflow. Figure \ref{fig:COdistribution2} shows the breakdown of the binary effect for different evolution channels. Here, we split the stellar population depending on their state prior to core collapse:
\begin{itemize}
    \item \textit{Nonmerger population}: binary systems where the stars do not merge before core collapse. We further split this into binary systems where both stars are core-collapse progenitors (double) and binary systems where only one of the stars are core-collapse progenitors (single). 
    \item \textit{Merger population}: binary systems where the stars merge before core collapse. This includes progenitors which become core-collapse progenitors as a result of rotational effects. We use three different treatments for this population, as described in Sec.~\ref{sec:binary:effects}. 
\end{itemize}
The upper panel of Fig.~\ref{fig:COdistribution2} shows the CO core mass distributions of nonmerger binaries, while the bottom panel shows the CO core mass distribution of the merger population, separately for each rotation scheme. We can observe from these panels that the highest CO core mass population arises from mass accretion due to the Roche lobe overflow in the nonmerger double and single systems. On the other hand, the distribution of lower CO core mass stars  depend on the rotation scheme (lower panel of Fig.~\ref{fig:COdistribution2}). In particular, the rate of increase of the CO core mass is very high for the extrapolated scheme for progenitors with $M_{\rm initial}<13M_{\odot}$ (see Fig.\ \ref{fig:coremass}). The effect is to almost double the CO core mass of $M_{\rm ZAMS} \sim 13\msun$ stars from its original $M_{\rm CO}\sim3\msun$, which explains the peak seen in the final CO core mass distribution at $\sim5\msun$ (bottom panel of Fig.\ \ref{fig:COdistribution2}).

\begin{figure*}
\begin{tabular}{cc}
   \begin{minipage}[t]{0.42\hsize}
\includegraphics[width=\hsize]{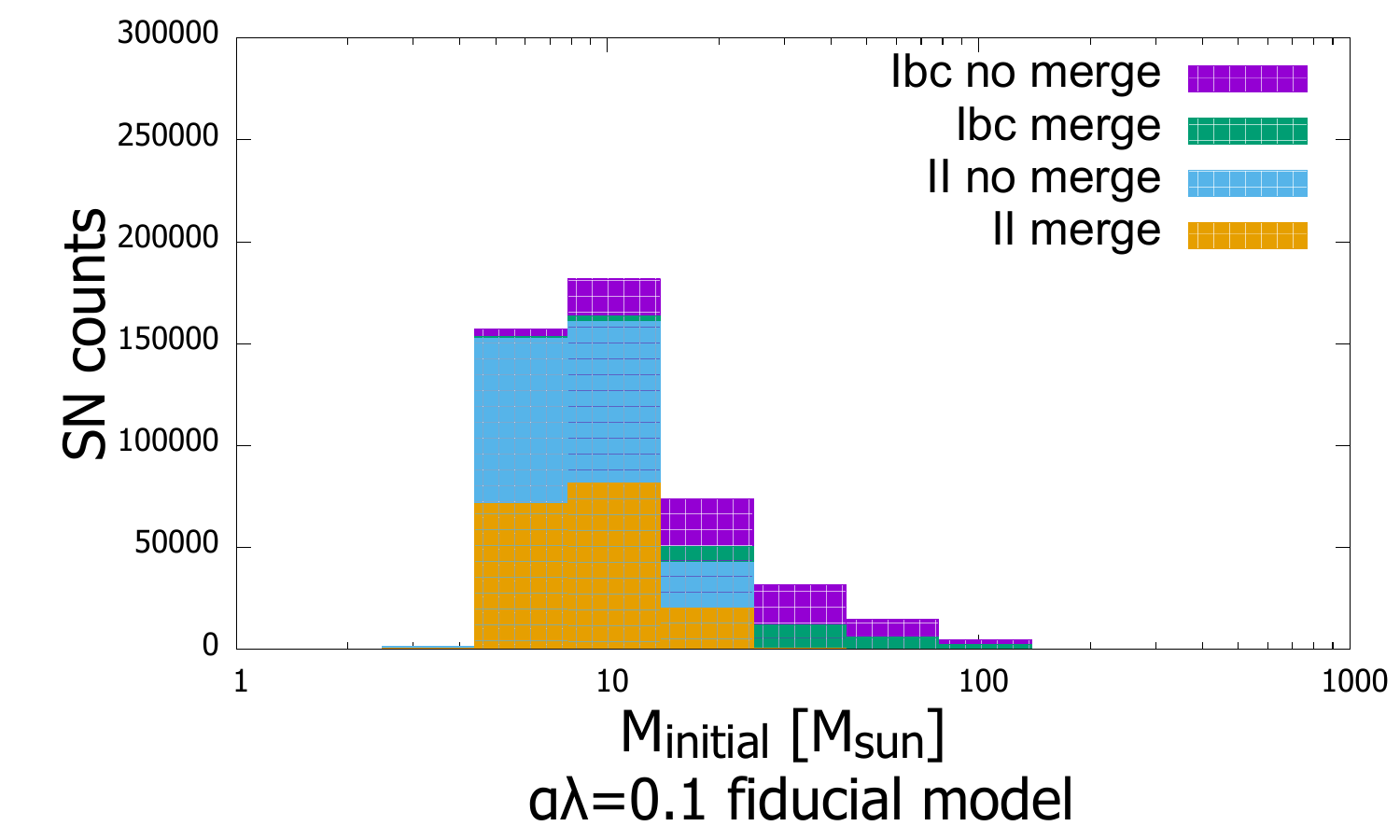}
\label{al01_fiducial}
\end{minipage} 
\begin{minipage}[t]{0.42\hsize}
\centering
\includegraphics[width=\hsize]{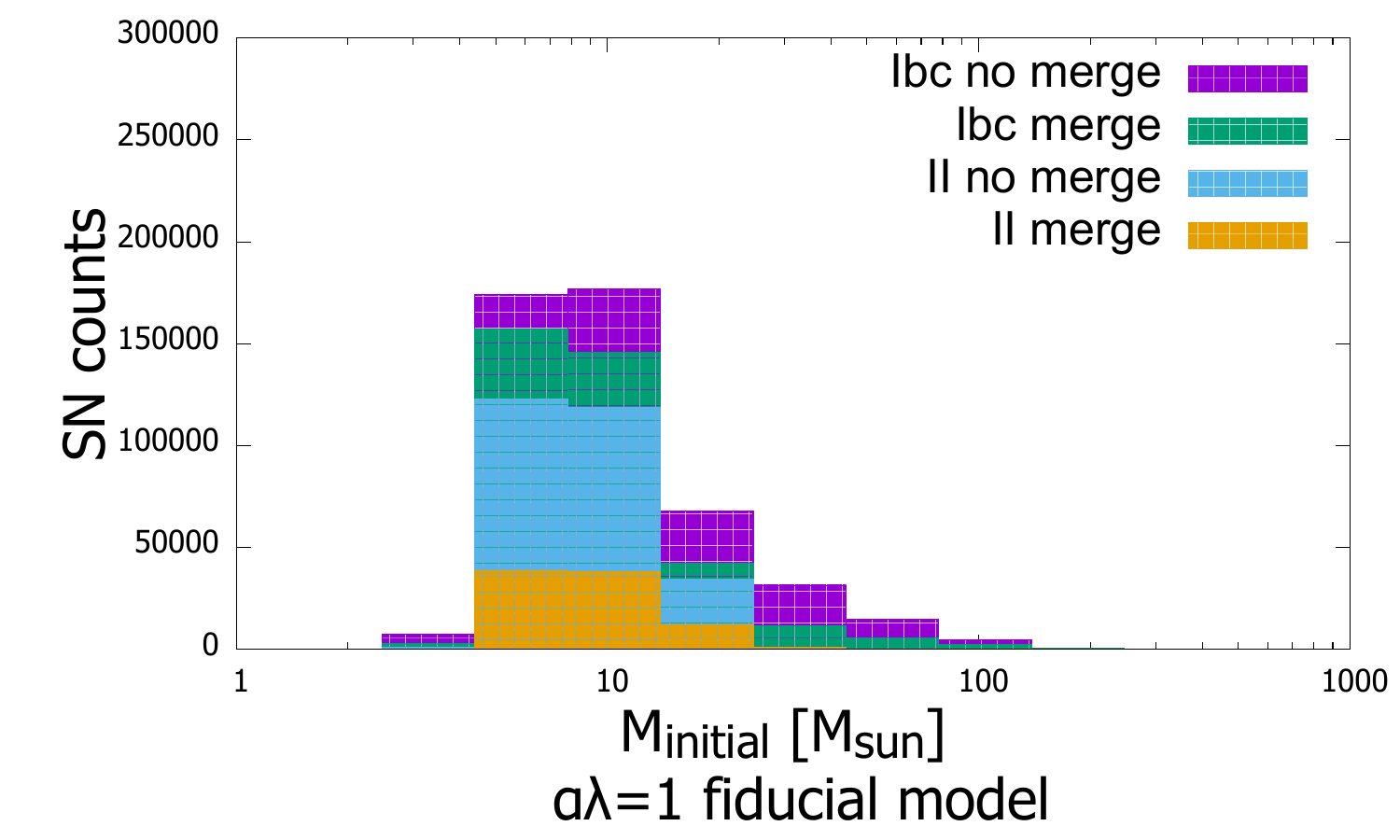}
\label{al1_fiducial}
\end{minipage} \\
 \begin{minipage}[t]{0.42\hsize}
 \centering
\includegraphics[width=\hsize]{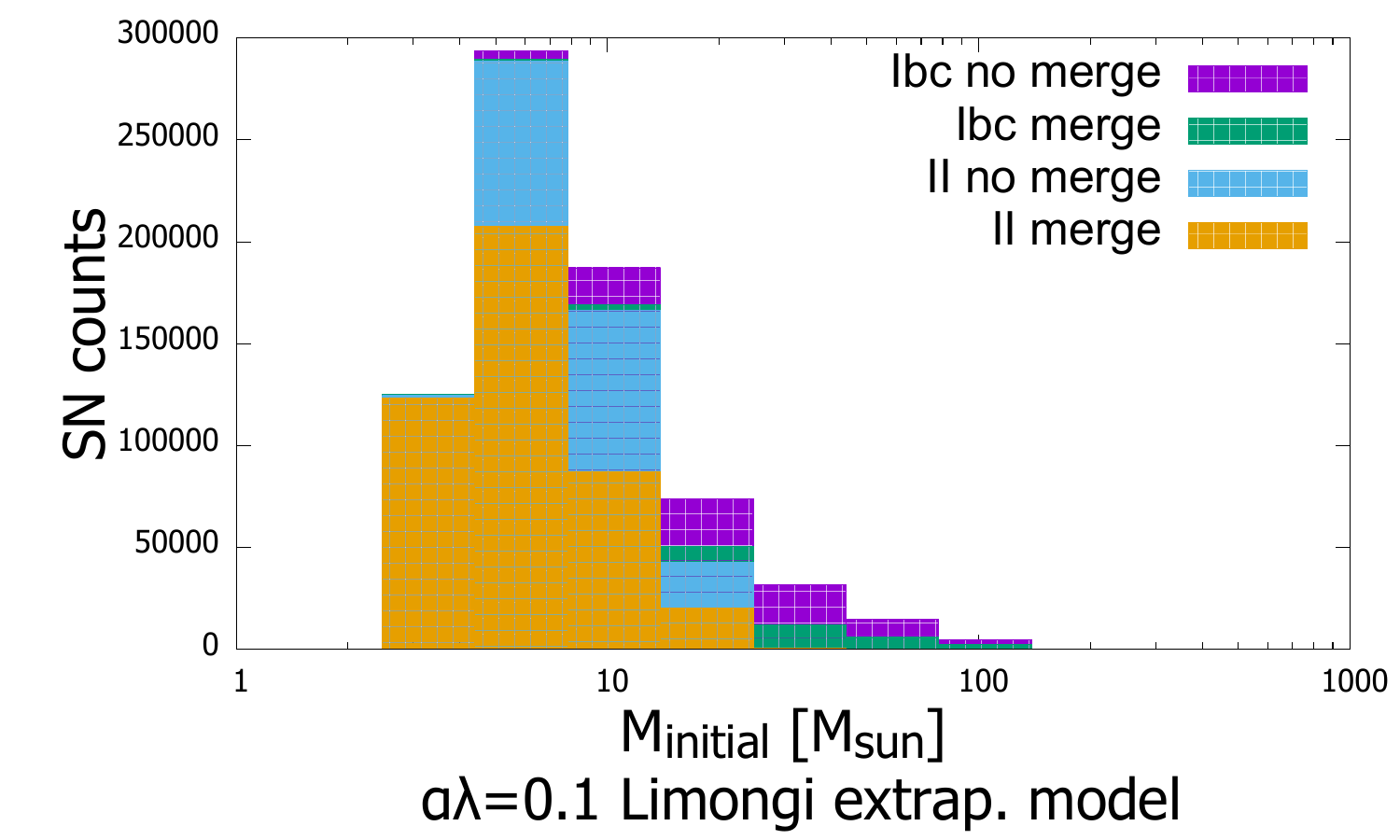}
\label{al01_LE}
\end{minipage} 
\begin{minipage}[t]{0.42\hsize}
\centering
\includegraphics[width=\hsize]{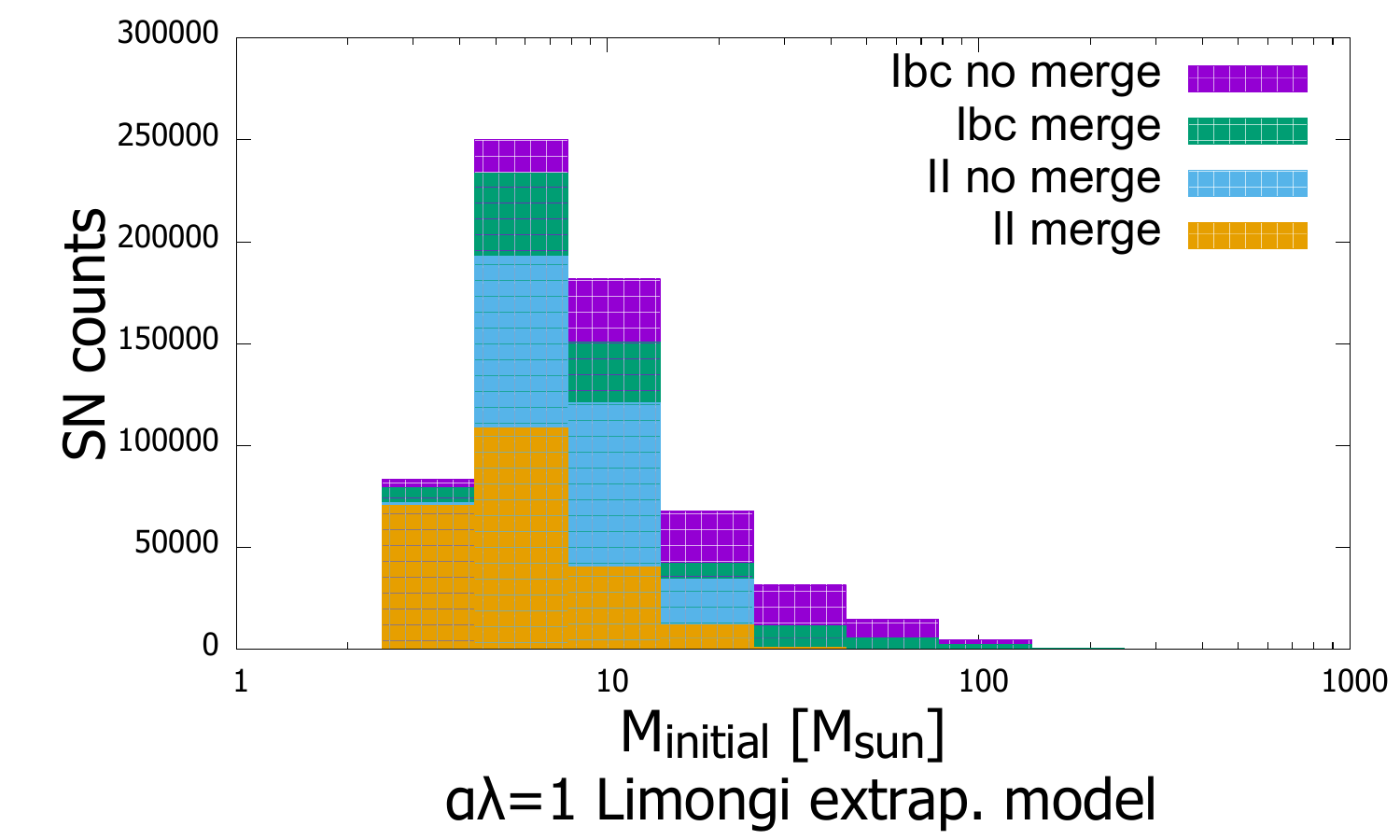}
\label{al1_LE}
\end{minipage} \\
\begin{minipage}[t]{0.42\hsize}
\centering
\includegraphics[width=\hsize]{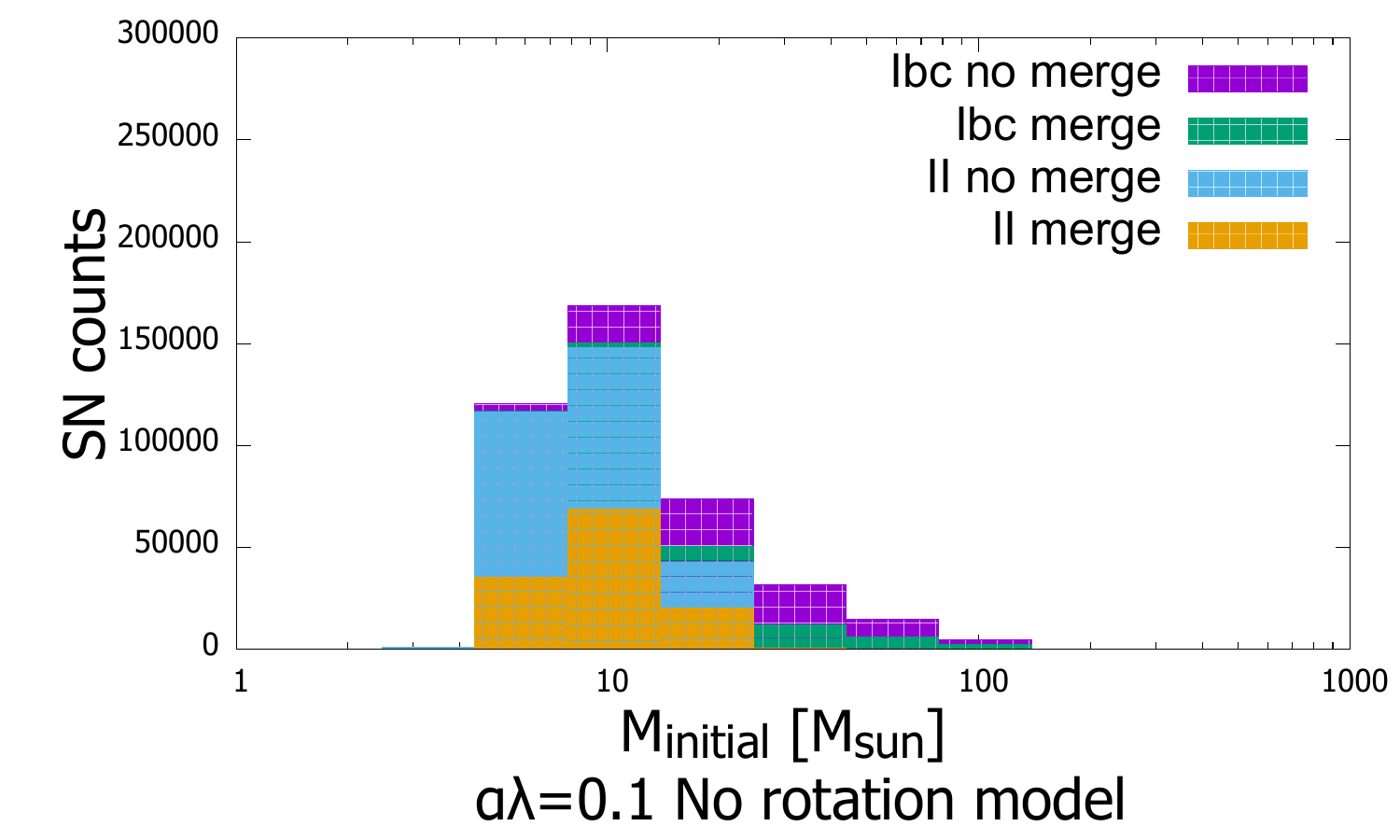}
\label{al01_NoRotation}
\end{minipage} 
      \begin{minipage}[t]{0.42\hsize}
        \centering
\includegraphics[width=\hsize]{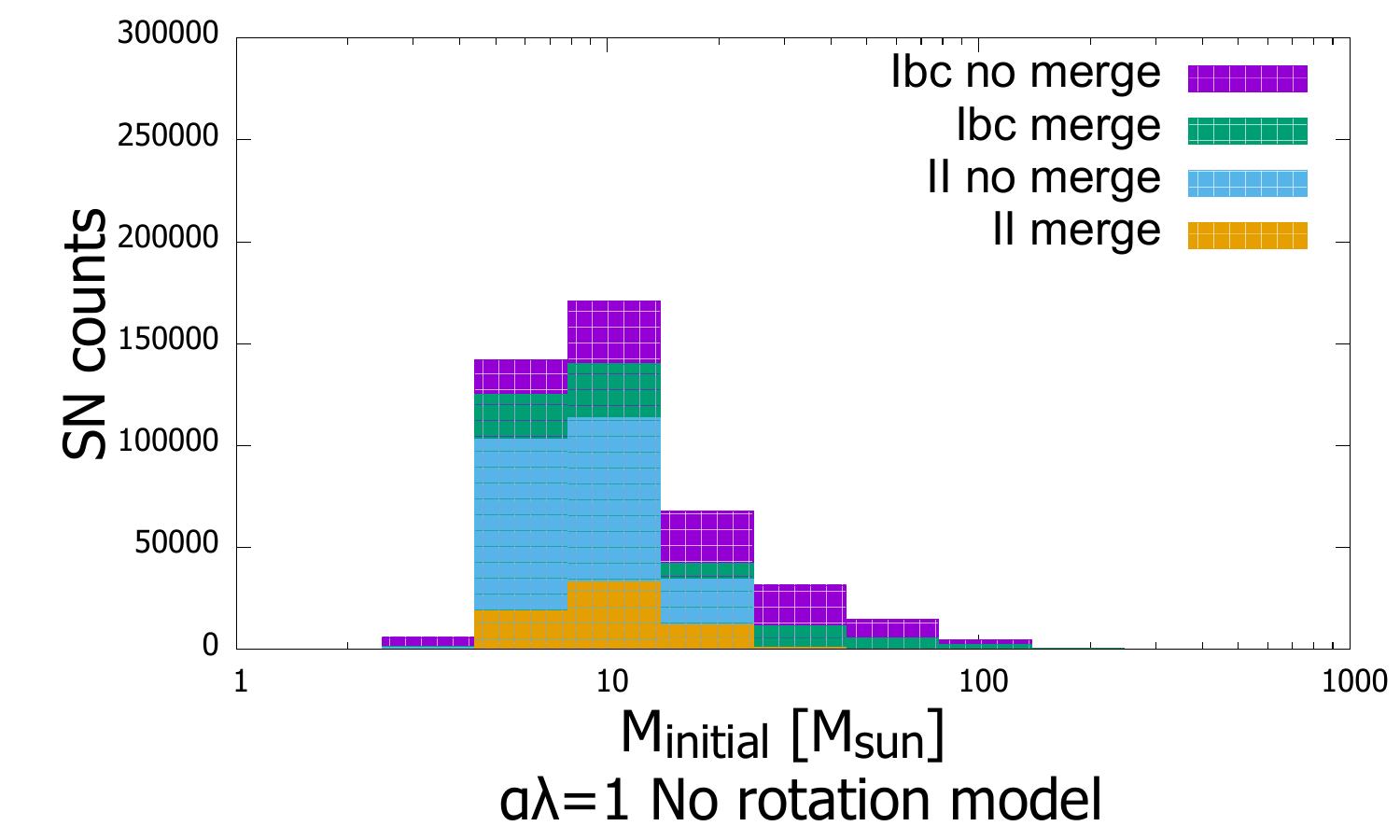}
\label{al1_NoRotation}
\end{minipage} \\
\end{tabular}
\caption{Cumulative bar charts of SN types, showing whether the progenitor merged or not for each. The horizontal axis shows the initial mass of progenitors. For merged progenitors, the horizontal axis shows the total initial mass of the binary. 
} 
\label{fig:SNtype}
\end{figure*}

In Table \ref{tab:counts}, we show the breakdown of the total number of binary systems for our 6 synthetic binary populations as well as the nonbinary population for comparison. For the merger population we show in a separate column the contribution from rotating stars, and for the nonmerger population we show the numbers separately for double and single systems. We see that in the no rotation scheme, the number of core-collapse progenitors is almost unchanged from the no binary case. Mergers or mass accretion by binary interactions can make low-mass stars become core-collapse progenitors and increase the tally. On the other hand, the mergers of massive stars will reduce the number of core-collapse progenitors. Even though the IMF is steeply falling with mass, the total number of stars capable of becoming core-collapse progenitors through mergers (i.e., say, with initial mass between 4--$8 M_\odot$) is not that much larger than the total number of stars above $8 M_\odot$. For our binary synthesis setup, with an effective IMF slope of $-2.2$ and upper mass $140 M_\odot$, this ratio is $\approx 4:3$. In the case of no rotation scheme, we therefore see that these effects are roughly balanced. 

However, the merged remnant should gain spin angular momentum from the orbital angular momentum. Figure \ref{fig:coremass} shows the CO core mass gain due to rotation for our fiducial scheme. This gain provides additional core-collapse progenitors. Thus, in our fiducial and extrapolated schemes, the ratio with respect to no binary is greater than one. Especially in the extrapolated scheme where the mass gain is large, it is easier to form core-collapse progenitors than our fiducial scheme. As for the CE parameter $\alpha\lambda$, a small $\alpha\lambda$ makes it easier to merge than a large $\alpha\lambda$. Thus, our $\alpha\lambda=0.1$ populations show more merger progenitors that our $\alpha\lambda=1$ populations.

Figure \ref{fig:SNtype} shows the cumulative bar charts for the CCSN type and whether the progenitor is a merger or nonmerger, separately for each rotation scheme. In no binary evolution (i.e., single stellar evolution), whether the SN type is II or Ibc switches at $M_{\rm ZAMS}\sim24\msun$.
However, Fig.~\ref{fig:SNtype} shows for binary populations there are some Type Ibc SNe whose progenitor ZAMS masses are less than $24\msun$; these are due to the envelope loss by a mass transfer or a CE phase. Especially, the number of Type Ibc SNe arising from merged remnants is higher in the $\alpha\lambda=1$ case than the $\alpha\lambda=0.1$ case. The reason is that a larger $\alpha\lambda$ means it is easier to unbind the envelope during a CE phase than a smaller $\alpha\lambda$.
On the other hand, there are still a few Type II SNe  whose progenitor ZAMS masses are more than $24\msun$, due to the mass accretion by a mass transfer or the remaining envelope after a CE phase. In this case, they often lose the envelope by stellar wind mass loss. Thus the number of such SNe is small.

Table \ref{tab:SNtype} shows the fractions of SN types for each of our population synthesis realizations. The fraction of SN type strongly depends on the combined CE parameter $\alpha\lambda$. In the case of $\alpha\lambda=0.1$, fractions of SN types are II:Ibc $\sim$ 75:25. On the other hand, in the case of $\alpha\lambda=1$, fractions of SN types are $\sim$ 60:40. If $\alpha\lambda$ is small, a large orbital energy is need to unbind the envelope. It is easier to merge during the CE phase and the merged remnants tend to maintain large envelopes. Thus, in the case of small $\alpha\lambda$, the fraction of merged type II SNe are high (Fig. \ref{fig:SNtype}). The fraction of SN types from a volume-limited sample of low-redshift SNe is II:Ibc=75:25 \cite{Li2011}, consistent with our models using $\alpha\lambda=0.1$.

\begin{table}[]
\begin{tabular}{|l|l|l|}
\hline
              & Ibc     & II      \\ \hline
$\alpha\lambda$=1 Fiducial        & 41.8\%  & 58.2\%  \\ \hline
$\alpha\lambda$=1 Extrapolated   & 33.6\% & 66.4\% \\ \hline
$\alpha\lambda$=1 NoRotation & 42.1\%  & 57.9\% \\ \hline
$\alpha\lambda$=0.1 Fiducial      & 23.1\%  & 76.8\%  \\ \hline
$\alpha\lambda$=0.1 Extrapolated & 25.9\% & 74.1\% \\ \hline
$\alpha\lambda$=0.1 NoRotation &  26.0\% & 74.0\% \\ \hline
\end{tabular}
\caption{Fractions of SN types for each of our population synthesis realizations.}
\label{tab:SNtype}
\end{table}

\section{Supernova neutrino predictions including binary treatment}

We now quantify the impact of binary interactions on the DSNB. Binaries cause three important changes relative single stellar populations: (1) due to mass transfer and mergers, stars initially below the core-collapse threshold can become a core-collapse progenitor, (2) mergers can cause two core-collapse progenitors to reduce to a single core-collapse progenitor, and (3) due to mass transfer, the distribution of final CO core masses is changed. Here, we include these effects in a new prediction of the DSNB. 

\subsection{DSNB formulation}
The DSNB is predicted by integrating over time the comoving volumetric occurrence rate of core collapses, $R_{\rm CC}$, weighted by the mean neutrino spectrum for a population of core-collapse progenitors, $dN_\nu/dE$, appropriately redshifted (see, e.g., reviews \cite{Beacom:2010kk,Lunardini:2010ab}), 
\begin{equation}\label{eq:dsnb}
    \frac{d\phi}{dE} = c \int R_{\rm CC}(z)
\frac{dN_\nu}{dE^\prime} (1+z) \left| \frac{dt}{dz} \right| dz,
\end{equation}
where $E^\prime=E(1+z)$ and $|dz/dt| = H_0 (1+z) [\Omega_m(1+z)^3 + \Omega_\Lambda]^{1/2}$. In what follows, we perform the integration out to $z_{\rm max}=5$, which is sufficiently large for flux predictions in the detectable range of neutrinos. 

We compute the cosmic core-collapse rate $R_{CC}(z)$ from the cosmic star formation rate (SFR) history, $\dot{\rho}_*(z)$. This is because while the uncertainties of the SFR are fairly large \cite{Hopkins:2006bw,Mathews:2014qba,Madau:2014bja}, they are smaller than those for direct CCSN rate measurements \cite{Horiuchi:2013bc,Strolger:2015kra,Mattila:2012zr,Botticella:2011nd,Taylor:2014rlo,Cappellaro:2015qia}. Furthermore, not all core collapse may yield CCSNe. We adopt the piecewise linear fit to $\dot{\rho}_*(z)$, i.e., Eq.~(5) of Ref.~ \cite{Yuksel:2008cu}, with shape parameters $a=3.4$, $b=-0.3$, and $c=-3.5$; redshift breaks at $z=1$ and $4$; and smoothing parameter $\eta=10$. We convert the SFR to a core-collapse rate by the conversion 
\begin{equation}\label{eq:rate}
    R_{\rm CC}(z) = \dot{\rho}_*(z)\frac{\int_{M_{\rm min} = 8}^{140}\psi(M)dM}{\int_{0.1}^{140} M \psi(M)dM},
\end{equation}
where $\psi(M)=dN/dM$ is the IMF, defined so that $\psi(M)dM$ gives the number of stars in the mass range $M$ to $M+dM$, and $M_{\rm min}$ is the minimum mass for a star to undergo core collapse. We adopt measurements of $\dot{\rho}_*(z)$ assuming a Baldry-Glazebrook IMF \cite{Baldry:2003xi}, and consistently use the same IMF for computing Eq.~(\ref{eq:rate}). The Baldry-Glazebrook IMF has a shallow high-mass slope of $-2.15$ and produces better agreements with other astronomical observations such as the stellar mass density buildup \cite{Hopkins:2006bw} and the extragalactic background light \cite{Horiuchi:2008jz}. Importantly, the Baldry-Glazebrook high-mass slope is also close to the effective IMF of our population synthesis setup (a power-law index of approximately $-2.2$). 

Fortunately, the IMF impacts the core-collapse rate only at the percent level. For example, Ref.~\cite{Hopkins:2006bw} explored the impacts of different IMFs, including the Baldry-Glazebrook and Salpeter IMFs. Since the same massive stars acting as tracers of star-formation activity are also core-collapse progenitors, the IMF is not needed to sample the entire stellar population mass range (see, also, \cite{Horiuchi:2008jz}). As a result, our CCSN rate normalization at z=0 is $R_{CC}(0) = (1.3 \pm 0.26) \times 10^{-4} {\rm Mpc^{-3} \, yr^{-1}}$, consistent with values obtained with other IMFs (e.g., \cite{Hopkins:2006bw,Horiuchi:2008jz,Moller:2018kpn}). 

We do not model the potential impact of binary interactions on the SFR calibration factors. While binary interactions can dramatically impact stellar evolution, their effects are strongly concentrated on the late evolutionary epochs. The main sequence, which dominate as SFR tracers, are relatively unaffected, so we do not expect the SFR calibrations to be strongly influenced. 

\subsection{Including binary effects}

We account for the binary effects on the mean neutrino emission by using the modifications to the distribution of CO core masses (Fig.~\ref{fig:COdistribution}). To this end, we first establish the connection between the CO core mass and the neutrino emission during core collapse. Then, we derive the mean neutrino emission based on the distribution of CO core mass as predicted by our population synthesis calculations. 

In order to obtain the neutrino emission from a wide range of massive stars, we adopt the systematic set of two-dimensional (2D) core-collapse simulations of Summa et al (2016) \cite{Summa:2015nyk}, hereafter S16, and augment them by additional simulations. The simulations of S16 include a total of 18 progenitor models in the ZAMS mass range $11.2$--$28 M_\odot$. Self-gravity is computed using the general relativistic monopole corrections as described in \cite{Marek:2005if}, and neutrino transport is solved with a ray-by-ray approximation along radial rays using a variable Eddington factor method (e.g., \cite{Buras:2005rp}) for all neutrinos: $\nu_e$, $\bar{\nu}_e$, and $\nu_x$ ($=\nu_\mu$, $\nu_\tau$, $\bar{\nu}_\mu$, $\bar{\nu}_\tau$). 

To the set of 18 simulations of S16 we add two core-collapse simulations: (i) a $9.6M_\odot$ progenitor of \cite{heger10} and (ii) an $8.8M_\odot$ ONeMg progenitor of \cite{Nomoto:1984,Nomoto:1987}. For the  $9.6M_\odot$ progenitor, the hydrodynamic simulation has been performed by the {\tt 3DnSNe} code (see the references \citep{Nakamura:2019,Kotake:2018, Takiwaki:2016} for recent applications). The method for the 2D supernova model is summarized in previous works \cite{Zaizen:2020,sasaki20}. The code provides consistent results on neutrino luminosities and average energies with more sophisticated schemes (see Ref.~\cite{OConnor:2018} for detailed comparison), whereas the code employs an idealized neutrino transport scheme of IDSA (isotropic diffusion source approximation) \cite{Liebendorfer:2009}. The simulation of the $8.8M_\odot$ progenitor is the same as adopted in the DSNB study of \cite{Horiuchi:2017qja} and is based on simulations of \cite{Huedepohl:2009wh}. Core collapse of the $8.8 M_\odot$ progenitor of \cite{Nomoto:1984,Nomoto:1987} is simulated in 1D with the {\tt PROMETHEUS-VERTEX} code long-term, until $\sim 25$ seconds. A variable Eddington-factor closure scheme is used to model Boltzmann transport \cite{Rampp:2002bq}.

\begin{figure}
\includegraphics[width=125mm]{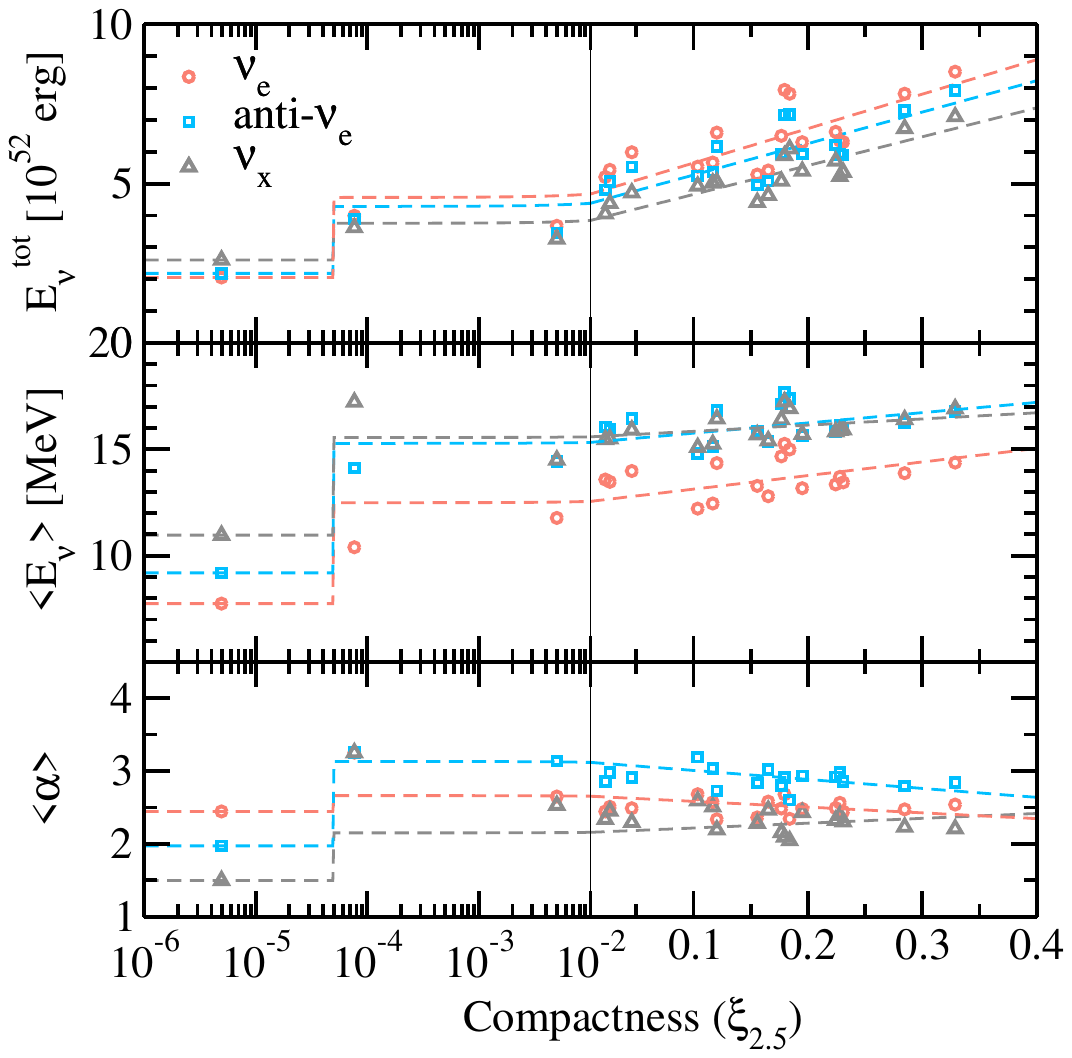}
\caption{Time-integrated neutrino spectral parameters: total neutrino energetic (top panel), mean energy (middle panel), and pinching parameter (bottom panel), shown separately for $\nu_e$ (red circles), $\bar{\nu}_e$ (blue squares), and $\nu_x$ (gray triangles). These are based on the 2D simulations of S16 augmented by simulations of a $9.6 M_\odot$ star and a $8.8 M_\odot$ ONeMg star. Note the axis change between the left and right, in logarithmic and linear, respectively. The dashed lines indicate our phenomenological fits through the simulations.}
\label{fig:parameters}
\end{figure}

At each simulation time snapshot, the neutrino energy spectrum is well described by a pinched Fermi-Dirac form described by three parameters, 
\begin{eqnarray}\label{pinchedFD}
    f(E) &=& \frac{(1+  \langle \alpha \rangle )^{(1+  \langle \alpha \rangle )}}{\Gamma(1+  \langle \alpha \rangle )}  \\ \nonumber 
    && \quad \times \frac{E^{\rm tot}_\nu E^{ \langle \alpha \rangle }}{\langle E_\nu \rangle^{2+  \langle \alpha \rangle }} {\rm exp}\left[ {-(1+  \langle \alpha \rangle ) \frac{E}{ \langle E_\nu \rangle }} \right],
\end{eqnarray}
where $\langle E \rangle$ is the mean neutrino energy, $\alpha$ is the pinching parameter, and $E_\nu^{\rm tot}$ is the total neutrino energy in the time bin. Since the models of S16 and the $9.6M_\odot$ simulation were terminated typically at $\sim 0.5$--1 seconds post bounce, we estimate the late-time neutrino emission using the following assumptions: (1) the protoneutron star radius contracts from the radius at the final time- step to a final radius of 15 km; (2) the protoneutron star mass grows according to $M(r) = M_0 + M_1(1 -{\rm e}^{-t/\tau_M})$, where $M_0$, $M_1$, and $\tau_M$ are found by fitting this function to the time evolution of the protoneutron mass during the simulated epoch; (3) the gravitational binding energy released after the final simulation time step is equipartitioned between all neutrino flavours; and (4) the average neutrino energies and pinching factors are fixed to the final value of each simulation. The simulation of the $8.8 M_\odot$ progenitor includes the late-time cooling phase and does not need such extrapolations. 

The neutrino emission is then summed over time, adopting the set of $(L,\langle E \rangle,\alpha)$ at each time step. The summed emission is then fitted to a functional form identical to Eq.~(\ref{pinchedFD}), except now the normalization $E_\nu^{\rm tot}$ represents the neutrino energetics over the entire duration of the CCSN. The results of the fits are shown in Fig.~\ref{fig:parameters}, where for each progenitor the neutrino emission parameters $E_\nu^{\rm tot}$, $\langle E \rangle$, and $\alpha$ are plotted against the compactness of the progenitor. Here, the compactness is simple parameter to capture the size of the core at the moment of core collapse,
\begin{equation} \label{eq:compactness}
\xi_M = \left. \frac{M/M_\odot}{R(M_{\rm bary}=M)/1000\,{\rm km}} \right\vert_{ t },
\end{equation}
where $R(M_{\rm bary}=M)$ is the radial coordinate that encloses a baryonic mass $M$ at epoch $t$. First described in \cite{O'Connor:2010tk} as a proxy for collapse to black holes, it has been shown to be also a versatile indicator including the neutrino emission \cite{Horiuchi:2017qlw,Horiuchi:2017sku}. We adopt $t$ as the onset of collapse as was done in Ref.~\cite{Sukhbold:2013yca}, rather than at core bounce as in Ref.~\cite{O'Connor:2010tk}. Figure \ref{fig:parameters} reveals a growing trend of higher compactness stars emitting more neutrinos with higher mean energies, consistent with previous findings in the literature. 

\begin{figure}
\includegraphics[width=125mm]{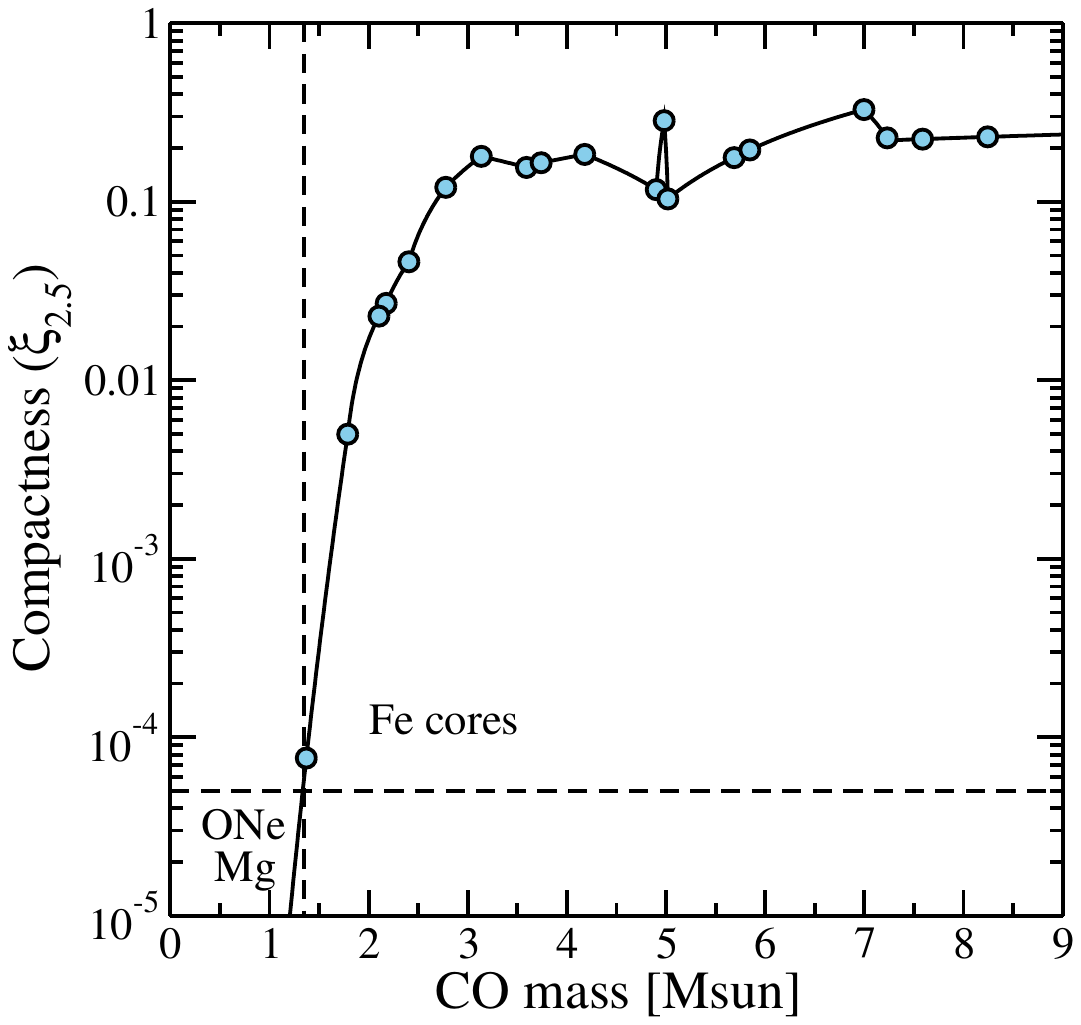}
\caption{The core compactness (defined by $M=2.5 M_\odot$) at moment of core-collapse, as a function of the CO core mass. The dashed lines separate the ONeMg regime and the Fe core regime in both compactness and CO mass planes.}
\label{fig:conversion}
\end{figure}

We fill in the gaps between our 20 progenitors by performing a linear fit through the simulation outputs versus compactness. When performing a global fit through both ONeMg and Fe core progenitors combined, we find a systematically lower neutrino output in the ONeMg case. Therefore, we perform a fit only on the iron core progenitors (19 in total), and maintain the ONeMg case separately. We put the separation at a compactness value of $\xi_{2.5} = 5 \times 10^{-5}$, which is between the $8.8M_\odot$ ONeMg core and the lowest compactness Fe core, $11.2 M_\odot$ progenitor. The final fit results are shown by the dashed lines in Fig.~\ref{fig:parameters}. 

To incorporate our new CO mass distributions, we map the core compactness of our adopted core-collapse progenitors to the CO core mass, as shown in Fig.~\ref{fig:conversion}. In general, the relation is non-monotonic, and recently has been shown to depend on also the ratio of C and O masses \cite{Patton:2020tiy}. However, for typical solar mass progenitors they should trace a trajectory within the large C/O mass plane \cite{Patton:2020tiy}. We interpolate between core-collapse progenitors, but for CO mass $<2 M_\odot$ and $>7 M_\odot$ we use a linear fit through the last few progenitors, as shown by the black curve. The dashed lines in Fig.~\ref{fig:conversion} illustrates the transition between ONeMg and Fe core regions, in both compactness and CO mass. The separation in CO mass is $M_{\rm CO} = 1.35 M_\odot$ \cite{Suwa:2018,2016MNRAS.463.3461S}. 

Finally, we compute the mean neutrino emissions by summing the neutrino emission for each CO mass bin, appropriate weighted by the CO core mass distribution. To account for changes in the number of core-collapse progenitors, we apply a re-normalization factor $f_b$ (see final column of Table \ref{tab:counts}) to our predictions. As discussed in Section \ref{sec:binary:progenitors}, this physically arises from the increased mass range for core-collapse progenitors, i.e., a smaller $M_{\rm min}$. While direct measurements of $R_{CC}$ in principle allows a data-driven approach to constrain $f_b$, in reality measurements are strongly affected by systematic uncertainties due to corrections for dust and missing CCSNe. Furthermore, older direct measurements of $R_{CC}$ and estimates from the SFR show hints of a systematic normalization discrepancy at a factor $\sim 2$ level \cite{Horiuchi:2011zz}. This discrepancy may be explained by new $R_{CC}$ measurements (e.g., \cite{Dahlen:2012cm,Perley:2020ajb}) and/or new SFR (e.g., \cite{Mathews:2014qba,Madau:2014bja}). We therefore do not attempt to constrain large binary enhancements based on rates alone. 

\begin{figure}
\includegraphics[width=125mm]{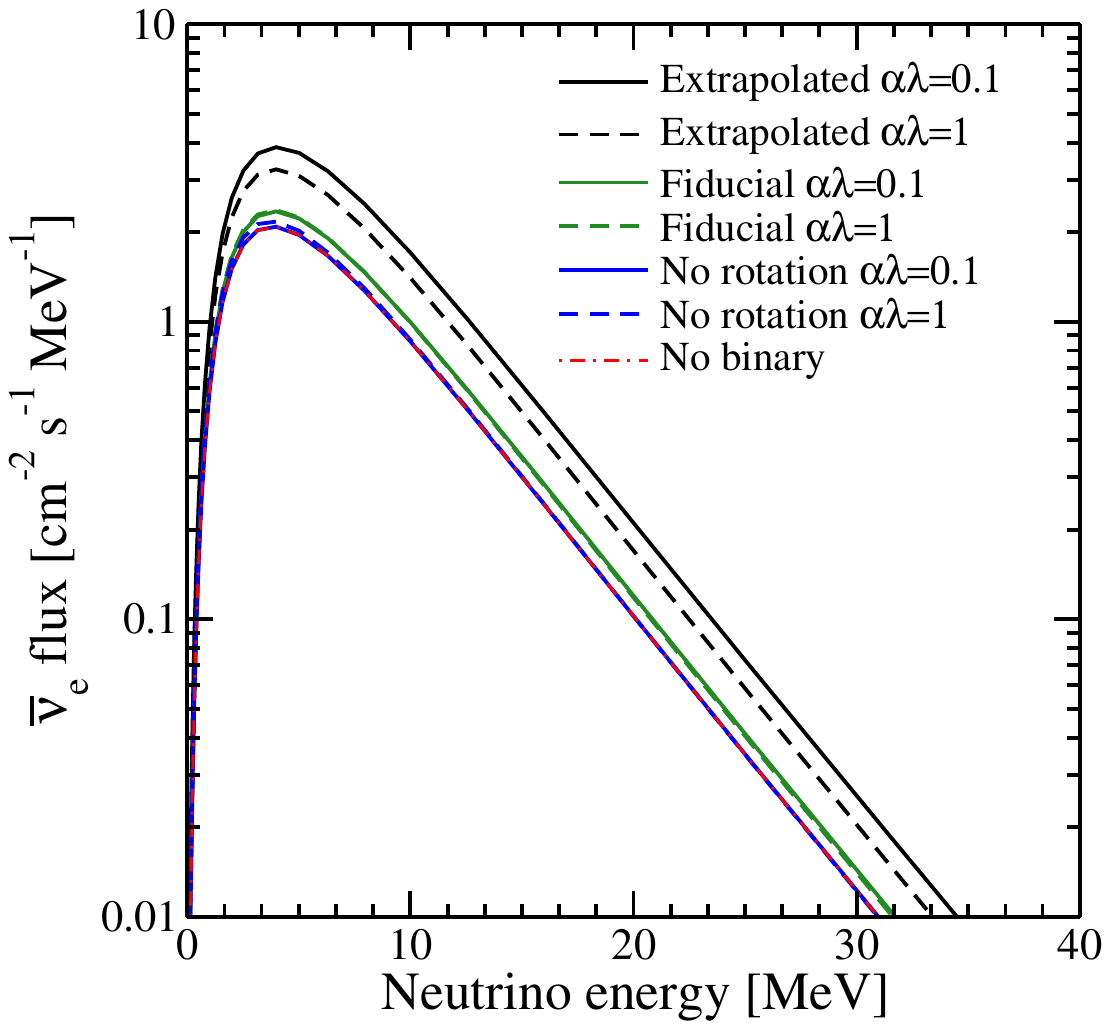}
\caption{The predicted DSNB flux of $\bar{\nu}_e$ for NH, for 6 different binary population synthesis models, compared with a single stellar evolution model. The binary population synthesis models include different treatments for the postmerger rapidly rotating stars and different CE modeling. We find that the minimal estimate, which neglects postmerger rotation effects (blue solid and dashed), yields a DSNB flux indistinguishable from a single stellar evolution model (red dot-dashed), while our fiducial scheme (green solid and dashed) and the extrapolated scheme (black solid and dashed) yield higher DSNB fluxes.}
\label{fig:DSNBflux}
\end{figure}

\subsection{DSNB predictions}

The neutrinos that are emitted by the collapsing cores of massive stars undergo flavor transformations during their propagation to Earth. We implement the well-established matter-induced Mikheyev-Smirnov-Wolfenstein (MSW) effect which occurs during the neutrino's propagation through the progenitor \cite{Mirizzi:2015eza}. The mixing of neutrino flavors depends on the neutrino mass hierarchy. For example, the $\bar{\nu}_e$ survival probability is $\cos^2\theta_{12}\approx0.7$ in the normal mass hierarchy (NH), and $\approx 0$ in the inverted mass hierarchy (IH). The terrestrial flux, $F^{\rm obs}_{\bar{\nu}_e}$, as a mixture of fluxes at core collapse, $F_i$, is then,
\begin{eqnarray}
{\rm(NH)} \quad F^{\rm obs}_{\bar{\nu}_e} &\simeq& \cos^2 \theta_{12} F_{\bar{\nu}_e} + \sin^2 \theta_{12} F_{{\nu}_x}~, \\
{\rm(IH)} \,\, \quad F^{\rm obs}_{\bar{\nu}_e} &\simeq& F_{{\nu}_x},
\end{eqnarray}
Additional flavor mixing can be induced by the coherent neutrino-neutrino forward scattering potential. However, this so-called collective oscillations are complex and the detailed oscillation predictions and their time-dependence are far from known (see, e.g., review \cite{Tamborra:2020cul}). Since the DSNB  samples the entire core-collapse population, it may be expected that any particular features (e.g., spectral splits caused by swaps) may be washed out, resulting in an overall minimal effect. Also, the differences between neutrino flavors are not large during the cooling phase. The effect for the DSNB has even been estimated to be less than $\sim 10$\% \cite{Lunardini:2012ne}. However, it is not clear to what extent this holds in light of recently discovered flavor instabilities, e.g., due to spontaneously broken symmetries \cite{Raffelt:2013rqa,Abbar:2015mca,Chakraborty:2014nma,Mirizzi:2015fva,Zaizen:2020xum} or so-called fast conversion \cite{Sawyer:2005jk,Sawyer:2008zs}. Therefore, we do not consider mixing effects beyond MSW.

\begin{table*}[t]
\vspace{1em}
\begin{tabular}{|l|l|l|l|l|}
\hline
                        & \multicolumn{2}{c|}{SK-Gd [/yr]}     &   \multicolumn{2}{c|}{HK [/yr]}      \\ 
                        & Normal    & Inverted     & Normal   & Inverted    \\ 
\hline \hline
No binary evolution                  & 2.3   & 2.4   & 5.5  & 6.2      \\ 
\hline
Binary $\alpha\lambda=0.1$  Extrapolated & 4.7   & 4.6   & 11.4  & 12.0      \\ 
Binary $\alpha\lambda=0.1$  Fiducial  & 2.7   & 2.7   & 6.4   & 7.1      \\ 
Binary $\alpha\lambda=0.1$  No rotation & 2.3   & 2.4   & 5.5   & 6.2      \\ 
\hline
Binary $\alpha\lambda=1$  Extrapolated  & 3.8   & 3.8   & 9.1   & 9.9      \\ 
Binary $\alpha\lambda=1$  Fiducial    & 2.7   & 2.7   & 6.3   & 7.0      \\ 
Binary $\alpha\lambda=1$  No rotation  & 2.3   & 2.5   & 5.5   & 6.4      \\ 
\hline 
\end{tabular}
\caption{Annual DSNB event rates at Super-K (Hyper-K) employing 22.5 kton (187 kton) fiducial volume. Signals are computed over an energy range of $E_e = [10,28]$ MeV ($[18,28]$ MeV) for gadolinium-doped Super-K (pure-water Hyper-K). Perfect detection efficiencies are used and only MSW oscillations are included shown separately for normal/inverted mass hierarchies.}
\label{tab:dsnb}
\end{table*}

Figure \ref{fig:DSNBflux} shows the DSNB predictions for our three population synthesis models: fiducial, extrapolated, and no rotation, each for our two different choice of CE parameter $\alpha \lambda=0.1$ and 1. We also add the prediction of the single stellar evolution for comparison. For clarify, only the NH case is considered, and only the $\bar{\nu}_e$ flux is shown; results for the IH and other flavors show similar trends. We see that in our pessimistic no rotation scheme, where postmerger rotation effects are neglected, the DSNB flux remains to good accuracy the same as the no binary case. However, for both our fiducial and extrapolated schemes, the DSNB flux is noticeably increased. 

To estimate the DSNB event rates, we consider separately the NH and IH, and compute the event rates at two detector setups: Super-K assuming a fiducial volume of 22.5 kton containing gadolinium doped water for a signal energy range of 10--28 MeV, and Hyper-Kamiokande (Hyper-K) assuming a fiducial volume of 187 kton containing pure water for a signal energy range of 18--28 MeV. The event rates for Hyper-K filled with gadolinium-doped water can be approximately made by scaling the Super-K-Gd numbers by the fiducial volume. We assume perfect detection efficiency for simplicity. The predicted DSNB event rates are summarized in Table \ref{tab:dsnb}. 

In the absence of binary interactions, we predict annual event rates of 2.3--2.4  yr$^{-1}$ at Super-K-Gd and 5.5--6.2 yr$^{-1}$ at Hyper-K. While Hyper-K is $\sim 8$ times the volume of Super-K, its predicted signal rate is only $\sim 2.4$ that of Super-K. This is because we assume a narrower signal energy window for Hyper-K due to larger backgrounds. 

When binary interactions are included, we observe increases in the predicted DSNB event rates. The largest increase is seen in our extrapolated scheme, based on extrapolating the Limongi models, where the predicted rates at Super-K-Gd and Hyper-K are 4.6--4.7 yr$^{-1}$ and 11.4--12.0 yr$^{-1}$, respectively, i.e., a factor $\sim 2$ increase. The increase is more modest in our fiducial scheme, generally in the 15\%--20\% range compared to the no binary case. In our most conservative case where we neglect the effect of increased rotation in mergers, the change is at the percent level. It is important to stress that quantitatively these rates serve primarily to compare the impacts of with and without binary interactions, rather than an attempt at deriving the most accurate rate predictions, the latter of which will be affected by, e.g., contributions from collapse to black holes, more detailed binary modeling, more sophisticated long-term supernova collapse simulations, neutrino oscillations, and other input physics.

The rate changes can be understood by the balance of two competing effects. On the one hand, the mergers of stars whose individual masses were high enough to each undergo core collapse reduce the DSNB. While the merged star will be more massive than the individual stars and therefore will have a stronger neutrino emission, it cannot compensate for the reduction from two to one progenitor. On the other hand, the mergers of light stars whose individual masses are below the core collapse threshold can increase the DSNB if the combined mass is above the core collapse threshold. The net effect can be seen in the counts in Table \ref{tab:counts}. In the no rotation case, we find that these effects nearly cancel each other out, with a net change in the total number of core-collapse progenitors of less than percent (for $\alpha\lambda=0.1$) and $+5$\% (for $\alpha\lambda=1.0$) compared with the no binary case. More impressively, both in our fiducial and extrapolated schemes, the mergers of even light stars can form a core-collapse progenitor owing to rotational effects, reducing the minimal core mass for core collapse. The increase through this additional population results in net changes in the total number of core-collapse progenitors of $+24$\% and $+53$--$75$\% for the fiducial and extrapolated schemes, respectively. These largely explain the major trends seen in the DSNB rates. 

A secondary effect occurs due to mass transfer in binary systems. Since our mass transfer prescription conserves mass, any mass lost by a star is accreted onto its companion. We can quantify the impact of this by comparing the \emph{shape} of CO mass among different population syntheses, i.e., we renormalize the nonmerge populations to match the no binary counts and compare the DSNB event rates. We find that the DSNB event rate increases by 2\%--3\% for $\alpha\lambda =0.1$, while the increase is smaller for $\alpha\lambda =1$. We therefore conclude that the effect of mass stripping is more or less compensated by the effect of mass gained by its companion. 

\section{Discussion and conclusions}

In this paper, we explored the impacts of binary interactions on the predicted DSNB detection rates. We incorporate the key effects of binary interactions---mass transfer and mergers---through binary population synthesis calculations to predict the final distributions of CO core masses of progenitors that undergo core collapse at the end of stellar evolution. Using a suite of 20 core-collapse supernova simulations, we compute the neutrino emissions from such progenitor distributions and predict the DSNB detection rate at present and future neutrino detectors. 

\begin{figure}
\includegraphics[width=125mm]{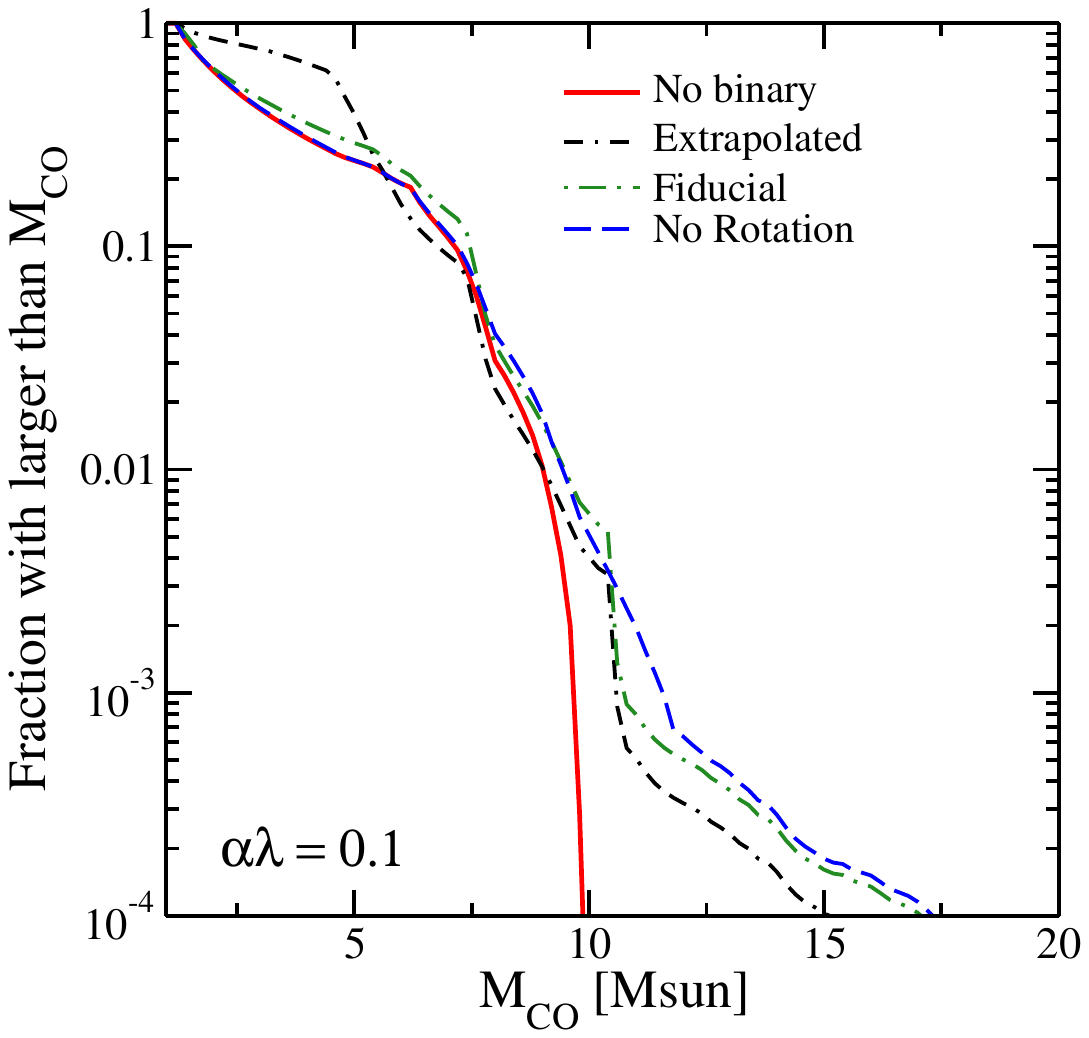}
\caption{Reverse cumulative plot showing the fraction of progenitors with higher CO mass. Shown are the results for no binary treatment (solid red), the extrapolated scheme (black dot-dashed-dashed), our fiducial scheme (green dot-dashed), and a scheme neglecting postmerger rotation (dashed blue). Binary populations have higher fractions of CO cores massive enough to likely collapse to black holes ($M_{\rm CO} \gtrsim 10 M_\odot$) than a no binary population. CE parameter is $\alpha\lambda=0.1$.}
\label{fig:cumulative}
\end{figure}

We find that binary effects can be large. Compared to predictions without binary considerations, we find increases of 15--20\% for our fiducial scheme. However, it is as small as 1\% in our estimate ignoring rotation effects, or as large as 70\% in our extrapolated scheme. Binary effects can be separated into three effects: change in mass due to mass transfer, reduced progenitor counts due to mergers of stars originally massive enough to core collapse, and increased progenitor counts due to mergers of stars originally not massive enough to core collapse (and massive enough after merger). Of these, the latter has the strongest effect. However, its contribution to the DSNB depends on the evolution of the postmerger stars which would be rapidly rotating as a result of the merger. Completely ignoring the rotation or extrapolating the results of Limongi et al.~\cite{Limongi2017} yields increases of 1--70\%. Therefore, it is important to further investigate the evolution of such postmerger, rapidly rotating stars near the white dwarf / core collapse threshold.

Our results differ from those of Ref.~\cite{Kresse:2020nto} who conclude that binary interactions yield a reduction of the DSNB. However, their study uses the suite of hydrogen-stripped progenitors of Ref.~\cite{Woosley2019} as a proxy for stars undergoing close-binary interactions. Therefore, it includes the effect of mass stripping from stars, but no consideration for mass gained by its binary companion. The study also did not include the impacts of stellar mergers. Our study differs in that we perform binary population syntheses incorporating mass transfer and mergers. We find that mergers can cause a sizable increase in the core-collapse progenitor population numbers and hence the DSNB rates. 

In this study, we have not included other contributions to the DSNB, most notably the collapse to black holes. This is because our intent is to focus on the impacts of binary interactions on the primary population: collapse to neutron stars. Binary interactions would also affect the fraction of massive stars that collapse to black holes. Indeed, our binary population synthesis is already being applied to interpret compact object merger rates (e.g., Refs.\ \cite{Kinugawa2014,Yu2015,Ablimit2018}). However, the emission of neutrinos from black holes comes with additional uncertainties including the equation of state of hot dense matter and the fraction of massive stars which collapse to black hole. We can still speculate what the impact may be, based on the relative fractions of potential black hole progenitors. Figure \ref{fig:cumulative} shows the reverse cumulative plot of CO mass, i.e., the fraction of stars with higher CO mass. It is evident that all our binary population syntheses have a long tail of high CO mass progenitors, but these remain too rare to make substantial impact on the DSNB. More important would be the stars near the neutron star/black hole boundary, around CO core mass of $\sim 10 M_\odot$. Compared to the no binary model, our binary models have larger fractions, implying any consideration of collapse to black holes will likely enhance our results due to their harder neutrino spectra.

It should be emphasized that there are various sources of uncertainties affecting predictions of the DSNB which we have neglected, e.g., from the cosmic core-collapse rate, collective neutrino oscillation effects, as well as uncertainties arising from different simulation suites. Our intent here is not to make the most accurate or robust error estimates of the DSNB. Rather, we aim to highlight the potential impacts of binary interactions on the DSNB flux. Thorough discussions of the DSNB uncertainties be found in the literature, e.g., \cite{Kresse:2020nto}.  

While the DSNB has not yet been detected, this is expected to change in the coming decade. Super-K has completed its gadolinium upgrade, and next-generation detectors such as Hyper-K and DUNE are under construction. Interpreting the results of DSNB searches will require a commensurate effort in theoretical modeling of the DSNB, including binary effects which massive stars are subject to. As simulations advance, DSNB predictions will continue to improve and allow future DSNB detections to reveal the diversity in massive stellar core collapse and their neutrino emissions.

\begin{acknowledgments}
We thank Shin'ichiro Ando, Takami Kuroda, and Masamichi Zaizen for useful discussions, and Hans-Thomas Janka for sharing simulations data of S16. We also thank the hospitality of the organizers of the workshop ``Workshop on core-collapse supernova explosions and related physics'' where parts of the discussion took place. S.H.\ is supported by the U.S.\ Department of Energy Office of Science under award No.~DE-SC0020262 and NSF Grants No.~AST-1908960 and No.~PHY-1914409. This study was supported in part by the Grants-in-Aid for the Scientific Research of Japan Society for the Promotion of Science (JSPS) KAKENHI Grant Number (JP17H05206, JP17K14306, and JP17H01130, JP17H06364, JP18H01212 (T.T.\ and K.K.), by the Central Research Institute of Explosive Stellar Phenomena (REISEP) of Fukuoka University and the associated project (No.\ 207002), and by Joint Institute for Computational Fundamental Science (JICFuS) as ``Program for Promoting Researches on the Supercomputer Fugaku'' (Toward a unified view of the universe: from large scale structures to planets).
The numerical simulations were done using XC50 at Center for Computational Astrophysics at National Astronomical Observatory of Japan. T.K.\ acknowledges support from University of Tokyo Young Excellent Researcher program.
\end{acknowledgments}

\bibliography{ref.bib}

\begin{thebibliography}{126}%
\makeatletter
\providecommand \@ifxundefined [1]{%
 \@ifx{#1\undefined}
}%
\providecommand \@ifnum [1]{%
 \ifnum #1\expandafter \@firstoftwo
 \else \expandafter \@secondoftwo
 \fi
}%
\providecommand \@ifx [1]{%
 \ifx #1\expandafter \@firstoftwo
 \else \expandafter \@secondoftwo
 \fi
}%
\providecommand \natexlab [1]{#1}%
\providecommand \enquote  [1]{``#1''}%
\providecommand \bibnamefont  [1]{#1}%
\providecommand \bibfnamefont [1]{#1}%
\providecommand \citenamefont [1]{#1}%
\providecommand \href@noop [0]{\@secondoftwo}%
\providecommand \href [0]{\begingroup \@sanitize@url \@href}%
\providecommand \@href[1]{\@@startlink{#1}\@@href}%
\providecommand \@@href[1]{\endgroup#1\@@endlink}%
\providecommand \@sanitize@url [0]{\catcode `\\12\catcode `\$12\catcode
  `\&12\catcode `\#12\catcode `\^12\catcode `\_12\catcode `\%12\relax}%
\providecommand \@@startlink[1]{}%
\providecommand \@@endlink[0]{}%
\providecommand \url  [0]{\begingroup\@sanitize@url \@url }%
\providecommand \@url [1]{\endgroup\@href {#1}{\urlprefix }}%
\providecommand \urlprefix  [0]{URL }%
\providecommand \Eprint [0]{\href }%
\providecommand \doibase [0]{http://dx.doi.org/}%
\providecommand \selectlanguage [0]{\@gobble}%
\providecommand \bibinfo  [0]{\@secondoftwo}%
\providecommand \bibfield  [0]{\@secondoftwo}%
\providecommand \translation [1]{[#1]}%
\providecommand \BibitemOpen [0]{}%
\providecommand \bibitemStop [0]{}%
\providecommand \bibitemNoStop [0]{.\EOS\space}%
\providecommand \EOS [0]{\spacefactor3000\relax}%
\providecommand \BibitemShut  [1]{\csname bibitem#1\endcsname}%
\let\auto@bib@innerbib\@empty
\bibitem [{\citenamefont {Langanke}\ and\ \citenamefont
  {Mart{\'i}nez-Pinedo}(2003)}]{Langanke:2002ab}%
  \BibitemOpen
  \bibfield  {author} {\bibinfo {author} {\bibfnamefont {K.}~\bibnamefont
  {Langanke}}\ and\ \bibinfo {author} {\bibfnamefont {G.}~\bibnamefont
  {Mart{\'i}nez-Pinedo}},\ }\href {\doibase 10.1103/RevModPhys.75.819}
  {\bibfield  {journal} {\bibinfo  {journal} {Rev. Mod. Phys.}\ }\textbf
  {\bibinfo {volume} {75}},\ \bibinfo {pages} {819} (\bibinfo {year} {2003})},\
  \Eprint {http://arxiv.org/abs/nucl-th/0203071} {arXiv:nucl-th/0203071
  [nucl-th]} \BibitemShut {NoStop}%
\bibitem [{\citenamefont {Mezzacappa}(2005)}]{Mezzacappa:2005ju}%
  \BibitemOpen
  \bibfield  {author} {\bibinfo {author} {\bibfnamefont {A.}~\bibnamefont
  {Mezzacappa}},\ }\href {\doibase 10.1146/annurev.nucl.55.090704.151608}
  {\bibfield  {journal} {\bibinfo  {journal} {Ann. Rev. Nucl. Part. Sci.}\
  }\textbf {\bibinfo {volume} {55}},\ \bibinfo {pages} {467} (\bibinfo {year}
  {2005})}\BibitemShut {NoStop}%
\bibitem [{\citenamefont {Kotake}\ \emph {et~al.}(2006)\citenamefont {Kotake},
  \citenamefont {Sato},\ and\ \citenamefont {Takahashi}}]{Kotake:2005zn}%
  \BibitemOpen
  \bibfield  {author} {\bibinfo {author} {\bibfnamefont {K.}~\bibnamefont
  {Kotake}}, \bibinfo {author} {\bibfnamefont {K.}~\bibnamefont {Sato}}, \ and\
  \bibinfo {author} {\bibfnamefont {K.}~\bibnamefont {Takahashi}},\ }\href
  {\doibase 10.1088/0034-4885/69/4/R03} {\bibfield  {journal} {\bibinfo
  {journal} {Rept. Prog. Phys.}\ }\textbf {\bibinfo {volume} {69}},\ \bibinfo
  {pages} {971} (\bibinfo {year} {2006})},\ \Eprint
  {http://arxiv.org/abs/astro-ph/0509456} {arXiv:astro-ph/0509456 [astro-ph]}
  \BibitemShut {NoStop}%
\bibitem [{\citenamefont {Woosley}\ and\ \citenamefont
  {Janka}(2005)}]{Woosley:2006ie}%
  \BibitemOpen
  \bibfield  {author} {\bibinfo {author} {\bibfnamefont {S.}~\bibnamefont
  {Woosley}}\ and\ \bibinfo {author} {\bibfnamefont {T.}~\bibnamefont
  {Janka}},\ }\href {\doibase 10.1038/nphys172} {\bibfield  {journal} {\bibinfo
   {journal} {Nature Phys.}\ }\textbf {\bibinfo {volume} {1}},\ \bibinfo
  {pages} {147} (\bibinfo {year} {2005})},\ \Eprint
  {http://arxiv.org/abs/astro-ph/0601261} {arXiv:astro-ph/0601261 [astro-ph]}
  \BibitemShut {NoStop}%
\bibitem [{\citenamefont {Burrows}(2013)}]{Burrows:2012ew}%
  \BibitemOpen
  \bibfield  {author} {\bibinfo {author} {\bibfnamefont {A.}~\bibnamefont
  {Burrows}},\ }\href {\doibase 10.1103/RevModPhys.85.245} {\bibfield
  {journal} {\bibinfo  {journal} {Rev. Mod. Phys.}\ }\textbf {\bibinfo {volume}
  {85}},\ \bibinfo {pages} {245} (\bibinfo {year} {2013})},\ \Eprint
  {http://arxiv.org/abs/1210.4921} {arXiv:1210.4921 [astro-ph.SR]} \BibitemShut
  {NoStop}%
\bibitem [{\citenamefont {Janka}(2012)}]{Janka:2012wk}%
  \BibitemOpen
  \bibfield  {author} {\bibinfo {author} {\bibfnamefont {H.-T.}\ \bibnamefont
  {Janka}},\ }\href {\doibase 10.1146/annurev-nucl-102711-094901} {\bibfield
  {journal} {\bibinfo  {journal} {Ann. Rev. Nucl. Part. Sci.}\ }\textbf
  {\bibinfo {volume} {62}},\ \bibinfo {pages} {407} (\bibinfo {year} {2012})},\
  \Eprint {http://arxiv.org/abs/1206.2503} {arXiv:1206.2503 [astro-ph.SR]}
  \BibitemShut {NoStop}%
\bibitem [{\citenamefont {Foglizzo}\ \emph {et~al.}(2015)\citenamefont
  {Foglizzo} \emph {et~al.}}]{Foglizzo:2015dma}%
  \BibitemOpen
  \bibfield  {author} {\bibinfo {author} {\bibfnamefont {T.}~\bibnamefont
  {Foglizzo}} \emph {et~al.},\ }\href {\doibase 10.1017/pasa.2015.9} {\bibfield
   {journal} {\bibinfo  {journal} {Publ. Astron. Soc. Austral.}\ }\textbf
  {\bibinfo {volume} {32}},\ \bibinfo {pages} {9} (\bibinfo {year} {2015})},\
  \Eprint {http://arxiv.org/abs/1501.01334} {arXiv:1501.01334 [astro-ph.HE]}
  \BibitemShut {NoStop}%
\bibitem [{\citenamefont {Janka}(2017)}]{Janka:2017vlw}%
  \BibitemOpen
  \bibfield  {author} {\bibinfo {author} {\bibfnamefont {H.~T.}\ \bibnamefont
  {Janka}},\ }\href@noop {} {\  (\bibinfo {year} {2017})},\ \Eprint
  {http://arxiv.org/abs/1702.08713} {arXiv:1702.08713 [astro-ph.HE]}
  \BibitemShut {NoStop}%
\bibitem [{\citenamefont {Burrows}\ and\ \citenamefont
  {Vartanyan}(2020)}]{Burrows:2020qrp}%
  \BibitemOpen
  \bibfield  {author} {\bibinfo {author} {\bibfnamefont {A.}~\bibnamefont
  {Burrows}}\ and\ \bibinfo {author} {\bibfnamefont {D.}~\bibnamefont
  {Vartanyan}},\ }\href@noop {} {\  (\bibinfo {year} {2020})},\ \Eprint
  {http://arxiv.org/abs/2009.14157} {arXiv:2009.14157 [astro-ph.SR]}
  \BibitemShut {NoStop}%
\bibitem [{\citenamefont {Scholberg}(2012)}]{Scholberg:2012id}%
  \BibitemOpen
  \bibfield  {author} {\bibinfo {author} {\bibfnamefont {K.}~\bibnamefont
  {Scholberg}},\ }\href {\doibase 10.1146/annurev-nucl-102711-095006}
  {\bibfield  {journal} {\bibinfo  {journal} {Ann. Rev. Nucl. Part. Sci.}\
  }\textbf {\bibinfo {volume} {62}},\ \bibinfo {pages} {81} (\bibinfo {year}
  {2012})},\ \Eprint {http://arxiv.org/abs/1205.6003} {arXiv:1205.6003
  [astro-ph.IM]} \BibitemShut {NoStop}%
\bibitem [{\citenamefont {Mirizzi}\ \emph {et~al.}(2016)\citenamefont
  {Mirizzi}, \citenamefont {Tamborra}, \citenamefont {Janka}, \citenamefont
  {Saviano}, \citenamefont {Scholberg}, \citenamefont {Bollig}, \citenamefont
  {Hudepohl},\ and\ \citenamefont {Chakraborty}}]{Mirizzi:2015eza}%
  \BibitemOpen
  \bibfield  {author} {\bibinfo {author} {\bibfnamefont {A.}~\bibnamefont
  {Mirizzi}}, \bibinfo {author} {\bibfnamefont {I.}~\bibnamefont {Tamborra}},
  \bibinfo {author} {\bibfnamefont {H.-T.}\ \bibnamefont {Janka}}, \bibinfo
  {author} {\bibfnamefont {N.}~\bibnamefont {Saviano}}, \bibinfo {author}
  {\bibfnamefont {K.}~\bibnamefont {Scholberg}}, \bibinfo {author}
  {\bibfnamefont {R.}~\bibnamefont {Bollig}}, \bibinfo {author} {\bibfnamefont
  {L.}~\bibnamefont {Hudepohl}}, \ and\ \bibinfo {author} {\bibfnamefont
  {S.}~\bibnamefont {Chakraborty}},\ }\href {\doibase
  10.1393/ncr/i2016-10120-8} {\bibfield  {journal} {\bibinfo  {journal} {Riv.
  Nuovo Cim.}\ }\textbf {\bibinfo {volume} {39}},\ \bibinfo {pages} {1}
  (\bibinfo {year} {2016})},\ \Eprint {http://arxiv.org/abs/1508.00785}
  {arXiv:1508.00785 [astro-ph.HE]} \BibitemShut {NoStop}%
\bibitem [{\citenamefont {Horiuchi}\ and\ \citenamefont
  {Kneller}(2018)}]{Horiuchi:2017sku}%
  \BibitemOpen
  \bibfield  {author} {\bibinfo {author} {\bibfnamefont {S.}~\bibnamefont
  {Horiuchi}}\ and\ \bibinfo {author} {\bibfnamefont {J.~P.}\ \bibnamefont
  {Kneller}},\ }\href {\doibase 10.1088/1361-6471/aaa90a} {\bibfield  {journal}
  {\bibinfo  {journal} {J. Phys. G}\ }\textbf {\bibinfo {volume} {45}},\
  \bibinfo {pages} {043002} (\bibinfo {year} {2018})},\ \Eprint
  {http://arxiv.org/abs/1709.01515} {arXiv:1709.01515 [astro-ph.HE]}
  \BibitemShut {NoStop}%
\bibitem [{\citenamefont {Diehl}\ \emph {et~al.}(2006)\citenamefont {Diehl}
  \emph {et~al.}}]{Diehl:2006cf}%
  \BibitemOpen
  \bibfield  {author} {\bibinfo {author} {\bibfnamefont {R.}~\bibnamefont
  {Diehl}} \emph {et~al.},\ }\href {\doibase 10.1038/nature04364} {\bibfield
  {journal} {\bibinfo  {journal} {Nature}\ }\textbf {\bibinfo {volume} {439}},\
  \bibinfo {pages} {45} (\bibinfo {year} {2006})},\ \Eprint
  {http://arxiv.org/abs/astro-ph/0601015} {arXiv:astro-ph/0601015 [astro-ph]}
  \BibitemShut {NoStop}%
\bibitem [{\citenamefont {Rozwadowska}\ \emph {et~al.}(2021)\citenamefont
  {Rozwadowska}, \citenamefont {Vissani},\ and\ \citenamefont
  {Cappellaro}}]{Rozwadowska:2021lll}%
  \BibitemOpen
  \bibfield  {author} {\bibinfo {author} {\bibfnamefont {K.}~\bibnamefont
  {Rozwadowska}}, \bibinfo {author} {\bibfnamefont {F.}~\bibnamefont
  {Vissani}}, \ and\ \bibinfo {author} {\bibfnamefont {E.}~\bibnamefont
  {Cappellaro}},\ }\href {\doibase 10.1016/j.newast.2020.101498} {\bibfield
  {journal} {\bibinfo  {journal} {New Astron.}\ }\textbf {\bibinfo {volume}
  {83}},\ \bibinfo {pages} {101498} (\bibinfo {year} {2021})},\ \Eprint
  {http://arxiv.org/abs/2009.03438} {arXiv:2009.03438 [astro-ph.HE]}
  \BibitemShut {NoStop}%
\bibitem [{\citenamefont {Ando}\ \emph {et~al.}(2005)\citenamefont {Ando},
  \citenamefont {Beacom},\ and\ \citenamefont {Yuksel}}]{Ando:2005ka}%
  \BibitemOpen
  \bibfield  {author} {\bibinfo {author} {\bibfnamefont {S.}~\bibnamefont
  {Ando}}, \bibinfo {author} {\bibfnamefont {J.~F.}\ \bibnamefont {Beacom}}, \
  and\ \bibinfo {author} {\bibfnamefont {H.}~\bibnamefont {Yuksel}},\ }\href
  {\doibase 10.1103/PhysRevLett.95.171101} {\bibfield  {journal} {\bibinfo
  {journal} {Phys. Rev. Lett.}\ }\textbf {\bibinfo {volume} {95}},\ \bibinfo
  {pages} {171101} (\bibinfo {year} {2005})},\ \Eprint
  {http://arxiv.org/abs/astro-ph/0503321} {arXiv:astro-ph/0503321 [astro-ph]}
  \BibitemShut {NoStop}%
\bibitem [{\citenamefont {Kistler}\ \emph {et~al.}(2011)\citenamefont
  {Kistler}, \citenamefont {Yuksel}, \citenamefont {Ando}, \citenamefont
  {Beacom},\ and\ \citenamefont {Suzuki}}]{Kistler:2008us}%
  \BibitemOpen
  \bibfield  {author} {\bibinfo {author} {\bibfnamefont {M.~D.}\ \bibnamefont
  {Kistler}}, \bibinfo {author} {\bibfnamefont {H.}~\bibnamefont {Yuksel}},
  \bibinfo {author} {\bibfnamefont {S.}~\bibnamefont {Ando}}, \bibinfo {author}
  {\bibfnamefont {J.~F.}\ \bibnamefont {Beacom}}, \ and\ \bibinfo {author}
  {\bibfnamefont {Y.}~\bibnamefont {Suzuki}},\ }\href {\doibase
  10.1103/PhysRevD.83.123008} {\bibfield  {journal} {\bibinfo  {journal} {Phys.
  Rev.}\ }\textbf {\bibinfo {volume} {D83}},\ \bibinfo {pages} {123008}
  (\bibinfo {year} {2011})},\ \Eprint {http://arxiv.org/abs/0810.1959}
  {arXiv:0810.1959 [astro-ph]} \BibitemShut {NoStop}%
\bibitem [{\citenamefont {Horiuchi}\ \emph {et~al.}(2013)\citenamefont
  {Horiuchi}, \citenamefont {Beacom}, \citenamefont {Bothwell},\ and\
  \citenamefont {Thompson}}]{Horiuchi:2013bc}%
  \BibitemOpen
  \bibfield  {author} {\bibinfo {author} {\bibfnamefont {S.}~\bibnamefont
  {Horiuchi}}, \bibinfo {author} {\bibfnamefont {J.~F.}\ \bibnamefont
  {Beacom}}, \bibinfo {author} {\bibfnamefont {M.~S.}\ \bibnamefont
  {Bothwell}}, \ and\ \bibinfo {author} {\bibfnamefont {T.~A.}\ \bibnamefont
  {Thompson}},\ }\href {\doibase 10.1088/0004-637X/769/2/113} {\bibfield
  {journal} {\bibinfo  {journal} {Astrophys. J.}\ }\textbf {\bibinfo {volume}
  {769}},\ \bibinfo {pages} {113} (\bibinfo {year} {2013})},\ \Eprint
  {http://arxiv.org/abs/1302.0287} {arXiv:1302.0287 [astro-ph.SR]} \BibitemShut
  {NoStop}%
\bibitem [{\citenamefont {Nakamura}\ \emph {et~al.}(2016)\citenamefont
  {Nakamura}, \citenamefont {Horiuchi}, \citenamefont {Tanaka}, \citenamefont
  {Hayama}, \citenamefont {Takiwaki},\ and\ \citenamefont
  {Kotake}}]{Nakamura:2016kkl}%
  \BibitemOpen
  \bibfield  {author} {\bibinfo {author} {\bibfnamefont {K.}~\bibnamefont
  {Nakamura}}, \bibinfo {author} {\bibfnamefont {S.}~\bibnamefont {Horiuchi}},
  \bibinfo {author} {\bibfnamefont {M.}~\bibnamefont {Tanaka}}, \bibinfo
  {author} {\bibfnamefont {K.}~\bibnamefont {Hayama}}, \bibinfo {author}
  {\bibfnamefont {T.}~\bibnamefont {Takiwaki}}, \ and\ \bibinfo {author}
  {\bibfnamefont {K.}~\bibnamefont {Kotake}},\ }\href {\doibase
  10.1093/mnras/stw1453} {\bibfield  {journal} {\bibinfo  {journal} {Mon. Not.
  Roy. Astron. Soc.}\ }\textbf {\bibinfo {volume} {461}},\ \bibinfo {pages}
  {3296} (\bibinfo {year} {2016})},\ \Eprint {http://arxiv.org/abs/1602.03028}
  {arXiv:1602.03028 [astro-ph.HE]} \BibitemShut {NoStop}%
\bibitem [{\citenamefont {Beacom}(2010)}]{Beacom:2010kk}%
  \BibitemOpen
  \bibfield  {author} {\bibinfo {author} {\bibfnamefont {J.~F.}\ \bibnamefont
  {Beacom}},\ }\href {\doibase 10.1146/annurev.nucl.010909.083331} {\bibfield
  {journal} {\bibinfo  {journal} {Ann. Rev. Nucl. Part. Sci.}\ }\textbf
  {\bibinfo {volume} {60}},\ \bibinfo {pages} {439} (\bibinfo {year} {2010})},\
  \Eprint {http://arxiv.org/abs/1004.3311} {arXiv:1004.3311 [astro-ph.HE]}
  \BibitemShut {NoStop}%
\bibitem [{\citenamefont {Lunardini}(2016)}]{Lunardini:2010ab}%
  \BibitemOpen
  \bibfield  {author} {\bibinfo {author} {\bibfnamefont {C.}~\bibnamefont
  {Lunardini}},\ }\href {\doibase 10.1016/j.astropartphys.2016.02.005}
  {\bibfield  {journal} {\bibinfo  {journal} {Astropart. Phys.}\ }\textbf
  {\bibinfo {volume} {79}},\ \bibinfo {pages} {49} (\bibinfo {year} {2016})},\
  \Eprint {http://arxiv.org/abs/1007.3252} {arXiv:1007.3252 [astro-ph.CO]}
  \BibitemShut {NoStop}%
\bibitem [{\citenamefont {Bays}\ \emph {et~al.}(2012)\citenamefont {Bays} \emph
  {et~al.}}]{Bays:2011si}%
  \BibitemOpen
  \bibfield  {author} {\bibinfo {author} {\bibfnamefont {K.}~\bibnamefont
  {Bays}} \emph {et~al.} (\bibinfo {collaboration} {Super-Kamiokande}),\ }\href
  {\doibase 10.1103/PhysRevD.85.052007} {\bibfield  {journal} {\bibinfo
  {journal} {Phys. Rev.}\ }\textbf {\bibinfo {volume} {D85}},\ \bibinfo {pages}
  {052007} (\bibinfo {year} {2012})},\ \Eprint {http://arxiv.org/abs/1111.5031}
  {arXiv:1111.5031 [hep-ex]} \BibitemShut {NoStop}%
\bibitem [{\citenamefont {Watanabe}\ \emph {et~al.}(2009)\citenamefont
  {Watanabe} \emph {et~al.}}]{Watanabe:2008ru}%
  \BibitemOpen
  \bibfield  {author} {\bibinfo {author} {\bibfnamefont {H.}~\bibnamefont
  {Watanabe}} \emph {et~al.} (\bibinfo {collaboration} {Super-Kamiokande}),\
  }\href {\doibase 10.1016/j.astropartphys.2009.03.002} {\bibfield  {journal}
  {\bibinfo  {journal} {Astropart. Phys.}\ }\textbf {\bibinfo {volume} {31}},\
  \bibinfo {pages} {320} (\bibinfo {year} {2009})},\ \Eprint
  {http://arxiv.org/abs/0811.0735} {arXiv:0811.0735 [hep-ex]} \BibitemShut
  {NoStop}%
\bibitem [{\citenamefont {Beacom}\ and\ \citenamefont
  {Vagins}(2004)}]{Beacom:2003nk}%
  \BibitemOpen
  \bibfield  {author} {\bibinfo {author} {\bibfnamefont {J.~F.}\ \bibnamefont
  {Beacom}}\ and\ \bibinfo {author} {\bibfnamefont {M.~R.}\ \bibnamefont
  {Vagins}},\ }\href {\doibase 10.1103/PhysRevLett.93.171101} {\bibfield
  {journal} {\bibinfo  {journal} {Phys. Rev. Lett.}\ }\textbf {\bibinfo
  {volume} {93}},\ \bibinfo {pages} {171101} (\bibinfo {year} {2004})},\
  \Eprint {http://arxiv.org/abs/hep-ph/0309300} {arXiv:hep-ph/0309300 [hep-ph]}
  \BibitemShut {NoStop}%
\bibitem [{\citenamefont {Krauss}\ \emph {et~al.}(1984)\citenamefont {Krauss},
  \citenamefont {Glashow},\ and\ \citenamefont {Schramm}}]{Krauss:1983zn}%
  \BibitemOpen
  \bibfield  {author} {\bibinfo {author} {\bibfnamefont {L.~M.}\ \bibnamefont
  {Krauss}}, \bibinfo {author} {\bibfnamefont {S.~L.}\ \bibnamefont {Glashow}},
  \ and\ \bibinfo {author} {\bibfnamefont {D.~N.}\ \bibnamefont {Schramm}},\
  }\href {\doibase 10.1038/310191a0} {\bibfield  {journal} {\bibinfo  {journal}
  {Nature}\ }\textbf {\bibinfo {volume} {310}},\ \bibinfo {pages} {191}
  (\bibinfo {year} {1984})}\BibitemShut {NoStop}%
\bibitem [{\citenamefont {Dar}(1985)}]{Dar:1984aj}%
  \BibitemOpen
  \bibfield  {author} {\bibinfo {author} {\bibfnamefont {A.}~\bibnamefont
  {Dar}},\ }\bibfield  {booktitle} {\emph {\bibinfo {booktitle} {{Phys. Rev.
  Lett. 55 ( 1985) 1422-1425 and Haifa Isr. Inst. Technol. - TECHNION-PH-84-54
  (84,REC.JAN.85) 15 P. (500879)}}},\ }\href {\doibase
  10.1103/PhysRevLett.55.1422} {\bibfield  {journal} {\bibinfo  {journal}
  {Phys. Rev. Lett.}\ }\textbf {\bibinfo {volume} {55}},\ \bibinfo {pages}
  {1422} (\bibinfo {year} {1985})}\BibitemShut {NoStop}%
\bibitem [{\citenamefont {Totani}\ and\ \citenamefont
  {Sato}(1995)}]{Totani:1995rg}%
  \BibitemOpen
  \bibfield  {author} {\bibinfo {author} {\bibfnamefont {T.}~\bibnamefont
  {Totani}}\ and\ \bibinfo {author} {\bibfnamefont {K.}~\bibnamefont {Sato}},\
  }\href {\doibase 10.1016/0927-6505(95)00015-9} {\bibfield  {journal}
  {\bibinfo  {journal} {Astropart. Phys.}\ }\textbf {\bibinfo {volume} {3}},\
  \bibinfo {pages} {367} (\bibinfo {year} {1995})},\ \Eprint
  {http://arxiv.org/abs/astro-ph/9504015} {arXiv:astro-ph/9504015 [astro-ph]}
  \BibitemShut {NoStop}%
\bibitem [{\citenamefont {Totani}\ \emph {et~al.}(1996)\citenamefont {Totani},
  \citenamefont {Sato},\ and\ \citenamefont {Yoshii}}]{Totani:1995dw}%
  \BibitemOpen
  \bibfield  {author} {\bibinfo {author} {\bibfnamefont {T.}~\bibnamefont
  {Totani}}, \bibinfo {author} {\bibfnamefont {K.}~\bibnamefont {Sato}}, \ and\
  \bibinfo {author} {\bibfnamefont {Y.}~\bibnamefont {Yoshii}},\ }\href
  {\doibase 10.1086/176970} {\bibfield  {journal} {\bibinfo  {journal}
  {Astrophys. J.}\ }\textbf {\bibinfo {volume} {460}},\ \bibinfo {pages} {303}
  (\bibinfo {year} {1996})},\ \Eprint {http://arxiv.org/abs/astro-ph/9509130}
  {arXiv:astro-ph/9509130 [astro-ph]} \BibitemShut {NoStop}%
\bibitem [{\citenamefont {Malaney}(1997)}]{Malaney:1996ar}%
  \BibitemOpen
  \bibfield  {author} {\bibinfo {author} {\bibfnamefont {R.~A.}\ \bibnamefont
  {Malaney}},\ }\href {\doibase 10.1016/S0927-6505(97)00012-1} {\bibfield
  {journal} {\bibinfo  {journal} {Astropart. Phys.}\ }\textbf {\bibinfo
  {volume} {7}},\ \bibinfo {pages} {125} (\bibinfo {year} {1997})},\ \Eprint
  {http://arxiv.org/abs/astro-ph/9612012} {arXiv:astro-ph/9612012 [astro-ph]}
  \BibitemShut {NoStop}%
\bibitem [{\citenamefont {Hartmann}\ and\ \citenamefont
  {Woosley}(1997)}]{Hartmann:1997qe}%
  \BibitemOpen
  \bibfield  {author} {\bibinfo {author} {\bibfnamefont {D.~H.}\ \bibnamefont
  {Hartmann}}\ and\ \bibinfo {author} {\bibfnamefont {S.~E.}\ \bibnamefont
  {Woosley}},\ }\href {\doibase 10.1016/S0927-6505(97)00018-2} {\bibfield
  {journal} {\bibinfo  {journal} {Astropart. Phys.}\ }\textbf {\bibinfo
  {volume} {7}},\ \bibinfo {pages} {137} (\bibinfo {year} {1997})}\BibitemShut
  {NoStop}%
\bibitem [{\citenamefont {Kaplinghat}\ \emph {et~al.}(2000)\citenamefont
  {Kaplinghat}, \citenamefont {Steigman},\ and\ \citenamefont
  {Walker}}]{Kaplinghat:1999xi}%
  \BibitemOpen
  \bibfield  {author} {\bibinfo {author} {\bibfnamefont {M.}~\bibnamefont
  {Kaplinghat}}, \bibinfo {author} {\bibfnamefont {G.}~\bibnamefont
  {Steigman}}, \ and\ \bibinfo {author} {\bibfnamefont {T.~P.}\ \bibnamefont
  {Walker}},\ }\href {\doibase 10.1103/PhysRevD.62.043001} {\bibfield
  {journal} {\bibinfo  {journal} {Phys. Rev.}\ }\textbf {\bibinfo {volume}
  {D62}},\ \bibinfo {pages} {043001} (\bibinfo {year} {2000})},\ \Eprint
  {http://arxiv.org/abs/astro-ph/9912391} {arXiv:astro-ph/9912391 [astro-ph]}
  \BibitemShut {NoStop}%
\bibitem [{\citenamefont {Ando}\ and\ \citenamefont
  {Sato}(2003)}]{Ando:2002zj}%
  \BibitemOpen
  \bibfield  {author} {\bibinfo {author} {\bibfnamefont {S.}~\bibnamefont
  {Ando}}\ and\ \bibinfo {author} {\bibfnamefont {K.}~\bibnamefont {Sato}},\
  }\href {\doibase 10.1016/S0370-2693(03)00374-5} {\bibfield  {journal}
  {\bibinfo  {journal} {Phys. Lett.}\ }\textbf {\bibinfo {volume} {B559}},\
  \bibinfo {pages} {113} (\bibinfo {year} {2003})},\ \Eprint
  {http://arxiv.org/abs/astro-ph/0210502} {arXiv:astro-ph/0210502 [astro-ph]}
  \BibitemShut {NoStop}%
\bibitem [{\citenamefont {Fukugita}\ and\ \citenamefont
  {Kawasaki}(2003)}]{Fukugita:2002qw}%
  \BibitemOpen
  \bibfield  {author} {\bibinfo {author} {\bibfnamefont {M.}~\bibnamefont
  {Fukugita}}\ and\ \bibinfo {author} {\bibfnamefont {M.}~\bibnamefont
  {Kawasaki}},\ }\href {\doibase 10.1046/j.1365-8711.2003.06507.x} {\bibfield
  {journal} {\bibinfo  {journal} {Mon. Not. Roy. Astron. Soc.}\ }\textbf
  {\bibinfo {volume} {340}},\ \bibinfo {pages} {L7} (\bibinfo {year} {2003})},\
  \Eprint {http://arxiv.org/abs/astro-ph/0204376} {arXiv:astro-ph/0204376
  [astro-ph]} \BibitemShut {NoStop}%
\bibitem [{\citenamefont {Strigari}\ \emph {et~al.}(2004)\citenamefont
  {Strigari}, \citenamefont {Kaplinghat}, \citenamefont {Steigman},\ and\
  \citenamefont {Walker}}]{Strigari:2003ig}%
  \BibitemOpen
  \bibfield  {author} {\bibinfo {author} {\bibfnamefont {L.~E.}\ \bibnamefont
  {Strigari}}, \bibinfo {author} {\bibfnamefont {M.}~\bibnamefont
  {Kaplinghat}}, \bibinfo {author} {\bibfnamefont {G.}~\bibnamefont
  {Steigman}}, \ and\ \bibinfo {author} {\bibfnamefont {T.~P.}\ \bibnamefont
  {Walker}},\ }\href {\doibase 10.1088/1475-7516/2004/03/007} {\bibfield
  {journal} {\bibinfo  {journal} {JCAP}\ }\textbf {\bibinfo {volume} {0403}},\
  \bibinfo {pages} {007} (\bibinfo {year} {2004})},\ \Eprint
  {http://arxiv.org/abs/astro-ph/0312346} {arXiv:astro-ph/0312346 [astro-ph]}
  \BibitemShut {NoStop}%
\bibitem [{\citenamefont {Iocco}\ \emph {et~al.}(2005)\citenamefont {Iocco},
  \citenamefont {Mangano}, \citenamefont {Miele}, \citenamefont {Raffelt},\
  and\ \citenamefont {Serpico}}]{Iocco:2004wd}%
  \BibitemOpen
  \bibfield  {author} {\bibinfo {author} {\bibfnamefont {F.}~\bibnamefont
  {Iocco}}, \bibinfo {author} {\bibfnamefont {G.}~\bibnamefont {Mangano}},
  \bibinfo {author} {\bibfnamefont {G.}~\bibnamefont {Miele}}, \bibinfo
  {author} {\bibfnamefont {G.~G.}\ \bibnamefont {Raffelt}}, \ and\ \bibinfo
  {author} {\bibfnamefont {P.~D.}\ \bibnamefont {Serpico}},\ }\href {\doibase
  10.1016/j.astropartphys.2005.01.004} {\bibfield  {journal} {\bibinfo
  {journal} {Astropart. Phys.}\ }\textbf {\bibinfo {volume} {23}},\ \bibinfo
  {pages} {303} (\bibinfo {year} {2005})},\ \Eprint
  {http://arxiv.org/abs/astro-ph/0411545} {arXiv:astro-ph/0411545 [astro-ph]}
  \BibitemShut {NoStop}%
\bibitem [{\citenamefont {Strigari}\ \emph {et~al.}(2005)\citenamefont
  {Strigari}, \citenamefont {Beacom}, \citenamefont {Walker},\ and\
  \citenamefont {Zhang}}]{Strigari:2005hu}%
  \BibitemOpen
  \bibfield  {author} {\bibinfo {author} {\bibfnamefont {L.~E.}\ \bibnamefont
  {Strigari}}, \bibinfo {author} {\bibfnamefont {J.~F.}\ \bibnamefont
  {Beacom}}, \bibinfo {author} {\bibfnamefont {T.~P.}\ \bibnamefont {Walker}},
  \ and\ \bibinfo {author} {\bibfnamefont {P.}~\bibnamefont {Zhang}},\ }\href
  {\doibase 10.1088/1475-7516/2005/04/017} {\bibfield  {journal} {\bibinfo
  {journal} {JCAP}\ }\textbf {\bibinfo {volume} {0504}},\ \bibinfo {pages}
  {017} (\bibinfo {year} {2005})},\ \Eprint
  {http://arxiv.org/abs/astro-ph/0502150} {arXiv:astro-ph/0502150 [astro-ph]}
  \BibitemShut {NoStop}%
\bibitem [{\citenamefont {Lunardini}(2006)}]{Lunardini:2005jf}%
  \BibitemOpen
  \bibfield  {author} {\bibinfo {author} {\bibfnamefont {C.}~\bibnamefont
  {Lunardini}},\ }\href {\doibase 10.1016/j.astropartphys.2006.06.008}
  {\bibfield  {journal} {\bibinfo  {journal} {Astropart. Phys.}\ }\textbf
  {\bibinfo {volume} {26}},\ \bibinfo {pages} {190} (\bibinfo {year} {2006})},\
  \Eprint {http://arxiv.org/abs/astro-ph/0509233} {arXiv:astro-ph/0509233
  [astro-ph]} \BibitemShut {NoStop}%
\bibitem [{\citenamefont {Daigne}\ \emph {et~al.}(2005)\citenamefont {Daigne},
  \citenamefont {Olive}, \citenamefont {Sandick},\ and\ \citenamefont
  {Vangioni}}]{Daigne:2005xi}%
  \BibitemOpen
  \bibfield  {author} {\bibinfo {author} {\bibfnamefont {F.}~\bibnamefont
  {Daigne}}, \bibinfo {author} {\bibfnamefont {K.~A.}\ \bibnamefont {Olive}},
  \bibinfo {author} {\bibfnamefont {P.}~\bibnamefont {Sandick}}, \ and\
  \bibinfo {author} {\bibfnamefont {E.}~\bibnamefont {Vangioni}},\ }\href
  {\doibase 10.1103/PhysRevD.72.103007} {\bibfield  {journal} {\bibinfo
  {journal} {Phys. Rev.}\ }\textbf {\bibinfo {volume} {D72}},\ \bibinfo {pages}
  {103007} (\bibinfo {year} {2005})},\ \Eprint
  {http://arxiv.org/abs/astro-ph/0509404} {arXiv:astro-ph/0509404 [astro-ph]}
  \BibitemShut {NoStop}%
\bibitem [{\citenamefont {Yuksel}\ \emph {et~al.}(2006)\citenamefont {Yuksel},
  \citenamefont {Ando},\ and\ \citenamefont {Beacom}}]{Yuksel:2005ae}%
  \BibitemOpen
  \bibfield  {author} {\bibinfo {author} {\bibfnamefont {H.}~\bibnamefont
  {Yuksel}}, \bibinfo {author} {\bibfnamefont {S.}~\bibnamefont {Ando}}, \ and\
  \bibinfo {author} {\bibfnamefont {J.~F.}\ \bibnamefont {Beacom}},\ }\href
  {\doibase 10.1103/PhysRevC.74.015803} {\bibfield  {journal} {\bibinfo
  {journal} {Phys. Rev.}\ }\textbf {\bibinfo {volume} {C74}},\ \bibinfo {pages}
  {015803} (\bibinfo {year} {2006})},\ \Eprint
  {http://arxiv.org/abs/astro-ph/0509297} {arXiv:astro-ph/0509297 [astro-ph]}
  \BibitemShut {NoStop}%
\bibitem [{\citenamefont {Horiuchi}\ \emph {et~al.}(2009)\citenamefont
  {Horiuchi}, \citenamefont {Beacom},\ and\ \citenamefont
  {Dwek}}]{Horiuchi:2008jz}%
  \BibitemOpen
  \bibfield  {author} {\bibinfo {author} {\bibfnamefont {S.}~\bibnamefont
  {Horiuchi}}, \bibinfo {author} {\bibfnamefont {J.~F.}\ \bibnamefont
  {Beacom}}, \ and\ \bibinfo {author} {\bibfnamefont {E.}~\bibnamefont
  {Dwek}},\ }\href {\doibase 10.1103/PhysRevD.79.083013} {\bibfield  {journal}
  {\bibinfo  {journal} {Phys. Rev.}\ }\textbf {\bibinfo {volume} {D79}},\
  \bibinfo {pages} {083013} (\bibinfo {year} {2009})},\ \Eprint
  {http://arxiv.org/abs/0812.3157} {arXiv:0812.3157 [astro-ph]} \BibitemShut
  {NoStop}%
\bibitem [{\citenamefont {Lunardini}(2009)}]{Lunardini:2009ya}%
  \BibitemOpen
  \bibfield  {author} {\bibinfo {author} {\bibfnamefont {C.}~\bibnamefont
  {Lunardini}},\ }\href {\doibase 10.1103/PhysRevLett.102.231101} {\bibfield
  {journal} {\bibinfo  {journal} {Phys. Rev. Lett.}\ }\textbf {\bibinfo
  {volume} {102}},\ \bibinfo {pages} {231101} (\bibinfo {year} {2009})},\
  \Eprint {http://arxiv.org/abs/0901.0568} {arXiv:0901.0568 [astro-ph.SR]}
  \BibitemShut {NoStop}%
\bibitem [{\citenamefont {Lien}\ \emph {et~al.}(2010)\citenamefont {Lien},
  \citenamefont {Fields},\ and\ \citenamefont {Beacom}}]{Lien:2010yb}%
  \BibitemOpen
  \bibfield  {author} {\bibinfo {author} {\bibfnamefont {A.}~\bibnamefont
  {Lien}}, \bibinfo {author} {\bibfnamefont {B.~D.}\ \bibnamefont {Fields}}, \
  and\ \bibinfo {author} {\bibfnamefont {J.~F.}\ \bibnamefont {Beacom}},\
  }\href {\doibase 10.1103/PhysRevD.81.083001} {\bibfield  {journal} {\bibinfo
  {journal} {Phys. Rev.}\ }\textbf {\bibinfo {volume} {D81}},\ \bibinfo {pages}
  {083001} (\bibinfo {year} {2010})},\ \Eprint {http://arxiv.org/abs/1001.3678}
  {arXiv:1001.3678 [astro-ph.CO]} \BibitemShut {NoStop}%
\bibitem [{\citenamefont {Keehn}\ and\ \citenamefont
  {Lunardini}(2012)}]{Keehn:2010pn}%
  \BibitemOpen
  \bibfield  {author} {\bibinfo {author} {\bibfnamefont {J.~G.}\ \bibnamefont
  {Keehn}}\ and\ \bibinfo {author} {\bibfnamefont {C.}~\bibnamefont
  {Lunardini}},\ }\href {\doibase 10.1103/PhysRevD.85.043011} {\bibfield
  {journal} {\bibinfo  {journal} {Phys. Rev.}\ }\textbf {\bibinfo {volume}
  {D85}},\ \bibinfo {pages} {043011} (\bibinfo {year} {2012})},\ \Eprint
  {http://arxiv.org/abs/1012.1274} {arXiv:1012.1274 [astro-ph.CO]} \BibitemShut
  {NoStop}%
\bibitem [{\citenamefont {Vissani}\ and\ \citenamefont
  {Pagliaroli}(2011)}]{Vissani:2011kx}%
  \BibitemOpen
  \bibfield  {author} {\bibinfo {author} {\bibfnamefont {F.}~\bibnamefont
  {Vissani}}\ and\ \bibinfo {author} {\bibfnamefont {G.}~\bibnamefont
  {Pagliaroli}},\ }\href {\doibase 10.1051/0004-6361/201016109} {\bibfield
  {journal} {\bibinfo  {journal} {Astron. Astrophys.}\ }\textbf {\bibinfo
  {volume} {528}},\ \bibinfo {pages} {L1} (\bibinfo {year} {2011})},\ \Eprint
  {http://arxiv.org/abs/1102.0447} {arXiv:1102.0447 [astro-ph.HE]} \BibitemShut
  {NoStop}%
\bibitem [{\citenamefont {Lunardini}\ and\ \citenamefont
  {Tamborra}(2012)}]{Lunardini:2012ne}%
  \BibitemOpen
  \bibfield  {author} {\bibinfo {author} {\bibfnamefont {C.}~\bibnamefont
  {Lunardini}}\ and\ \bibinfo {author} {\bibfnamefont {I.}~\bibnamefont
  {Tamborra}},\ }\href {\doibase 10.1088/1475-7516/2012/07/012} {\bibfield
  {journal} {\bibinfo  {journal} {JCAP}\ }\textbf {\bibinfo {volume} {1207}},\
  \bibinfo {pages} {012} (\bibinfo {year} {2012})},\ \Eprint
  {http://arxiv.org/abs/1205.6292} {arXiv:1205.6292 [astro-ph.SR]} \BibitemShut
  {NoStop}%
\bibitem [{\citenamefont {Nakazato}(2013)}]{Nakazato:2013maa}%
  \BibitemOpen
  \bibfield  {author} {\bibinfo {author} {\bibfnamefont {K.}~\bibnamefont
  {Nakazato}},\ }\href {\doibase 10.1103/PhysRevD.88.083012} {\bibfield
  {journal} {\bibinfo  {journal} {Phys\cite{Nomoto:1984,Nomoto:1987}. Rev.}\
  }\textbf {\bibinfo {volume} {D88}},\ \bibinfo {pages} {083012} (\bibinfo
  {year} {2013})},\ \Eprint {http://arxiv.org/abs/1306.4526} {arXiv:1306.4526
  [astro-ph.HE]} \BibitemShut {NoStop}%
\bibitem [{\citenamefont {Mathews}\ \emph {et~al.}(2014)\citenamefont
  {Mathews}, \citenamefont {Hidaka}, \citenamefont {Kajino},\ and\
  \citenamefont {Suzuki}}]{Mathews:2014qba}%
  \BibitemOpen
  \bibfield  {author} {\bibinfo {author} {\bibfnamefont {G.~J.}\ \bibnamefont
  {Mathews}}, \bibinfo {author} {\bibfnamefont {J.}~\bibnamefont {Hidaka}},
  \bibinfo {author} {\bibfnamefont {T.}~\bibnamefont {Kajino}}, \ and\ \bibinfo
  {author} {\bibfnamefont {J.}~\bibnamefont {Suzuki}},\ }\href {\doibase
  10.1088/0004-637X/790/2/115} {\bibfield  {journal} {\bibinfo  {journal}
  {Astrophys. J.}\ }\textbf {\bibinfo {volume} {790}},\ \bibinfo {pages} {115}
  (\bibinfo {year} {2014})},\ \Eprint {http://arxiv.org/abs/1405.0458}
  {arXiv:1405.0458 [astro-ph.CO]} \BibitemShut {NoStop}%
\bibitem [{\citenamefont {Yuksel}\ and\ \citenamefont
  {Kistler}(2015)}]{Yuksel:2012zy}%
  \BibitemOpen
  \bibfield  {author} {\bibinfo {author} {\bibfnamefont {H.}~\bibnamefont
  {Yuksel}}\ and\ \bibinfo {author} {\bibfnamefont {M.~D.}\ \bibnamefont
  {Kistler}},\ }\href {\doibase 10.1016/j.physletb.2015.10.055} {\bibfield
  {journal} {\bibinfo  {journal} {Phys. Lett.}\ }\textbf {\bibinfo {volume}
  {B751}},\ \bibinfo {pages} {413} (\bibinfo {year} {2015})},\ \Eprint
  {http://arxiv.org/abs/1212.4844} {arXiv:1212.4844 [astro-ph.HE]} \BibitemShut
  {NoStop}%
\bibitem [{\citenamefont {Nakazato}\ \emph {et~al.}(2015)\citenamefont
  {Nakazato}, \citenamefont {Mochida}, \citenamefont {Niino},\ and\
  \citenamefont {Suzuki}}]{Nakazato:2015rya}%
  \BibitemOpen
  \bibfield  {author} {\bibinfo {author} {\bibfnamefont {K.}~\bibnamefont
  {Nakazato}}, \bibinfo {author} {\bibfnamefont {E.}~\bibnamefont {Mochida}},
  \bibinfo {author} {\bibfnamefont {Y.}~\bibnamefont {Niino}}, \ and\ \bibinfo
  {author} {\bibfnamefont {H.}~\bibnamefont {Suzuki}},\ }\href {\doibase
  10.1088/0004-637X/804/1/75} {\bibfield  {journal} {\bibinfo  {journal}
  {Astrophys. J.}\ }\textbf {\bibinfo {volume} {804}},\ \bibinfo {pages} {75}
  (\bibinfo {year} {2015})},\ \Eprint {http://arxiv.org/abs/1503.01236}
  {arXiv:1503.01236 [astro-ph.HE]} \BibitemShut {NoStop}%
\bibitem [{\citenamefont {Hidaka}\ \emph {et~al.}(2016)\citenamefont {Hidaka},
  \citenamefont {Kajino},\ and\ \citenamefont {Mathews}}]{Hidaka:2016zei}%
  \BibitemOpen
  \bibfield  {author} {\bibinfo {author} {\bibfnamefont {J.}~\bibnamefont
  {Hidaka}}, \bibinfo {author} {\bibfnamefont {T.}~\bibnamefont {Kajino}}, \
  and\ \bibinfo {author} {\bibfnamefont {G.~J.}\ \bibnamefont {Mathews}},\
  }\href {\doibase 10.3847/0004-637X/827/1/85} {\bibfield  {journal} {\bibinfo
  {journal} {Astrophys. J.}\ }\textbf {\bibinfo {volume} {827}},\ \bibinfo
  {pages} {85} (\bibinfo {year} {2016})}\BibitemShut {NoStop}%
\bibitem [{\citenamefont {Priya}\ and\ \citenamefont
  {Lunardini}(2017)}]{Priya:2017bmm}%
  \BibitemOpen
  \bibfield  {author} {\bibinfo {author} {\bibfnamefont {A.}~\bibnamefont
  {Priya}}\ and\ \bibinfo {author} {\bibfnamefont {C.}~\bibnamefont
  {Lunardini}},\ }\href@noop {} {\  (\bibinfo {year} {2017})},\ \Eprint
  {http://arxiv.org/abs/1705.02122} {arXiv:1705.02122 [astro-ph.HE]}
  \BibitemShut {NoStop}%
\bibitem [{\citenamefont {Horiuchi}\ \emph {et~al.}(2018)\citenamefont
  {Horiuchi}, \citenamefont {Sumiyoshi}, \citenamefont {Nakamura},
  \citenamefont {Fischer}, \citenamefont {Summa}, \citenamefont {Takiwaki},
  \citenamefont {Janka},\ and\ \citenamefont {Kotake}}]{Horiuchi:2017qja}%
  \BibitemOpen
  \bibfield  {author} {\bibinfo {author} {\bibfnamefont {S.}~\bibnamefont
  {Horiuchi}}, \bibinfo {author} {\bibfnamefont {K.}~\bibnamefont {Sumiyoshi}},
  \bibinfo {author} {\bibfnamefont {K.}~\bibnamefont {Nakamura}}, \bibinfo
  {author} {\bibfnamefont {T.}~\bibnamefont {Fischer}}, \bibinfo {author}
  {\bibfnamefont {A.}~\bibnamefont {Summa}}, \bibinfo {author} {\bibfnamefont
  {T.}~\bibnamefont {Takiwaki}}, \bibinfo {author} {\bibfnamefont {H.-T.}\
  \bibnamefont {Janka}}, \ and\ \bibinfo {author} {\bibfnamefont
  {K.}~\bibnamefont {Kotake}},\ }\href {\doibase 10.1093/mnras/stx3271}
  {\bibfield  {journal} {\bibinfo  {journal} {Mon. Not. Roy. Astron. Soc.}\
  }\textbf {\bibinfo {volume} {475}},\ \bibinfo {pages} {1363} (\bibinfo {year}
  {2018})},\ \Eprint {http://arxiv.org/abs/1709.06567} {arXiv:1709.06567
  [astro-ph.HE]} \BibitemShut {NoStop}%
\bibitem [{\citenamefont {Møller}\ \emph {et~al.}(2018)\citenamefont
  {Møller}, \citenamefont {Suliga}, \citenamefont {Tamborra},\ and\
  \citenamefont {Denton}}]{Moller:2018kpn}%
  \BibitemOpen
  \bibfield  {author} {\bibinfo {author} {\bibfnamefont {K.}~\bibnamefont
  {Møller}}, \bibinfo {author} {\bibfnamefont {A.~M.}\ \bibnamefont {Suliga}},
  \bibinfo {author} {\bibfnamefont {I.}~\bibnamefont {Tamborra}}, \ and\
  \bibinfo {author} {\bibfnamefont {P.~B.}\ \bibnamefont {Denton}},\ }\href
  {\doibase 10.1088/1475-7516/2018/05/066} {\bibfield  {journal} {\bibinfo
  {journal} {JCAP}\ }\textbf {\bibinfo {volume} {05}},\ \bibinfo {pages} {066}
  (\bibinfo {year} {2018})},\ \Eprint {http://arxiv.org/abs/1804.03157}
  {arXiv:1804.03157 [astro-ph.HE]} \BibitemShut {NoStop}%
\bibitem [{\citenamefont {Riya}\ and\ \citenamefont
  {Rentala}(2020)}]{Riya:2020wpw}%
  \BibitemOpen
  \bibfield  {author} {\bibinfo {author} {\bibnamefont {Riya}}\ and\ \bibinfo
  {author} {\bibfnamefont {V.}~\bibnamefont {Rentala}},\ }\href@noop {} {\
  (\bibinfo {year} {2020})},\ \Eprint {http://arxiv.org/abs/2007.02951}
  {arXiv:2007.02951 [astro-ph.HE]} \BibitemShut {NoStop}%
\bibitem [{\citenamefont {Kresse}\ \emph {et~al.}(2020)\citenamefont {Kresse},
  \citenamefont {Ertl},\ and\ \citenamefont {Janka}}]{Kresse:2020nto}%
  \BibitemOpen
  \bibfield  {author} {\bibinfo {author} {\bibfnamefont {D.}~\bibnamefont
  {Kresse}}, \bibinfo {author} {\bibfnamefont {T.}~\bibnamefont {Ertl}}, \ and\
  \bibinfo {author} {\bibfnamefont {H.-T.}\ \bibnamefont {Janka}},\ }\href@noop
  {} {\  (\bibinfo {year} {2020})},\ \Eprint {http://arxiv.org/abs/2010.04728}
  {arXiv:2010.04728 [astro-ph.HE]} \BibitemShut {NoStop}%
\bibitem [{\citenamefont {{Sana}}\ \emph {et~al.}(2012)\citenamefont {{Sana}},
  \citenamefont {{de Mink}}, \citenamefont {{de Koter}}, \citenamefont
  {{Langer}}, \citenamefont {{Evans}}, \citenamefont {{Gieles}}, \citenamefont
  {{Gosset}}, \citenamefont {{Izzard}}, \citenamefont {{Le Bouquin}},\ and\
  \citenamefont {{Schneider}}}]{sana12}%
  \BibitemOpen
  \bibfield  {author} {\bibinfo {author} {\bibfnamefont {H.}~\bibnamefont
  {{Sana}}}, \bibinfo {author} {\bibfnamefont {S.~E.}\ \bibnamefont {{de
  Mink}}}, \bibinfo {author} {\bibfnamefont {A.}~\bibnamefont {{de Koter}}},
  \bibinfo {author} {\bibfnamefont {N.}~\bibnamefont {{Langer}}}, \bibinfo
  {author} {\bibfnamefont {C.~J.}\ \bibnamefont {{Evans}}}, \bibinfo {author}
  {\bibfnamefont {M.}~\bibnamefont {{Gieles}}}, \bibinfo {author}
  {\bibfnamefont {E.}~\bibnamefont {{Gosset}}}, \bibinfo {author}
  {\bibfnamefont {R.~G.}\ \bibnamefont {{Izzard}}}, \bibinfo {author}
  {\bibfnamefont {J.~B.}\ \bibnamefont {{Le Bouquin}}}, \ and\ \bibinfo
  {author} {\bibfnamefont {F.~R.~N.}\ \bibnamefont {{Schneider}}},\ }\href
  {\doibase 10.1126/science.1223344} {\bibfield  {journal} {\bibinfo  {journal}
  {Science}\ }\textbf {\bibinfo {volume} {337}},\ \bibinfo {pages} {444}
  (\bibinfo {year} {2012})},\ \Eprint {http://arxiv.org/abs/1207.6397}
  {arXiv:1207.6397 [astro-ph.SR]} \BibitemShut {NoStop}%
\bibitem [{\citenamefont {{Zapartas}}\ \emph {et~al.}(2020)\citenamefont
  {{Zapartas}}, \citenamefont {{de Mink}}, \citenamefont {{Justham}},
  \citenamefont {{Smith}}, \citenamefont {{Renzo}},\ and\ \citenamefont {{de
  Koter}}}]{zapa20}%
  \BibitemOpen
  \bibfield  {author} {\bibinfo {author} {\bibfnamefont {E.}~\bibnamefont
  {{Zapartas}}}, \bibinfo {author} {\bibfnamefont {S.~E.}\ \bibnamefont {{de
  Mink}}}, \bibinfo {author} {\bibfnamefont {S.}~\bibnamefont {{Justham}}},
  \bibinfo {author} {\bibfnamefont {N.}~\bibnamefont {{Smith}}}, \bibinfo
  {author} {\bibfnamefont {M.}~\bibnamefont {{Renzo}}}, \ and\ \bibinfo
  {author} {\bibfnamefont {A.}~\bibnamefont {{de Koter}}},\ }\href@noop {}
  {\bibfield  {journal} {\bibinfo  {journal} {arXiv e-prints}\ ,\ \bibinfo
  {eid} {arXiv:2002.07230}} (\bibinfo {year} {2020})},\ \Eprint
  {http://arxiv.org/abs/2002.07230} {arXiv:2002.07230 [astro-ph.HE]}
  \BibitemShut {NoStop}%
\bibitem [{\citenamefont {{Podsiadlowski}}\ \emph {et~al.}(1992)\citenamefont
  {{Podsiadlowski}}, \citenamefont {{Joss}},\ and\ \citenamefont
  {{Hsu}}}]{Pod1992}%
  \BibitemOpen
  \bibfield  {author} {\bibinfo {author} {\bibfnamefont {P.}~\bibnamefont
  {{Podsiadlowski}}}, \bibinfo {author} {\bibfnamefont {P.~C.}\ \bibnamefont
  {{Joss}}}, \ and\ \bibinfo {author} {\bibfnamefont {J.~J.~L.}\ \bibnamefont
  {{Hsu}}},\ }\href {\doibase 10.1086/171341} {\bibfield  {journal} {\bibinfo
  {journal} {\apj}\ }\textbf {\bibinfo {volume} {391}},\ \bibinfo {pages} {246}
  (\bibinfo {year} {1992})}\BibitemShut {NoStop}%
\bibitem [{\citenamefont {Woosley}\ \emph {et~al.}(2002)\citenamefont
  {Woosley}, \citenamefont {Heger},\ and\ \citenamefont
  {Weaver}}]{Woosley:2002zz}%
  \BibitemOpen
  \bibfield  {author} {\bibinfo {author} {\bibfnamefont {S.~E.}\ \bibnamefont
  {Woosley}}, \bibinfo {author} {\bibfnamefont {A.}~\bibnamefont {Heger}}, \
  and\ \bibinfo {author} {\bibfnamefont {T.~A.}\ \bibnamefont {Weaver}},\
  }\href {\doibase 10.1103/RevModPhys.74.1015} {\bibfield  {journal} {\bibinfo
  {journal} {Rev. Mod. Phys.}\ }\textbf {\bibinfo {volume} {74}},\ \bibinfo
  {pages} {1015} (\bibinfo {year} {2002})}\BibitemShut {NoStop}%
\bibitem [{\citenamefont {Woosley}(2019)}]{Woosley2019}%
  \BibitemOpen
  \bibfield  {author} {\bibinfo {author} {\bibfnamefont {S.~E.}\ \bibnamefont
  {Woosley}},\ }\href@noop {} {\bibfield  {journal} {\bibinfo  {journal}
  {\apj}\ }\textbf {\bibinfo {volume} {878}},\ \bibinfo {pages} {49} (\bibinfo
  {year} {2019})}\BibitemShut {NoStop}%
\bibitem [{\citenamefont {{Eggleton}}(2006)}]{2006epbm.book.....E}%
  \BibitemOpen
  \bibfield  {author} {\bibinfo {author} {\bibfnamefont {P.}~\bibnamefont
  {{Eggleton}}},\ }\href@noop {} {\emph {\bibinfo {title} {{Evolutionary
  Processes in Binary and Multiple Stars}}}}\ (\bibinfo {year}
  {2006})\BibitemShut {NoStop}%
\bibitem [{\citenamefont {{Podsiadlowski}}(2010)}]{Pod2010}%
  \BibitemOpen
  \bibfield  {author} {\bibinfo {author} {\bibfnamefont {P.}~\bibnamefont
  {{Podsiadlowski}}},\ }\href {\doibase 10.1016/j.newar.2010.09.023} {\bibfield
   {journal} {\bibinfo  {journal} {New Astronomy Reviews}\ }\textbf {\bibinfo
  {volume} {54}},\ \bibinfo {pages} {39} (\bibinfo {year} {2010})}\BibitemShut
  {NoStop}%
\bibitem [{\citenamefont {{Eggleton}}(1983)}]{Eggleton_1983}%
  \BibitemOpen
  \bibfield  {author} {\bibinfo {author} {\bibfnamefont {P.~P.}\ \bibnamefont
  {{Eggleton}}},\ }\href {\doibase 10.1086/160960} {\bibfield  {journal}
  {\bibinfo  {journal} {\apj}\ }\textbf {\bibinfo {volume} {268}},\ \bibinfo
  {pages} {368} (\bibinfo {year} {1983})}\BibitemShut {NoStop}%
\bibitem [{\citenamefont {{Paczynski}}(1976)}]{Paczynski_1976}%
  \BibitemOpen
  \bibfield  {author} {\bibinfo {author} {\bibfnamefont {B.}~\bibnamefont
  {{Paczynski}}},\ }in\ \href@noop {} {\emph {\bibinfo {booktitle} {Structure
  and Evolution of Close Binary Systems}}},\ \bibinfo {series} {IAU Symposium},
  Vol.~\bibinfo {volume} {73},\ \bibinfo {editor} {edited by\ \bibinfo {editor}
  {\bibfnamefont {P.}~\bibnamefont {{Eggleton}}}, \bibinfo {editor}
  {\bibfnamefont {S.}~\bibnamefont {{Mitton}}}, \ and\ \bibinfo {editor}
  {\bibfnamefont {J.}~\bibnamefont {{Whelan}}}}\ (\bibinfo {year} {1976})\
  p.~\bibinfo {pages} {75}\BibitemShut {NoStop}%
\bibitem [{\citenamefont {{Tout}}\ \emph {et~al.}(1997)\citenamefont {{Tout}},
  \citenamefont {{Aarseth}}, \citenamefont {{Pols}},\ and\ \citenamefont
  {{Eggleton}}}]{Tout_1997}%
  \BibitemOpen
  \bibfield  {author} {\bibinfo {author} {\bibfnamefont {C.~A.}\ \bibnamefont
  {{Tout}}}, \bibinfo {author} {\bibfnamefont {S.~J.}\ \bibnamefont
  {{Aarseth}}}, \bibinfo {author} {\bibfnamefont {O.~R.}\ \bibnamefont
  {{Pols}}}, \ and\ \bibinfo {author} {\bibfnamefont {P.~P.}\ \bibnamefont
  {{Eggleton}}},\ }\href {\doibase 10.1093/mnras/291.4.732} {\bibfield
  {journal} {\bibinfo  {journal} {Mon. Not. Roy. Astron. Soc.}\ }\textbf
  {\bibinfo {volume} {291}},\ \bibinfo {pages} {732} (\bibinfo {year}
  {1997})}\BibitemShut {NoStop}%
\bibitem [{\citenamefont {{Hjellming}}(1989)}]{Hjellming_1989}%
  \BibitemOpen
  \bibfield  {author} {\bibinfo {author} {\bibfnamefont {M.~S.}\ \bibnamefont
  {{Hjellming}}},\ }\emph {\bibinfo {title} {{Rapid mass transfer in binary
  systems}}},\ \href@noop {} {Ph.D. thesis},\ \bibinfo  {school} {Illinois
  Univ.~at Urbana-Champaign, Savoy.} (\bibinfo {year} {1989})\BibitemShut
  {NoStop}%
\bibitem [{\citenamefont {{Ivanova}}\ \emph {et~al.}(2002)\citenamefont
  {{Ivanova}}, \citenamefont {{Podsiadlowski}},\ and\ \citenamefont
  {{Spruit}}}]{Ivanova_2002}%
  \BibitemOpen
  \bibfield  {author} {\bibinfo {author} {\bibfnamefont {N.}~\bibnamefont
  {{Ivanova}}}, \bibinfo {author} {\bibfnamefont {P.}~\bibnamefont
  {{Podsiadlowski}}}, \ and\ \bibinfo {author} {\bibfnamefont {H.}~\bibnamefont
  {{Spruit}}},\ }\href {\doibase 10.1046/j.1365-8711.2002.05543.x} {\bibfield
  {journal} {\bibinfo  {journal} {Mon. Not. Roy. Astron. Soc.}\ }\textbf
  {\bibinfo {volume} {334}},\ \bibinfo {pages} {819} (\bibinfo {year}
  {2002})},\ \Eprint {http://arxiv.org/abs/astro-ph/0109524}
  {arXiv:astro-ph/0109524 [astro-ph]} \BibitemShut {NoStop}%
\bibitem [{\citenamefont {{Belczynski}}\ \emph {et~al.}(2008)\citenamefont
  {{Belczynski}}, \citenamefont {{Kalogera}}, \citenamefont {{Rasio}},
  \citenamefont {{Taam}}, \citenamefont {{Zezas}}, \citenamefont {{Bulik}},
  \citenamefont {{Maccarone}},\ and\ \citenamefont
  {{Ivanova}}}]{Belczynski_2008}%
  \BibitemOpen
  \bibfield  {author} {\bibinfo {author} {\bibfnamefont {K.}~\bibnamefont
  {{Belczynski}}}, \bibinfo {author} {\bibfnamefont {V.}~\bibnamefont
  {{Kalogera}}}, \bibinfo {author} {\bibfnamefont {F.~A.}\ \bibnamefont
  {{Rasio}}}, \bibinfo {author} {\bibfnamefont {R.~E.}\ \bibnamefont {{Taam}}},
  \bibinfo {author} {\bibfnamefont {A.}~\bibnamefont {{Zezas}}}, \bibinfo
  {author} {\bibfnamefont {T.}~\bibnamefont {{Bulik}}}, \bibinfo {author}
  {\bibfnamefont {T.~J.}\ \bibnamefont {{Maccarone}}}, \ and\ \bibinfo {author}
  {\bibfnamefont {N.}~\bibnamefont {{Ivanova}}},\ }\href {\doibase
  10.1086/521026} {\bibfield  {journal} {\bibinfo  {journal} {The Astrophysical
  Journal Supplement Series}\ }\textbf {\bibinfo {volume} {174}},\ \bibinfo
  {pages} {223} (\bibinfo {year} {2008})},\ \Eprint
  {http://arxiv.org/abs/astro-ph/0511811} {arXiv:astro-ph/0511811 [astro-ph]}
  \BibitemShut {NoStop}%
\bibitem [{\citenamefont {{Hurley}}\ \emph {et~al.}(2002)\citenamefont
  {{Hurley}}, \citenamefont {{Tout}},\ and\ \citenamefont
  {{Pols}}}]{Hurley_2002}%
  \BibitemOpen
  \bibfield  {author} {\bibinfo {author} {\bibfnamefont {J.~R.}\ \bibnamefont
  {{Hurley}}}, \bibinfo {author} {\bibfnamefont {C.~A.}\ \bibnamefont
  {{Tout}}}, \ and\ \bibinfo {author} {\bibfnamefont {O.~R.}\ \bibnamefont
  {{Pols}}},\ }\href {\doibase 10.1046/j.1365-8711.2002.05038.x} {\bibfield
  {journal} {\bibinfo  {journal} {Mon. Not. Roy. Astron. Soc.}\ }\textbf
  {\bibinfo {volume} {329}},\ \bibinfo {pages} {897} (\bibinfo {year}
  {2002})},\ \Eprint {http://arxiv.org/abs/astro-ph/0201220} {astro-ph/0201220}
  \BibitemShut {NoStop}%
\bibitem [{\citenamefont {{Webbink}}(1984)}]{Webbink1984}%
  \BibitemOpen
  \bibfield  {author} {\bibinfo {author} {\bibfnamefont {R.~F.}\ \bibnamefont
  {{Webbink}}},\ }\href {\doibase 10.1086/161701} {\bibfield  {journal}
  {\bibinfo  {journal} {\apj}\ }\textbf {\bibinfo {volume} {277}},\ \bibinfo
  {pages} {355} (\bibinfo {year} {1984})}\BibitemShut {NoStop}%
\bibitem [{\citenamefont {{Nelemans}}\ \emph {et~al.}(2001)\citenamefont
  {{Nelemans}}, \citenamefont {{Yungelson}}, \citenamefont {{Portegies
  Zwart}},\ and\ \citenamefont {{Verbunt}}}]{Nelemans2001}%
  \BibitemOpen
  \bibfield  {author} {\bibinfo {author} {\bibfnamefont {G.}~\bibnamefont
  {{Nelemans}}}, \bibinfo {author} {\bibfnamefont {L.~R.}\ \bibnamefont
  {{Yungelson}}}, \bibinfo {author} {\bibfnamefont {S.~F.}\ \bibnamefont
  {{Portegies Zwart}}}, \ and\ \bibinfo {author} {\bibfnamefont
  {F.}~\bibnamefont {{Verbunt}}},\ }\href {\doibase 10.1051/0004-6361:20000147}
  {\bibfield  {journal} {\bibinfo  {journal} {Astronomy and Astrophysics}\
  }\textbf {\bibinfo {volume} {365}},\ \bibinfo {pages} {491} (\bibinfo {year}
  {2001})},\ \Eprint {http://arxiv.org/abs/astro-ph/0010457}
  {arXiv:astro-ph/0010457 [astro-ph]} \BibitemShut {NoStop}%
\bibitem [{\citenamefont {{Ivanova}}\ \emph {et~al.}(2013)\citenamefont
  {{Ivanova}}, \citenamefont {{Justham}}, \citenamefont {{Chen}}, \citenamefont
  {{De Marco}}, \citenamefont {{Fryer}}, \citenamefont {{Gaburov}},
  \citenamefont {{Ge}}, \citenamefont {{Glebbeek}}, \citenamefont {{Han}},
  \citenamefont {{Li}}, \citenamefont {{Lu}}, \citenamefont {{Marsh}},
  \citenamefont {{Podsiadlowski}}, \citenamefont {{Potter}}, \citenamefont
  {{Soker}}, \citenamefont {{Taam}}, \citenamefont {{Tauris}}, \citenamefont
  {{van den Heuvel}},\ and\ \citenamefont {{Webbink}}}]{Ivanova_2013}%
  \BibitemOpen
  \bibfield  {author} {\bibinfo {author} {\bibfnamefont {N.}~\bibnamefont
  {{Ivanova}}}, \bibinfo {author} {\bibfnamefont {S.}~\bibnamefont
  {{Justham}}}, \bibinfo {author} {\bibfnamefont {X.}~\bibnamefont {{Chen}}},
  \bibinfo {author} {\bibfnamefont {O.}~\bibnamefont {{De Marco}}}, \bibinfo
  {author} {\bibfnamefont {C.~L.}\ \bibnamefont {{Fryer}}}, \bibinfo {author}
  {\bibfnamefont {E.}~\bibnamefont {{Gaburov}}}, \bibinfo {author}
  {\bibfnamefont {H.}~\bibnamefont {{Ge}}}, \bibinfo {author} {\bibfnamefont
  {E.}~\bibnamefont {{Glebbeek}}}, \bibinfo {author} {\bibfnamefont
  {Z.}~\bibnamefont {{Han}}}, \bibinfo {author} {\bibfnamefont {X.-D.}\
  \bibnamefont {{Li}}}, \bibinfo {author} {\bibfnamefont {G.}~\bibnamefont
  {{Lu}}}, \bibinfo {author} {\bibfnamefont {T.}~\bibnamefont {{Marsh}}},
  \bibinfo {author} {\bibfnamefont {P.}~\bibnamefont {{Podsiadlowski}}},
  \bibinfo {author} {\bibfnamefont {A.}~\bibnamefont {{Potter}}}, \bibinfo
  {author} {\bibfnamefont {N.}~\bibnamefont {{Soker}}}, \bibinfo {author}
  {\bibfnamefont {R.}~\bibnamefont {{Taam}}}, \bibinfo {author} {\bibfnamefont
  {T.~M.}\ \bibnamefont {{Tauris}}}, \bibinfo {author} {\bibfnamefont
  {E.~P.~J.}\ \bibnamefont {{van den Heuvel}}}, \ and\ \bibinfo {author}
  {\bibfnamefont {R.~F.}\ \bibnamefont {{Webbink}}},\ }\href {\doibase
  10.1007/s00159-013-0059-2} {\bibfield  {journal} {\bibinfo  {journal}
  {Astronomy and Astrophysics Review}\ }\textbf {\bibinfo {volume} {21}},\
  \bibinfo {eid} {59} (\bibinfo {year} {2013})},\ \Eprint
  {http://arxiv.org/abs/1209.4302} {arXiv:1209.4302 [astro-ph.HE]} \BibitemShut
  {NoStop}%
\bibitem [{\citenamefont {{Belczynski}}\ \emph {et~al.}(2007)\citenamefont
  {{Belczynski}}, \citenamefont {{Taam}}, \citenamefont {{Kalogera}},
  \citenamefont {{Rasio}},\ and\ \citenamefont {{Bulik}}}]{Belczynski2007}%
  \BibitemOpen
  \bibfield  {author} {\bibinfo {author} {\bibfnamefont {K.}~\bibnamefont
  {{Belczynski}}}, \bibinfo {author} {\bibfnamefont {R.~E.}\ \bibnamefont
  {{Taam}}}, \bibinfo {author} {\bibfnamefont {V.}~\bibnamefont {{Kalogera}}},
  \bibinfo {author} {\bibfnamefont {F.~A.}\ \bibnamefont {{Rasio}}}, \ and\
  \bibinfo {author} {\bibfnamefont {T.}~\bibnamefont {{Bulik}}},\ }\href
  {\doibase 10.1086/513562} {\bibfield  {journal} {\bibinfo  {journal} {\apj}\
  }\textbf {\bibinfo {volume} {662}},\ \bibinfo {pages} {504} (\bibinfo {year}
  {2007})},\ \Eprint {http://arxiv.org/abs/astro-ph/0612032}
  {arXiv:astro-ph/0612032 [astro-ph]} \BibitemShut {NoStop}%
\bibitem [{\citenamefont {{Kinugawa}}\ \emph {et~al.}(2014)\citenamefont
  {{Kinugawa}}, \citenamefont {{Inayoshi}}, \citenamefont {{Hotokezaka}},
  \citenamefont {{Nakauchi}},\ and\ \citenamefont {{Nakamura}}}]{Kinugawa2014}%
  \BibitemOpen
  \bibfield  {author} {\bibinfo {author} {\bibfnamefont {T.}~\bibnamefont
  {{Kinugawa}}}, \bibinfo {author} {\bibfnamefont {K.}~\bibnamefont
  {{Inayoshi}}}, \bibinfo {author} {\bibfnamefont {K.}~\bibnamefont
  {{Hotokezaka}}}, \bibinfo {author} {\bibfnamefont {D.}~\bibnamefont
  {{Nakauchi}}}, \ and\ \bibinfo {author} {\bibfnamefont {T.}~\bibnamefont
  {{Nakamura}}},\ }\href {\doibase 10.1093/mnras/stu1022} {\bibfield  {journal}
  {\bibinfo  {journal} {Mon. Not. Roy. Astron. Soc.}\ }\textbf {\bibinfo
  {volume} {442}},\ \bibinfo {pages} {2963} (\bibinfo {year} {2014})},\ \Eprint
  {http://arxiv.org/abs/1402.6672} {arXiv:1402.6672 [astro-ph.HE]} \BibitemShut
  {NoStop}%
\bibitem [{\citenamefont {{Taam}}\ and\ \citenamefont
  {{Sandquist}}(2000)}]{Taam2000}%
  \BibitemOpen
  \bibfield  {author} {\bibinfo {author} {\bibfnamefont {R.~E.}\ \bibnamefont
  {{Taam}}}\ and\ \bibinfo {author} {\bibfnamefont {E.~L.}\ \bibnamefont
  {{Sandquist}}},\ }\href {\doibase 10.1146/annurev.astro.38.1.113} {\bibfield
  {journal} {\bibinfo  {journal} {Annual Review of Astronomy and Astrophysics}\
  }\textbf {\bibinfo {volume} {38}},\ \bibinfo {pages} {113} (\bibinfo {year}
  {2000})}\BibitemShut {NoStop}%
\bibitem [{\citenamefont {{Ivanova}}\ and\ \citenamefont
  {{Taam}}(2004)}]{Ivanova2004}%
  \BibitemOpen
  \bibfield  {author} {\bibinfo {author} {\bibfnamefont {N.}~\bibnamefont
  {{Ivanova}}}\ and\ \bibinfo {author} {\bibfnamefont {R.~E.}\ \bibnamefont
  {{Taam}}},\ }\href {\doibase 10.1086/380561} {\bibfield  {journal} {\bibinfo
  {journal} {\apj}\ }\textbf {\bibinfo {volume} {601}},\ \bibinfo {pages}
  {1058} (\bibinfo {year} {2004})},\ \Eprint
  {http://arxiv.org/abs/astro-ph/0310126} {arXiv:astro-ph/0310126 [astro-ph]}
  \BibitemShut {NoStop}%
\bibitem [{\citenamefont {{Limongi}}(2017)}]{Limongi2017}%
  \BibitemOpen
  \bibfield  {author} {\bibinfo {author} {\bibfnamefont {M.}~\bibnamefont
  {{Limongi}}},\ }\enquote {\bibinfo {title} {{Supernovae from Massive
  Stars}},}\ in\ \href {\doibase 10.1007/978-3-319-21846-5_119} {\emph
  {\bibinfo {booktitle} {Handbook of Supernovae}}},\ \bibinfo {editor} {edited
  by\ \bibinfo {editor} {\bibfnamefont {A.~W.}\ \bibnamefont {{Alsabti}}}\ and\
  \bibinfo {editor} {\bibfnamefont {P.}~\bibnamefont {{Murdin}}}}\ (\bibinfo
  {year} {2017})\ p.\ \bibinfo {pages} {513}\BibitemShut {NoStop}%
\bibitem [{\citenamefont {{Takahashi}}\ \emph {et~al.}(2014)\citenamefont
  {{Takahashi}}, \citenamefont {{Umeda}},\ and\ \citenamefont
  {{Yoshida}}}]{Takahashi2014}%
  \BibitemOpen
  \bibfield  {author} {\bibinfo {author} {\bibfnamefont {K.}~\bibnamefont
  {{Takahashi}}}, \bibinfo {author} {\bibfnamefont {H.}~\bibnamefont
  {{Umeda}}}, \ and\ \bibinfo {author} {\bibfnamefont {T.}~\bibnamefont
  {{Yoshida}}},\ }\href {\doibase 10.1088/0004-637X/794/1/40} {\bibfield
  {journal} {\bibinfo  {journal} {\apj}\ }\textbf {\bibinfo {volume} {794}},\
  \bibinfo {eid} {40} (\bibinfo {year} {2014})},\ \Eprint
  {http://arxiv.org/abs/1406.5305} {arXiv:1406.5305 [astro-ph.SR]} \BibitemShut
  {NoStop}%
\bibitem [{\citenamefont {{Salpeter}}(1955)}]{Salpeter1955}%
  \BibitemOpen
  \bibfield  {author} {\bibinfo {author} {\bibfnamefont {E.~E.}\ \bibnamefont
  {{Salpeter}}},\ }\href {\doibase 10.1086/145971} {\bibfield  {journal}
  {\bibinfo  {journal} {\apj}\ }\textbf {\bibinfo {volume} {121}},\ \bibinfo
  {pages} {161} (\bibinfo {year} {1955})}\BibitemShut {NoStop}%
\bibitem [{\citenamefont {{Kobulnicky}}\ and\ \citenamefont
  {{Fryer}}(2007)}]{Kobulnicky2007}%
  \BibitemOpen
  \bibfield  {author} {\bibinfo {author} {\bibfnamefont {H.~A.}\ \bibnamefont
  {{Kobulnicky}}}\ and\ \bibinfo {author} {\bibfnamefont {C.~L.}\ \bibnamefont
  {{Fryer}}},\ }\href {\doibase 10.1086/522073} {\bibfield  {journal} {\bibinfo
   {journal} {\apj}\ }\textbf {\bibinfo {volume} {670}},\ \bibinfo {pages}
  {747} (\bibinfo {year} {2007})}\BibitemShut {NoStop}%
\bibitem [{\citenamefont {{Abt}}(1983)}]{Abt_1983}%
  \BibitemOpen
  \bibfield  {author} {\bibinfo {author} {\bibfnamefont {H.~A.}\ \bibnamefont
  {{Abt}}},\ }\href {\doibase 10.1146/annurev.aa.21.090183.002015} {\bibfield
  {journal} {\bibinfo  {journal} {Annual Review of Astronomy and Astrophysics}\
  }\textbf {\bibinfo {volume} {21}},\ \bibinfo {pages} {343} (\bibinfo {year}
  {1983})}\BibitemShut {NoStop}%
\bibitem [{\citenamefont {{Heggie}}(1975)}]{Heggie_1975}%
  \BibitemOpen
  \bibfield  {author} {\bibinfo {author} {\bibfnamefont {D.~C.}\ \bibnamefont
  {{Heggie}}},\ }\href {\doibase 10.1093/mnras/173.3.729} {\bibfield  {journal}
  {\bibinfo  {journal} {Mon. Not. Roy. Astron. Soc.}\ }\textbf {\bibinfo
  {volume} {173}},\ \bibinfo {pages} {729} (\bibinfo {year}
  {1975})}\BibitemShut {NoStop}%
\bibitem [{\citenamefont {{Hurley}}\ \emph {et~al.}(2000)\citenamefont
  {{Hurley}}, \citenamefont {{Pols}},\ and\ \citenamefont
  {{Tout}}}]{Hurley2000}%
  \BibitemOpen
  \bibfield  {author} {\bibinfo {author} {\bibfnamefont {J.~R.}\ \bibnamefont
  {{Hurley}}}, \bibinfo {author} {\bibfnamefont {O.~R.}\ \bibnamefont
  {{Pols}}}, \ and\ \bibinfo {author} {\bibfnamefont {C.~A.}\ \bibnamefont
  {{Tout}}},\ }\href {\doibase 10.1046/j.1365-8711.2000.03426.x} {\bibfield
  {journal} {\bibinfo  {journal} {Mon. Not. Roy. Astron. Soc.}\ }\textbf
  {\bibinfo {volume} {315}},\ \bibinfo {pages} {543} (\bibinfo {year}
  {2000})},\ \Eprint {http://arxiv.org/abs/astro-ph/0001295}
  {arXiv:astro-ph/0001295 [astro-ph]} \BibitemShut {NoStop}%
\bibitem [{\citenamefont {{Li}}\ \emph {et~al.}(2011)\citenamefont {{Li}},
  \citenamefont {{Leaman}}, \citenamefont {{Chornock}}, \citenamefont
  {{Filippenko}}, \citenamefont {{Poznanski}}, \citenamefont {{Ganeshalingam}},
  \citenamefont {{Wang}}, \citenamefont {{Modjaz}}, \citenamefont {{Jha}},
  \citenamefont {{Foley}},\ and\ \citenamefont {{Smith}}}]{Li2011}%
  \BibitemOpen
  \bibfield  {author} {\bibinfo {author} {\bibfnamefont {W.}~\bibnamefont
  {{Li}}}, \bibinfo {author} {\bibfnamefont {J.}~\bibnamefont {{Leaman}}},
  \bibinfo {author} {\bibfnamefont {R.}~\bibnamefont {{Chornock}}}, \bibinfo
  {author} {\bibfnamefont {A.~V.}\ \bibnamefont {{Filippenko}}}, \bibinfo
  {author} {\bibfnamefont {D.}~\bibnamefont {{Poznanski}}}, \bibinfo {author}
  {\bibfnamefont {M.}~\bibnamefont {{Ganeshalingam}}}, \bibinfo {author}
  {\bibfnamefont {X.}~\bibnamefont {{Wang}}}, \bibinfo {author} {\bibfnamefont
  {M.}~\bibnamefont {{Modjaz}}}, \bibinfo {author} {\bibfnamefont
  {S.}~\bibnamefont {{Jha}}}, \bibinfo {author} {\bibfnamefont {R.~J.}\
  \bibnamefont {{Foley}}}, \ and\ \bibinfo {author} {\bibfnamefont
  {N.}~\bibnamefont {{Smith}}},\ }\href {\doibase
  10.1111/j.1365-2966.2011.18160.x} {\bibfield  {journal} {\bibinfo  {journal}
  {Monthly Notices of the Royal Astronomical Society}\ }\textbf {\bibinfo
  {volume} {412}},\ \bibinfo {pages} {1441} (\bibinfo {year} {2011})},\ \Eprint
  {http://arxiv.org/abs/1006.4612} {arXiv:1006.4612 [astro-ph.SR]} \BibitemShut
  {NoStop}%
\bibitem [{\citenamefont {Hopkins}\ and\ \citenamefont
  {Beacom}(2006)}]{Hopkins:2006bw}%
  \BibitemOpen
  \bibfield  {author} {\bibinfo {author} {\bibfnamefont {A.~M.}\ \bibnamefont
  {Hopkins}}\ and\ \bibinfo {author} {\bibfnamefont {J.~F.}\ \bibnamefont
  {Beacom}},\ }\href {\doibase 10.1086/506610} {\bibfield  {journal} {\bibinfo
  {journal} {Astrophys. J.}\ }\textbf {\bibinfo {volume} {651}},\ \bibinfo
  {pages} {142} (\bibinfo {year} {2006})},\ \Eprint
  {http://arxiv.org/abs/astro-ph/0601463} {arXiv:astro-ph/0601463 [astro-ph]}
  \BibitemShut {NoStop}%
\bibitem [{\citenamefont {Madau}\ and\ \citenamefont
  {Dickinson}(2014)}]{Madau:2014bja}%
  \BibitemOpen
  \bibfield  {author} {\bibinfo {author} {\bibfnamefont {P.}~\bibnamefont
  {Madau}}\ and\ \bibinfo {author} {\bibfnamefont {M.}~\bibnamefont
  {Dickinson}},\ }\href {\doibase 10.1146/annurev-astro-081811-125615}
  {\bibfield  {journal} {\bibinfo  {journal} {Ann. Rev. Astron. Astrophys.}\
  }\textbf {\bibinfo {volume} {52}},\ \bibinfo {pages} {415} (\bibinfo {year}
  {2014})},\ \Eprint {http://arxiv.org/abs/1403.0007} {arXiv:1403.0007
  [astro-ph.CO]} \BibitemShut {NoStop}%
\bibitem [{\citenamefont {Strolger}\ \emph {et~al.}(2015)\citenamefont
  {Strolger}, \citenamefont {Dahlen}, \citenamefont {Rodney}, \citenamefont
  {Graur}, \citenamefont {Riess}, \citenamefont {McCully}, \citenamefont
  {Ravindranath}, \citenamefont {Mobasher},\ and\ \citenamefont
  {Shahady}}]{Strolger:2015kra}%
  \BibitemOpen
  \bibfield  {author} {\bibinfo {author} {\bibfnamefont {L.-G.}\ \bibnamefont
  {Strolger}}, \bibinfo {author} {\bibfnamefont {T.}~\bibnamefont {Dahlen}},
  \bibinfo {author} {\bibfnamefont {S.~A.}\ \bibnamefont {Rodney}}, \bibinfo
  {author} {\bibfnamefont {O.}~\bibnamefont {Graur}}, \bibinfo {author}
  {\bibfnamefont {A.~G.}\ \bibnamefont {Riess}}, \bibinfo {author}
  {\bibfnamefont {C.}~\bibnamefont {McCully}}, \bibinfo {author} {\bibfnamefont
  {S.}~\bibnamefont {Ravindranath}}, \bibinfo {author} {\bibfnamefont
  {B.}~\bibnamefont {Mobasher}}, \ and\ \bibinfo {author} {\bibfnamefont
  {A.~K.}\ \bibnamefont {Shahady}},\ }\href {\doibase
  10.1088/0004-637X/813/2/93} {\bibfield  {journal} {\bibinfo  {journal}
  {Astrophys. J.}\ }\textbf {\bibinfo {volume} {813}},\ \bibinfo {pages} {93}
  (\bibinfo {year} {2015})},\ \Eprint {http://arxiv.org/abs/1509.06574}
  {arXiv:1509.06574 [astro-ph.GA]} \BibitemShut {NoStop}%
\bibitem [{\citenamefont {Mattila}\ \emph {et~al.}(2012)\citenamefont
  {Mattila}, \citenamefont {Dahlen}, \citenamefont {Efstathiou}, \citenamefont
  {Kankare}, \citenamefont {Melinder}, \citenamefont {Alonso-Herrero},
  \citenamefont {Perez-Torres}, \citenamefont {Ryder}, \citenamefont
  {Vaisanen},\ and\ \citenamefont {Ostlin}}]{Mattila:2012zr}%
  \BibitemOpen
  \bibfield  {author} {\bibinfo {author} {\bibfnamefont {S.}~\bibnamefont
  {Mattila}}, \bibinfo {author} {\bibfnamefont {T.}~\bibnamefont {Dahlen}},
  \bibinfo {author} {\bibfnamefont {A.}~\bibnamefont {Efstathiou}}, \bibinfo
  {author} {\bibfnamefont {E.}~\bibnamefont {Kankare}}, \bibinfo {author}
  {\bibfnamefont {J.}~\bibnamefont {Melinder}}, \bibinfo {author}
  {\bibfnamefont {A.}~\bibnamefont {Alonso-Herrero}}, \bibinfo {author}
  {\bibfnamefont {M.}~\bibnamefont {Perez-Torres}}, \bibinfo {author}
  {\bibfnamefont {S.}~\bibnamefont {Ryder}}, \bibinfo {author} {\bibfnamefont
  {P.}~\bibnamefont {Vaisanen}}, \ and\ \bibinfo {author} {\bibfnamefont
  {G.}~\bibnamefont {Ostlin}},\ }\href {\doibase 10.1088/0004-637X/756/2/111}
  {\bibfield  {journal} {\bibinfo  {journal} {Astrophys. J.}\ }\textbf
  {\bibinfo {volume} {756}},\ \bibinfo {pages} {111} (\bibinfo {year}
  {2012})},\ \Eprint {http://arxiv.org/abs/1206.1314} {arXiv:1206.1314
  [astro-ph.CO]} \BibitemShut {NoStop}%
\bibitem [{\citenamefont {Botticella}\ \emph {et~al.}(2012)\citenamefont
  {Botticella}, \citenamefont {Smartt}, \citenamefont {Kennicutt},
  \citenamefont {Cappellaro}, \citenamefont {Sereno},\ and\ \citenamefont
  {Lee}}]{Botticella:2011nd}%
  \BibitemOpen
  \bibfield  {author} {\bibinfo {author} {\bibfnamefont {M.}~\bibnamefont
  {Botticella}}, \bibinfo {author} {\bibfnamefont {S.}~\bibnamefont {Smartt}},
  \bibinfo {author} {\bibfnamefont {J.}~\bibnamefont {Kennicutt}, \bibfnamefont
  {R.C.}}, \bibinfo {author} {\bibfnamefont {E.}~\bibnamefont {Cappellaro}},
  \bibinfo {author} {\bibfnamefont {M.}~\bibnamefont {Sereno}}, \ and\ \bibinfo
  {author} {\bibfnamefont {J.}~\bibnamefont {Lee}},\ }\href {\doibase
  10.1051/0004-6361/201117343} {\bibfield  {journal} {\bibinfo  {journal}
  {Astron. Astrophys.}\ }\textbf {\bibinfo {volume} {537}},\ \bibinfo {pages}
  {A132} (\bibinfo {year} {2012})},\ \Eprint {http://arxiv.org/abs/1111.1692}
  {arXiv:1111.1692 [astro-ph.CO]} \BibitemShut {NoStop}%
\bibitem [{\citenamefont {Taylor}\ \emph {et~al.}(2014)\citenamefont {Taylor}
  \emph {et~al.}}]{Taylor:2014rlo}%
  \BibitemOpen
  \bibfield  {author} {\bibinfo {author} {\bibfnamefont {M.}~\bibnamefont
  {Taylor}} \emph {et~al.},\ }\href {\doibase 10.1088/0004-637X/792/2/135}
  {\bibfield  {journal} {\bibinfo  {journal} {Astrophys. J.}\ }\textbf
  {\bibinfo {volume} {792}},\ \bibinfo {pages} {135} (\bibinfo {year}
  {2014})},\ \Eprint {http://arxiv.org/abs/1407.0999} {arXiv:1407.0999
  [astro-ph.SR]} \BibitemShut {NoStop}%
\bibitem [{\citenamefont {Cappellaro}\ \emph {et~al.}(2015)\citenamefont
  {Cappellaro} \emph {et~al.}}]{Cappellaro:2015qia}%
  \BibitemOpen
  \bibfield  {author} {\bibinfo {author} {\bibfnamefont {E.}~\bibnamefont
  {Cappellaro}} \emph {et~al.},\ }\href {\doibase 10.1051/0004-6361/201526712}
  {\bibfield  {journal} {\bibinfo  {journal} {Astron. Astrophys.}\ }\textbf
  {\bibinfo {volume} {584}},\ \bibinfo {pages} {A62} (\bibinfo {year}
  {2015})},\ \Eprint {http://arxiv.org/abs/1509.04496} {arXiv:1509.04496
  [astro-ph.CO]} \BibitemShut {NoStop}%
\bibitem [{\citenamefont {Yuksel}\ \emph {et~al.}(2008)\citenamefont {Yuksel},
  \citenamefont {Kistler}, \citenamefont {Beacom},\ and\ \citenamefont
  {Hopkins}}]{Yuksel:2008cu}%
  \BibitemOpen
  \bibfield  {author} {\bibinfo {author} {\bibfnamefont {H.}~\bibnamefont
  {Yuksel}}, \bibinfo {author} {\bibfnamefont {M.~D.}\ \bibnamefont {Kistler}},
  \bibinfo {author} {\bibfnamefont {J.~F.}\ \bibnamefont {Beacom}}, \ and\
  \bibinfo {author} {\bibfnamefont {A.~M.}\ \bibnamefont {Hopkins}},\ }\href
  {\doibase 10.1086/591449} {\bibfield  {journal} {\bibinfo  {journal}
  {Astrophys. J. Lett.}\ }\textbf {\bibinfo {volume} {683}},\ \bibinfo {pages}
  {L5} (\bibinfo {year} {2008})},\ \Eprint {http://arxiv.org/abs/0804.4008}
  {arXiv:0804.4008 [astro-ph]} \BibitemShut {NoStop}%
\bibitem [{\citenamefont {Baldry}\ and\ \citenamefont
  {Glazebrook}(2003)}]{Baldry:2003xi}%
  \BibitemOpen
  \bibfield  {author} {\bibinfo {author} {\bibfnamefont {I.~K.}\ \bibnamefont
  {Baldry}}\ and\ \bibinfo {author} {\bibfnamefont {K.}~\bibnamefont
  {Glazebrook}},\ }\href {\doibase 10.1086/376502} {\bibfield  {journal}
  {\bibinfo  {journal} {Astrophys. J.}\ }\textbf {\bibinfo {volume} {593}},\
  \bibinfo {pages} {258} (\bibinfo {year} {2003})},\ \Eprint
  {http://arxiv.org/abs/astro-ph/0304423} {arXiv:astro-ph/0304423} \BibitemShut
  {NoStop}%
\bibitem [{\citenamefont {Summa}\ \emph {et~al.}(2016)\citenamefont {Summa},
  \citenamefont {Hanke}, \citenamefont {Janka}, \citenamefont {Melson},
  \citenamefont {Marek},\ and\ \citenamefont {Müller}}]{Summa:2015nyk}%
  \BibitemOpen
  \bibfield  {author} {\bibinfo {author} {\bibfnamefont {A.}~\bibnamefont
  {Summa}}, \bibinfo {author} {\bibfnamefont {F.}~\bibnamefont {Hanke}},
  \bibinfo {author} {\bibfnamefont {H.-T.}\ \bibnamefont {Janka}}, \bibinfo
  {author} {\bibfnamefont {T.}~\bibnamefont {Melson}}, \bibinfo {author}
  {\bibfnamefont {A.}~\bibnamefont {Marek}}, \ and\ \bibinfo {author}
  {\bibfnamefont {B.}~\bibnamefont {Müller}},\ }\href {\doibase
  10.3847/0004-637X/825/1/6} {\bibfield  {journal} {\bibinfo  {journal}
  {Astrophys. J.}\ }\textbf {\bibinfo {volume} {825}},\ \bibinfo {pages} {6}
  (\bibinfo {year} {2016})},\ \Eprint {http://arxiv.org/abs/1511.07871}
  {arXiv:1511.07871 [astro-ph.SR]} \BibitemShut {NoStop}%
\bibitem [{\citenamefont {Marek}\ \emph {et~al.}(2006)\citenamefont {Marek},
  \citenamefont {Dimmelmeier}, \citenamefont {Janka}, \citenamefont {Muller},\
  and\ \citenamefont {Buras}}]{Marek:2005if}%
  \BibitemOpen
  \bibfield  {author} {\bibinfo {author} {\bibfnamefont {A.}~\bibnamefont
  {Marek}}, \bibinfo {author} {\bibfnamefont {H.}~\bibnamefont {Dimmelmeier}},
  \bibinfo {author} {\bibfnamefont {H.~T.}\ \bibnamefont {Janka}}, \bibinfo
  {author} {\bibfnamefont {E.}~\bibnamefont {Muller}}, \ and\ \bibinfo {author}
  {\bibfnamefont {R.}~\bibnamefont {Buras}},\ }\href {\doibase
  10.1051/0004-6361:20052840} {\bibfield  {journal} {\bibinfo  {journal}
  {Astron. Astrophys.}\ }\textbf {\bibinfo {volume} {445}},\ \bibinfo {pages}
  {273} (\bibinfo {year} {2006})},\ \Eprint
  {http://arxiv.org/abs/astro-ph/0502161} {arXiv:astro-ph/0502161 [astro-ph]}
  \BibitemShut {NoStop}%
\bibitem [{\citenamefont {Buras}\ \emph {et~al.}(2006)\citenamefont {Buras},
  \citenamefont {Rampp}, \citenamefont {Janka},\ and\ \citenamefont
  {Kifonidis}}]{Buras:2005rp}%
  \BibitemOpen
  \bibfield  {author} {\bibinfo {author} {\bibfnamefont {R.}~\bibnamefont
  {Buras}}, \bibinfo {author} {\bibfnamefont {M.}~\bibnamefont {Rampp}},
  \bibinfo {author} {\bibfnamefont {H.~T.}\ \bibnamefont {Janka}}, \ and\
  \bibinfo {author} {\bibfnamefont {K.}~\bibnamefont {Kifonidis}},\ }\href
  {\doibase 10.1051/0004-6361:20053783} {\bibfield  {journal} {\bibinfo
  {journal} {Astron. Astrophys.}\ }\textbf {\bibinfo {volume} {447}},\ \bibinfo
  {pages} {1049} (\bibinfo {year} {2006})},\ \Eprint
  {http://arxiv.org/abs/astro-ph/0507135} {arXiv:astro-ph/0507135 [astro-ph]}
  \BibitemShut {NoStop}%
\bibitem [{\citenamefont {{Heger}}\ and\ \citenamefont
  {{Woosley}}(2010)}]{heger10}%
  \BibitemOpen
  \bibfield  {author} {\bibinfo {author} {\bibfnamefont {A.}~\bibnamefont
  {{Heger}}}\ and\ \bibinfo {author} {\bibfnamefont {S.~E.}\ \bibnamefont
  {{Woosley}}},\ }\href {\doibase 10.1088/0004-637X/724/1/341} {\bibfield
  {journal} {\bibinfo  {journal} {\apj}\ }\textbf {\bibinfo {volume} {724}},\
  \bibinfo {pages} {341} (\bibinfo {year} {2010})},\ \Eprint
  {http://arxiv.org/abs/0803.3161} {arXiv:0803.3161 [astro-ph]} \BibitemShut
  {NoStop}%
\bibitem [{\citenamefont {Nomoto}(1984)}]{Nomoto:1984}%
  \BibitemOpen
  \bibfield  {author} {\bibinfo {author} {\bibfnamefont {K.}~\bibnamefont
  {Nomoto}},\ }\href@noop {} {\bibfield  {journal} {\bibinfo  {journal}
  {Astrophys. J.}\ }\textbf {\bibinfo {volume} {277}},\ \bibinfo {pages} {791}
  (\bibinfo {year} {1984})}\BibitemShut {NoStop}%
\bibitem [{\citenamefont {Nomoto}(1987)}]{Nomoto:1987}%
  \BibitemOpen
  \bibfield  {author} {\bibinfo {author} {\bibfnamefont {K.}~\bibnamefont
  {Nomoto}},\ }\href@noop {} {\bibfield  {journal} {\bibinfo  {journal}
  {Astrophys. J.}\ }\textbf {\bibinfo {volume} {322}},\ \bibinfo {pages} {206}
  (\bibinfo {year} {1987})}\BibitemShut {NoStop}%
\bibitem [{\citenamefont {Nakamura}\ \emph {et~al.}(2019)\citenamefont
  {Nakamura}, \citenamefont {Takiwaki},\ and\ \citenamefont
  {Kotake}}]{Nakamura:2019}%
  \BibitemOpen
  \bibfield  {author} {\bibinfo {author} {\bibfnamefont {K.}~\bibnamefont
  {Nakamura}}, \bibinfo {author} {\bibfnamefont {T.}~\bibnamefont {Takiwaki}},
  \ and\ \bibinfo {author} {\bibfnamefont {K.}~\bibnamefont {Kotake}},\ }\href
  {\doibase 10.1093/pasj/psz080} {\bibfield  {journal} {\bibinfo  {journal}
  {Publications of the Astronomical Society of Japan}\ }\textbf {\bibinfo
  {volume} {71}} (\bibinfo {year} {2019}),\ 10.1093/pasj/psz080}\BibitemShut
  {NoStop}%
\bibitem [{\citenamefont {Kotake}\ \emph {et~al.}(2018)\citenamefont {Kotake},
  \citenamefont {Takiwaki}, \citenamefont {Fischer}, \citenamefont {Nakamura},\
  and\ \citenamefont {{Mart{\'i}nez-Pinedo}}}]{Kotake:2018}%
  \BibitemOpen
  \bibfield  {author} {\bibinfo {author} {\bibfnamefont {K.}~\bibnamefont
  {Kotake}}, \bibinfo {author} {\bibfnamefont {T.}~\bibnamefont {Takiwaki}},
  \bibinfo {author} {\bibfnamefont {T.}~\bibnamefont {Fischer}}, \bibinfo
  {author} {\bibfnamefont {K.}~\bibnamefont {Nakamura}}, \ and\ \bibinfo
  {author} {\bibfnamefont {G.}~\bibnamefont {{Mart{\'i}nez-Pinedo}}},\ }\href
  {\doibase 10.3847/1538-4357/aaa716} {\bibfield  {journal} {\bibinfo
  {journal} {The Astrophysical Journal}\ }\textbf {\bibinfo {volume} {853}},\
  \bibinfo {pages} {170} (\bibinfo {year} {2018})}\BibitemShut {NoStop}%
\bibitem [{\citenamefont {Takiwaki}\ \emph {et~al.}(2016)\citenamefont
  {Takiwaki}, \citenamefont {Kotake},\ and\ \citenamefont
  {Suwa}}]{Takiwaki:2016}%
  \BibitemOpen
  \bibfield  {author} {\bibinfo {author} {\bibfnamefont {T.}~\bibnamefont
  {Takiwaki}}, \bibinfo {author} {\bibfnamefont {K.}~\bibnamefont {Kotake}}, \
  and\ \bibinfo {author} {\bibfnamefont {Y.}~\bibnamefont {Suwa}},\ }\href
  {\doibase 10.1093/mnrasl/slw105} {\bibfield  {journal} {\bibinfo  {journal}
  {Monthly Notices of the Royal Astronomical Society: Letters}\ }\textbf
  {\bibinfo {volume} {461}},\ \bibinfo {pages} {L112} (\bibinfo {year}
  {2016})}\BibitemShut {NoStop}%
\bibitem [{\citenamefont {Zaizen}\ \emph
  {et~al.}(2020{\natexlab{a}})\citenamefont {Zaizen}, \citenamefont {Cherry},
  \citenamefont {Takiwaki}, \citenamefont {Horiuchi}, \citenamefont {Kotake},
  \citenamefont {Umeda},\ and\ \citenamefont {Yoshida}}]{Zaizen:2020}%
  \BibitemOpen
  \bibfield  {author} {\bibinfo {author} {\bibfnamefont {M.}~\bibnamefont
  {Zaizen}}, \bibinfo {author} {\bibfnamefont {J.~F.}\ \bibnamefont {Cherry}},
  \bibinfo {author} {\bibfnamefont {T.}~\bibnamefont {Takiwaki}}, \bibinfo
  {author} {\bibfnamefont {S.}~\bibnamefont {Horiuchi}}, \bibinfo {author}
  {\bibfnamefont {K.}~\bibnamefont {Kotake}}, \bibinfo {author} {\bibfnamefont
  {H.}~\bibnamefont {Umeda}}, \ and\ \bibinfo {author} {\bibfnamefont
  {T.}~\bibnamefont {Yoshida}},\ }\href {\doibase
  10.1088/1475-7516/2020/06/011} {\bibfield  {journal} {\bibinfo  {journal}
  {Journal of Cosmology and Astroparticle Physics}\ }\textbf {\bibinfo {volume}
  {2020}},\ \bibinfo {pages} {011} (\bibinfo {year}
  {2020}{\natexlab{a}})}\BibitemShut {NoStop}%
\bibitem [{\citenamefont {{Sasaki}}\ \emph {et~al.}(2020)\citenamefont
  {{Sasaki}}, \citenamefont {{Takiwaki}}, \citenamefont {{Kawagoe}},
  \citenamefont {{Horiuchi}},\ and\ \citenamefont {{Ishidoshiro}}}]{sasaki20}%
  \BibitemOpen
  \bibfield  {author} {\bibinfo {author} {\bibfnamefont {H.}~\bibnamefont
  {{Sasaki}}}, \bibinfo {author} {\bibfnamefont {T.}~\bibnamefont
  {{Takiwaki}}}, \bibinfo {author} {\bibfnamefont {S.}~\bibnamefont
  {{Kawagoe}}}, \bibinfo {author} {\bibfnamefont {S.}~\bibnamefont
  {{Horiuchi}}}, \ and\ \bibinfo {author} {\bibfnamefont {K.}~\bibnamefont
  {{Ishidoshiro}}},\ }\href {\doibase 10.1103/PhysRevD.101.063027} {\bibfield
  {journal} {\bibinfo  {journal} {\prd}\ }\textbf {\bibinfo {volume} {101}},\
  \bibinfo {eid} {063027} (\bibinfo {year} {2020})},\ \Eprint
  {http://arxiv.org/abs/1907.01002} {arXiv:1907.01002 [astro-ph.HE]}
  \BibitemShut {NoStop}%
\bibitem [{\citenamefont {O'Connor}\ \emph {et~al.}(2018)\citenamefont
  {O'Connor}, \citenamefont {Bollig}, \citenamefont {Burrows}, \citenamefont
  {Couch}, \citenamefont {Fischer}, \citenamefont {Janka}, \citenamefont
  {Kotake}, \citenamefont {Lentz}, \citenamefont {Liebend{\"o}rfer},
  \citenamefont {Messer}, \citenamefont {Mezzacappa}, \citenamefont
  {Takiwaki},\ and\ \citenamefont {Vartanyan}}]{OConnor:2018}%
  \BibitemOpen
  \bibfield  {author} {\bibinfo {author} {\bibfnamefont {E.}~\bibnamefont
  {O'Connor}}, \bibinfo {author} {\bibfnamefont {R.}~\bibnamefont {Bollig}},
  \bibinfo {author} {\bibfnamefont {A.}~\bibnamefont {Burrows}}, \bibinfo
  {author} {\bibfnamefont {S.}~\bibnamefont {Couch}}, \bibinfo {author}
  {\bibfnamefont {T.}~\bibnamefont {Fischer}}, \bibinfo {author} {\bibfnamefont
  {H.-T.}\ \bibnamefont {Janka}}, \bibinfo {author} {\bibfnamefont
  {K.}~\bibnamefont {Kotake}}, \bibinfo {author} {\bibfnamefont {E.~J.}\
  \bibnamefont {Lentz}}, \bibinfo {author} {\bibfnamefont {M.}~\bibnamefont
  {Liebend{\"o}rfer}}, \bibinfo {author} {\bibfnamefont {O.~E.~B.}\
  \bibnamefont {Messer}}, \bibinfo {author} {\bibfnamefont {A.}~\bibnamefont
  {Mezzacappa}}, \bibinfo {author} {\bibfnamefont {T.}~\bibnamefont
  {Takiwaki}}, \ and\ \bibinfo {author} {\bibfnamefont {D.}~\bibnamefont
  {Vartanyan}},\ }\href {\doibase 10.1088/1361-6471/aadeae} {\bibfield
  {journal} {\bibinfo  {journal} {Journal of Physics G: Nuclear and Particle
  Physics}\ }\textbf {\bibinfo {volume} {45}},\ \bibinfo {pages} {104001}
  (\bibinfo {year} {2018})}\BibitemShut {NoStop}%
\bibitem [{\citenamefont {Liebend{\"o}rfer}\ \emph {et~al.}(2009)\citenamefont
  {Liebend{\"o}rfer}, \citenamefont {Whitehouse},\ and\ \citenamefont
  {Fischer}}]{Liebendorfer:2009}%
  \BibitemOpen
  \bibfield  {author} {\bibinfo {author} {\bibfnamefont {M.}~\bibnamefont
  {Liebend{\"o}rfer}}, \bibinfo {author} {\bibfnamefont {S.~C.}\ \bibnamefont
  {Whitehouse}}, \ and\ \bibinfo {author} {\bibfnamefont {T.}~\bibnamefont
  {Fischer}},\ }\href {\doibase 10.1088/0004-637X/698/2/1174} {\bibfield
  {journal} {\bibinfo  {journal} {The Astrophysical Journal}\ }\textbf
  {\bibinfo {volume} {698}},\ \bibinfo {pages} {1174} (\bibinfo {year}
  {2009})}\BibitemShut {NoStop}%
\bibitem [{\citenamefont {Hudepohl}\ \emph {et~al.}(2010)\citenamefont
  {Hudepohl}, \citenamefont {Muller}, \citenamefont {Janka}, \citenamefont
  {Marek},\ and\ \citenamefont {Raffelt}}]{Huedepohl:2009wh}%
  \BibitemOpen
  \bibfield  {author} {\bibinfo {author} {\bibfnamefont {L.}~\bibnamefont
  {Hudepohl}}, \bibinfo {author} {\bibfnamefont {B.}~\bibnamefont {Muller}},
  \bibinfo {author} {\bibfnamefont {H.~T.}\ \bibnamefont {Janka}}, \bibinfo
  {author} {\bibfnamefont {A.}~\bibnamefont {Marek}}, \ and\ \bibinfo {author}
  {\bibfnamefont {G.~G.}\ \bibnamefont {Raffelt}},\ }\href {\doibase
  10.1103/PhysRevLett.104.251101, 10.1103/PhysRevLett.105.249901} {\bibfield
  {journal} {\bibinfo  {journal} {Phys. Rev. Lett.}\ }\textbf {\bibinfo
  {volume} {104}},\ \bibinfo {pages} {251101} (\bibinfo {year} {2010})},\
  \bibinfo {note} {[Erratum: Phys. Rev. Lett.105,249901(2010)]},\ \Eprint
  {http://arxiv.org/abs/0912.0260} {arXiv:0912.0260 [astro-ph.SR]} \BibitemShut
  {NoStop}%
\bibitem [{\citenamefont {Rampp}\ and\ \citenamefont
  {Janka}(2002)}]{Rampp:2002bq}%
  \BibitemOpen
  \bibfield  {author} {\bibinfo {author} {\bibfnamefont {M.}~\bibnamefont
  {Rampp}}\ and\ \bibinfo {author} {\bibfnamefont {H.}~\bibnamefont {Janka}},\
  }\href {\doibase 10.1051/0004-6361:20021398} {\bibfield  {journal} {\bibinfo
  {journal} {Astron. Astrophys.}\ }\textbf {\bibinfo {volume} {396}},\ \bibinfo
  {pages} {361} (\bibinfo {year} {2002})},\ \Eprint
  {http://arxiv.org/abs/astro-ph/0203101} {arXiv:astro-ph/0203101} \BibitemShut
  {NoStop}%
\bibitem [{\citenamefont {O'Connor}\ and\ \citenamefont
  {Ott}(2011)}]{O'Connor:2010tk}%
  \BibitemOpen
  \bibfield  {author} {\bibinfo {author} {\bibfnamefont {E.}~\bibnamefont
  {O'Connor}}\ and\ \bibinfo {author} {\bibfnamefont {C.~D.}\ \bibnamefont
  {Ott}},\ }\href {\doibase 10.1088/0004-637X/730/2/70} {\bibfield  {journal}
  {\bibinfo  {journal} {Astrophys. J.}\ }\textbf {\bibinfo {volume} {730}},\
  \bibinfo {pages} {70} (\bibinfo {year} {2011})},\ \Eprint
  {http://arxiv.org/abs/1010.5550} {arXiv:1010.5550 [astro-ph.HE]} \BibitemShut
  {NoStop}%
\bibitem [{\citenamefont {Horiuchi}\ \emph {et~al.}(2017)\citenamefont
  {Horiuchi}, \citenamefont {Nakamura}, \citenamefont {Takiwaki},\ and\
  \citenamefont {Kotake}}]{Horiuchi:2017qlw}%
  \BibitemOpen
  \bibfield  {author} {\bibinfo {author} {\bibfnamefont {S.}~\bibnamefont
  {Horiuchi}}, \bibinfo {author} {\bibfnamefont {K.}~\bibnamefont {Nakamura}},
  \bibinfo {author} {\bibfnamefont {T.}~\bibnamefont {Takiwaki}}, \ and\
  \bibinfo {author} {\bibfnamefont {K.}~\bibnamefont {Kotake}},\ }\href
  {\doibase 10.1088/1361-6471/aa8f1f} {\bibfield  {journal} {\bibinfo
  {journal} {J. Phys. G}\ }\textbf {\bibinfo {volume} {44}},\ \bibinfo {pages}
  {114001} (\bibinfo {year} {2017})},\ \Eprint
  {http://arxiv.org/abs/1708.08513} {arXiv:1708.08513 [astro-ph.HE]}
  \BibitemShut {NoStop}%
\bibitem [{\citenamefont {Sukhbold}\ and\ \citenamefont
  {Woosley}(2014)}]{Sukhbold:2013yca}%
  \BibitemOpen
  \bibfield  {author} {\bibinfo {author} {\bibfnamefont {T.}~\bibnamefont
  {Sukhbold}}\ and\ \bibinfo {author} {\bibfnamefont {S.}~\bibnamefont
  {Woosley}},\ }\href {\doibase 10.1088/0004-637X/783/1/10} {\bibfield
  {journal} {\bibinfo  {journal} {Astrophys. J.}\ }\textbf {\bibinfo {volume}
  {783}},\ \bibinfo {pages} {10} (\bibinfo {year} {2014})},\ \Eprint
  {http://arxiv.org/abs/1311.6546} {arXiv:1311.6546 [astro-ph.SR]} \BibitemShut
  {NoStop}%
\bibitem [{\citenamefont {Patton}\ and\ \citenamefont
  {Sukhbold}(2020)}]{Patton:2020tiy}%
  \BibitemOpen
  \bibfield  {author} {\bibinfo {author} {\bibfnamefont {R.~A.}\ \bibnamefont
  {Patton}}\ and\ \bibinfo {author} {\bibfnamefont {T.}~\bibnamefont
  {Sukhbold}},\ }\href@noop {} {\  (\bibinfo {year} {2020})},\ \Eprint
  {http://arxiv.org/abs/2005.03055} {arXiv:2005.03055 [astro-ph.SR]}
  \BibitemShut {NoStop}%
\bibitem [{\citenamefont {{Suwa}}\ \emph {et~al.}(2018)\citenamefont {{Suwa}},
  \citenamefont {{Yoshida}}, \citenamefont {{Shibata}}, \citenamefont
  {{Umeda}},\ and\ \citenamefont {{Takahashi}}}]{Suwa:2018}%
  \BibitemOpen
  \bibfield  {author} {\bibinfo {author} {\bibfnamefont {Y.}~\bibnamefont
  {{Suwa}}}, \bibinfo {author} {\bibfnamefont {T.}~\bibnamefont {{Yoshida}}},
  \bibinfo {author} {\bibfnamefont {M.}~\bibnamefont {{Shibata}}}, \bibinfo
  {author} {\bibfnamefont {H.}~\bibnamefont {{Umeda}}}, \ and\ \bibinfo
  {author} {\bibfnamefont {K.}~\bibnamefont {{Takahashi}}},\ }\href {\doibase
  10.1093/mnras/sty2460} {\bibfield  {journal} {\bibinfo  {journal} {Mon. Not.
  Roy. Astron. Soc.}\ }\textbf {\bibinfo {volume} {481}},\ \bibinfo {pages}
  {3305} (\bibinfo {year} {2018})},\ \Eprint {http://arxiv.org/abs/1808.02328}
  {arXiv:1808.02328 [astro-ph.HE]} \BibitemShut {NoStop}%
\bibitem [{\citenamefont {{Schwab}}\ \emph {et~al.}(2016)\citenamefont
  {{Schwab}}, \citenamefont {{Quataert}},\ and\ \citenamefont
  {{Kasen}}}]{2016MNRAS.463.3461S}%
  \BibitemOpen
  \bibfield  {author} {\bibinfo {author} {\bibfnamefont {J.}~\bibnamefont
  {{Schwab}}}, \bibinfo {author} {\bibfnamefont {E.}~\bibnamefont
  {{Quataert}}}, \ and\ \bibinfo {author} {\bibfnamefont {D.}~\bibnamefont
  {{Kasen}}},\ }\href {\doibase 10.1093/mnras/stw2249} {\bibfield  {journal}
  {\bibinfo  {journal} {Mon. Not. Roy. Astron. Soc.}\ }\textbf {\bibinfo
  {volume} {463}},\ \bibinfo {pages} {3461} (\bibinfo {year} {2016})},\ \Eprint
  {http://arxiv.org/abs/1606.02300} {arXiv:1606.02300 [astro-ph.SR]}
  \BibitemShut {NoStop}%
\bibitem [{\citenamefont {Horiuchi}\ \emph {et~al.}(2011)\citenamefont
  {Horiuchi}, \citenamefont {Beacom}, \citenamefont {Kochanek}, \citenamefont
  {Prieto}, \citenamefont {Stanek},\ and\ \citenamefont
  {Thompson}}]{Horiuchi:2011zz}%
  \BibitemOpen
  \bibfield  {author} {\bibinfo {author} {\bibfnamefont {S.}~\bibnamefont
  {Horiuchi}}, \bibinfo {author} {\bibfnamefont {J.~F.}\ \bibnamefont
  {Beacom}}, \bibinfo {author} {\bibfnamefont {C.~S.}\ \bibnamefont
  {Kochanek}}, \bibinfo {author} {\bibfnamefont {J.~L.}\ \bibnamefont
  {Prieto}}, \bibinfo {author} {\bibfnamefont {K.~Z.}\ \bibnamefont {Stanek}},
  \ and\ \bibinfo {author} {\bibfnamefont {T.~A.}\ \bibnamefont {Thompson}},\
  }\href {\doibase 10.1088/0004-637X/738/2/154} {\bibfield  {journal} {\bibinfo
   {journal} {Astrophys. J.}\ }\textbf {\bibinfo {volume} {738}},\ \bibinfo
  {pages} {154} (\bibinfo {year} {2011})},\ \Eprint
  {http://arxiv.org/abs/1102.1977} {arXiv:1102.1977 [astro-ph.CO]} \BibitemShut
  {NoStop}%
\bibitem [{\citenamefont {Dahlen}\ \emph {et~al.}(2012)\citenamefont {Dahlen},
  \citenamefont {Strolger}, \citenamefont {Riess}, \citenamefont {Mattila},
  \citenamefont {Kankare},\ and\ \citenamefont {Mobasher}}]{Dahlen:2012cm}%
  \BibitemOpen
  \bibfield  {author} {\bibinfo {author} {\bibfnamefont {T.}~\bibnamefont
  {Dahlen}}, \bibinfo {author} {\bibfnamefont {L.-G.}\ \bibnamefont
  {Strolger}}, \bibinfo {author} {\bibfnamefont {A.~G.}\ \bibnamefont {Riess}},
  \bibinfo {author} {\bibfnamefont {S.}~\bibnamefont {Mattila}}, \bibinfo
  {author} {\bibfnamefont {E.}~\bibnamefont {Kankare}}, \ and\ \bibinfo
  {author} {\bibfnamefont {B.}~\bibnamefont {Mobasher}},\ }\href {\doibase
  10.1088/0004-637X/757/1/70} {\bibfield  {journal} {\bibinfo  {journal}
  {Astrophys. J.}\ }\textbf {\bibinfo {volume} {757}},\ \bibinfo {pages} {70}
  (\bibinfo {year} {2012})},\ \Eprint {http://arxiv.org/abs/1208.0342}
  {arXiv:1208.0342 [astro-ph.CO]} \BibitemShut {NoStop}%
\bibitem [{\citenamefont {Perley}\ \emph {et~al.}(2020)\citenamefont {Perley}
  \emph {et~al.}}]{Perley:2020ajb}%
  \BibitemOpen
  \bibfield  {author} {\bibinfo {author} {\bibfnamefont {D.~A.}\ \bibnamefont
  {Perley}} \emph {et~al.},\ }\href@noop {} {\  (\bibinfo {year} {2020})},\
  \Eprint {http://arxiv.org/abs/2009.01242} {arXiv:2009.01242 [astro-ph.HE]}
  \BibitemShut {NoStop}%
\bibitem [{\citenamefont {Tamborra}\ and\ \citenamefont
  {Shalgar}(2020)}]{Tamborra:2020cul}%
  \BibitemOpen
  \bibfield  {author} {\bibinfo {author} {\bibfnamefont {I.}~\bibnamefont
  {Tamborra}}\ and\ \bibinfo {author} {\bibfnamefont {S.}~\bibnamefont
  {Shalgar}},\ }\href {\doibase 10.1146/annurev-nucl-102920-050505} {\
  (\bibinfo {year} {2020}),\ 10.1146/annurev-nucl-102920-050505},\ \Eprint
  {http://arxiv.org/abs/2011.01948} {arXiv:2011.01948 [astro-ph.HE]}
  \BibitemShut {NoStop}%
\bibitem [{\citenamefont {Raffelt}\ \emph {et~al.}(2013)\citenamefont
  {Raffelt}, \citenamefont {Sarikas},\ and\ \citenamefont
  {de~Sousa~Seixas}}]{Raffelt:2013rqa}%
  \BibitemOpen
  \bibfield  {author} {\bibinfo {author} {\bibfnamefont {G.}~\bibnamefont
  {Raffelt}}, \bibinfo {author} {\bibfnamefont {S.}~\bibnamefont {Sarikas}}, \
  and\ \bibinfo {author} {\bibfnamefont {D.}~\bibnamefont {de~Sousa~Seixas}},\
  }\href {\doibase 10.1103/PhysRevLett.111.091101} {\bibfield  {journal}
  {\bibinfo  {journal} {Phys. Rev. Lett.}\ }\textbf {\bibinfo {volume} {111}},\
  \bibinfo {pages} {091101} (\bibinfo {year} {2013})},\ \bibinfo {note}
  {[Erratum: Phys.Rev.Lett. 113, 239903 (2014)]},\ \Eprint
  {http://arxiv.org/abs/1305.7140} {arXiv:1305.7140 [hep-ph]} \BibitemShut
  {NoStop}%
\bibitem [{\citenamefont {Abbar}\ \emph {et~al.}(2015)\citenamefont {Abbar},
  \citenamefont {Duan},\ and\ \citenamefont {Shalgar}}]{Abbar:2015mca}%
  \BibitemOpen
  \bibfield  {author} {\bibinfo {author} {\bibfnamefont {S.}~\bibnamefont
  {Abbar}}, \bibinfo {author} {\bibfnamefont {H.}~\bibnamefont {Duan}}, \ and\
  \bibinfo {author} {\bibfnamefont {S.}~\bibnamefont {Shalgar}},\ }\href
  {\doibase 10.1103/PhysRevD.92.065019} {\bibfield  {journal} {\bibinfo
  {journal} {Phys. Rev. D}\ }\textbf {\bibinfo {volume} {92}},\ \bibinfo
  {pages} {065019} (\bibinfo {year} {2015})},\ \Eprint
  {http://arxiv.org/abs/1507.08992} {arXiv:1507.08992 [hep-ph]} \BibitemShut
  {NoStop}%
\bibitem [{\citenamefont {Chakraborty}\ \emph {et~al.}(2014)\citenamefont
  {Chakraborty}, \citenamefont {Mirizzi}, \citenamefont {Saviano},\ and\
  \citenamefont {Seixas}}]{Chakraborty:2014nma}%
  \BibitemOpen
  \bibfield  {author} {\bibinfo {author} {\bibfnamefont {S.}~\bibnamefont
  {Chakraborty}}, \bibinfo {author} {\bibfnamefont {A.}~\bibnamefont
  {Mirizzi}}, \bibinfo {author} {\bibfnamefont {N.}~\bibnamefont {Saviano}}, \
  and\ \bibinfo {author} {\bibfnamefont {D.~d.~S.}\ \bibnamefont {Seixas}},\
  }\href {\doibase 10.1103/PhysRevD.89.093001} {\bibfield  {journal} {\bibinfo
  {journal} {Phys. Rev. D}\ }\textbf {\bibinfo {volume} {89}},\ \bibinfo
  {pages} {093001} (\bibinfo {year} {2014})},\ \Eprint
  {http://arxiv.org/abs/1402.1767} {arXiv:1402.1767 [hep-ph]} \BibitemShut
  {NoStop}%
\bibitem [{\citenamefont {Mirizzi}\ \emph {et~al.}(2015)\citenamefont
  {Mirizzi}, \citenamefont {Mangano},\ and\ \citenamefont
  {Saviano}}]{Mirizzi:2015fva}%
  \BibitemOpen
  \bibfield  {author} {\bibinfo {author} {\bibfnamefont {A.}~\bibnamefont
  {Mirizzi}}, \bibinfo {author} {\bibfnamefont {G.}~\bibnamefont {Mangano}}, \
  and\ \bibinfo {author} {\bibfnamefont {N.}~\bibnamefont {Saviano}},\ }\href
  {\doibase 10.1103/PhysRevD.92.021702} {\bibfield  {journal} {\bibinfo
  {journal} {Phys. Rev. D}\ }\textbf {\bibinfo {volume} {92}},\ \bibinfo
  {pages} {021702} (\bibinfo {year} {2015})},\ \Eprint
  {http://arxiv.org/abs/1503.03485} {arXiv:1503.03485 [hep-ph]} \BibitemShut
  {NoStop}%
\bibitem [{\citenamefont {Zaizen}\ \emph
  {et~al.}(2020{\natexlab{b}})\citenamefont {Zaizen}, \citenamefont {Horiuchi},
  \citenamefont {Takiwaki}, \citenamefont {Kotake}, \citenamefont {Yoshida},
  \citenamefont {Umeda},\ and\ \citenamefont {Cherry}}]{Zaizen:2020xum}%
  \BibitemOpen
  \bibfield  {author} {\bibinfo {author} {\bibfnamefont {M.}~\bibnamefont
  {Zaizen}}, \bibinfo {author} {\bibfnamefont {S.}~\bibnamefont {Horiuchi}},
  \bibinfo {author} {\bibfnamefont {T.}~\bibnamefont {Takiwaki}}, \bibinfo
  {author} {\bibfnamefont {K.}~\bibnamefont {Kotake}}, \bibinfo {author}
  {\bibfnamefont {T.}~\bibnamefont {Yoshida}}, \bibinfo {author} {\bibfnamefont
  {H.}~\bibnamefont {Umeda}}, \ and\ \bibinfo {author} {\bibfnamefont {J.~F.}\
  \bibnamefont {Cherry}},\ }\href@noop {} {\  (\bibinfo {year}
  {2020}{\natexlab{b}})},\ \Eprint {http://arxiv.org/abs/2011.09635}
  {arXiv:2011.09635 [astro-ph.HE]} \BibitemShut {NoStop}%
\bibitem [{\citenamefont {Sawyer}(2005)}]{Sawyer:2005jk}%
  \BibitemOpen
  \bibfield  {author} {\bibinfo {author} {\bibfnamefont {R.}~\bibnamefont
  {Sawyer}},\ }\href {\doibase 10.1103/PhysRevD.72.045003} {\bibfield
  {journal} {\bibinfo  {journal} {Phys. Rev. D}\ }\textbf {\bibinfo {volume}
  {72}},\ \bibinfo {pages} {045003} (\bibinfo {year} {2005})},\ \Eprint
  {http://arxiv.org/abs/hep-ph/0503013} {arXiv:hep-ph/0503013} \BibitemShut
  {NoStop}%
\bibitem [{\citenamefont {Sawyer}(2009)}]{Sawyer:2008zs}%
  \BibitemOpen
  \bibfield  {author} {\bibinfo {author} {\bibfnamefont {R.}~\bibnamefont
  {Sawyer}},\ }\href {\doibase 10.1103/PhysRevD.79.105003} {\bibfield
  {journal} {\bibinfo  {journal} {Phys. Rev. D}\ }\textbf {\bibinfo {volume}
  {79}},\ \bibinfo {pages} {105003} (\bibinfo {year} {2009})},\ \Eprint
  {http://arxiv.org/abs/0803.4319} {arXiv:0803.4319 [astro-ph]} \BibitemShut
  {NoStop}%
\bibitem [{\citenamefont {{Yu}}\ and\ \citenamefont
  {{Jeffery}}(2015)}]{Yu2015}%
  \BibitemOpen
  \bibfield  {author} {\bibinfo {author} {\bibfnamefont {S.}~\bibnamefont
  {{Yu}}}\ and\ \bibinfo {author} {\bibfnamefont {C.~S.}\ \bibnamefont
  {{Jeffery}}},\ }\href {\doibase 10.1093/mnras/stv059} {\bibfield  {journal}
  {\bibinfo  {journal} {Mon. Not. Roy. Astron. Soc.}\ }\textbf {\bibinfo
  {volume} {448}},\ \bibinfo {pages} {1078} (\bibinfo {year} {2015})},\ \Eprint
  {http://arxiv.org/abs/1501.02314} {arXiv:1501.02314 [astro-ph.SR]}
  \BibitemShut {NoStop}%
\bibitem [{\citenamefont {{Ablimit}}\ and\ \citenamefont
  {{Maeda}}(2018)}]{Ablimit2018}%
  \BibitemOpen
  \bibfield  {author} {\bibinfo {author} {\bibfnamefont {I.}~\bibnamefont
  {{Ablimit}}}\ and\ \bibinfo {author} {\bibfnamefont {K.}~\bibnamefont
  {{Maeda}}},\ }\href {\doibase 10.3847/1538-4357/aae378} {\bibfield  {journal}
  {\bibinfo  {journal} {\apj}\ }\textbf {\bibinfo {volume} {866}},\ \bibinfo
  {eid} {151} (\bibinfo {year} {2018})},\ \Eprint
  {http://arxiv.org/abs/1710.05504} {arXiv:1710.05504 [astro-ph.HE]}
  \BibitemShut {NoStop}%
\end{thebibliography}%

\end{document}